\documentclass[default]{aastex701}

\newcommand{\m}{\rm\thinspace m}
\newcommand{\cm}{\rm\thinspace cm}

\newcommand{\s}{\rm\thinspace s}
\newcommand{\ks}{\rm\thinspace ks}

\newcommand{\K}{\rm\thinspace K}

\newcommand{\g}{\rm\thinspace g}
\newcommand{\gpcm}{\hbox{$\g\cm^{-3}\,$}}

\newcommand{\Msun}{\hbox{$\rm\thinspace M_{\odot}$}}

\newcommand{\keV}{\rm\thinspace keV}
\newcommand{\eV}{\rm\thinspace eV}

\newcommand{\W}{\rm\thinspace W}

\newcommand{\cts}{\rm\thinspace ct}
\newcommand{\ctsps}{\hbox{$\cts\s^{-1}\,$}}

\newcommand{\rg}{\rm\thinspace $r_\mathrm{g}$}

\begin{document}


\title{Time-resolved XRISM spectroscopy reveals the evolution and structure of the corona in MCG--6-30-15}

\author[orcid=0000-0002-4794-5998,gname='Daniel R.',sname='Wilkins']{D.~R.~Wilkins}
\affiliation{Department of Astronomy, The Ohio State University, 4055 McPherson Laboratory, 140 W 18th Ave, Columbus, OH 43210, USA}
\affiliation{Kavli Institute for Particle Astrophysics and Cosmology, Stanford University, 452 Lomita Mall, Stanford, CA 94305, USA}
\email[show]{wilkins.401@osu.edu}

\author[0000-0003-2663-1954,gname='Laura W.',sname='Brenneman']{L.~W.~Brenneman}
\affiliation{Center for Astrophysics $|$ Harvard~\&~Smithsonian, 60 Garden St., Cambridge, MA 02138, USA}
\email[]{lbrenneman@cfa.harvard.edu}

\author[0000-0003-4504-2557,gname='Anna', sname='Ogorzalek']{A.~Ogorza{\l}ek}
\affiliation{Department of Astronomy, University of Maryland, College Park, MD 20742, USA} 
\affiliation{NASA Goddard Space Flight Center, Greenbelt, MD 20771, USA}
\affiliation{Center for Research and Exploration in Space Science and Technology, NASA/GSFC (CRESST II), Greenbelt, MD 20771, USA}
\email[]{ogoann@umd.edu}

\author[0000-0002-9378-4072,gname='Andrew C.',sname='Fabian']{A.~C.~Fabian}
\affiliation{Institute of Astronomy, Cambridge University, Madingley Road, Cambridge CB3 0HA, UK}
\email[]{acf@ast.cam.ac.uk}


\author[orcid=0000-0001-9735-4873,gname='Ehud',sname='Behar']{E.~Behar}
\affiliation{Department of Physics, Technion, Technion City, Haifa 3200003, Israel}
\email[]{behar@physics.technion.ac.il}

\author[orcid=0000-0003-2704-599X,gname=‘Rozenn’,sname=‘Boissay-Malaquin’]{R.~Boissay-Malaquin}
\affiliation{Center for Space Science and Technology, University of Maryland, Baltimore County (UMBC), Baltimore, MD 21250, USA}
\affiliation{NASA Goddard Space Flight Center, Greenbelt, MD 20771, USA}
\email[]{rozennbm@umbc.edu}

\author[orcid=0000-0003-3828-2448,gname=‘Javier A.',sname=‘Garc\'ia’]{J.~A.~Garc\'ia}
\affiliation{NASA Goddard Space Flight Center, Greenbelt, MD 20771, USA}
\email[]{javier.a.garciamartinez@nasa.gov}

\author[orcid=0000-0003-3496-8928,gname='Erika B.',sname='Hoffman']{E.~B.~Hoffman}
\affiliation{Department of Astronomy, University of Maryland, College Park, 20742, USA}
\email[]{ebhoff@umd.edu}

\author[orcid=0000-0002-7292-6852,gname='Anna',sname='Jur\'{a}\v{n}ov\'{a}']{A.~Jur\'{a}\v{n}ov\'{a}}
\affiliation{MIT Kavli Institute for Astrophysics and Space Research, Massachusetts Institute of Technology, Cambridge, MA 02139, USA}
\email[]{ajuran@mit.edu}

\author[orcid=0000-0002-0786-7307,gname='Daniele',sname='Rogantini']{D.~Rogantini}
\affiliation{Department of Astronomy and Astrophysics, University of Chicago, 5640 S Ellis Ave, Chicago, IL 60637, USA}
\email[]{danieler@uchicago.edu}

\setcounter{footnote}{0}

\collaboration{all}{on behalf of the XRISM Collaboration}

\begin{abstract}

We present a time-resolved analysis of high-resolution spectra of the AGN MCG--6-30-15 obtained by \textit{XRISM} alongside broadband spectra from \textit{NuSTAR} and \textit{XMM-Newton} during a coordinated observing campaign in February 2024. These observations provide some of the most detailed measurements of X-ray reflection from the innermost regions of the accretion disc around a supermassive black hole, and its evolution during periods of significant variability. We find that both the X-ray spectrum and its variability can be described by a self-consistent model of the reflection of the coronal X-ray emission from the accretion disc around a rapidly-spinning ($a_\star > 0.93$) black hole, in which the observed variability arises from underlying changes in the luminosity, spatial extent and motion of the corona. While the corona is compact, residing within 10\rg\ of the black hole for the majority of the observations, finite spatial extent is required to fully explain the shape of the reflection spectrum. A flare was observed in the X-ray emission during which the corona expanded to around 15\rg\ and was accelerated away from the black hole reaching a velocity of $0.27c$. Around the flare were short dips in the observed flux, during which the corona was found to have collapsed to a confined region, within 2.5\rg\ of the black hole, enhancing the relativistic effects observed from the inner accretion disc. We find it is necessary to account for such significant spectral variation in order to obtain accurate measurements of the spin of the black hole via X-ray reflection spectroscopy.

\end{abstract}



\section{Introduction} 
X-ray reflection from the innermost regions of the accretion disc provides us with a unique view of the extreme environment just outside the event horizons of supermassive black holes, and the mechanisms by which the accretion of matter onto the supermassive black holes in the centres of galaxies powers some of the most luminous systems we observe in the Universe: active galactic nuclei (AGN).

X-ray emission is produced in a corona of accelerated particles close to the black hole, however much remains unknown about the exact location and structure of the corona and the mechanisms by which energy is injected into it from either the accretion flow or from the rotational energy of the spinning black hole. Components of the corona may include a compact emission region within the magnetosphere of the black hole \citep[\textit{e.g.}][]{yuan_fluxtubes_2}, perhaps at the base of a jet, or at the site of a failed jet \citep{ghisellini+04}. Such a scenario can be modelled by a simplified point source of emission, sometimes referred to as a `lamppost' \citep{matt+91}. The corona may also be comprised of an extended particle acceleration region over the surface of the inner accretion disc \citep{haardt+91}.

X-rays emitted from the corona illuminate the inner regions of the accretion disc and are reprocessed, producing a characteristic reflection spectrum \citep{george_fabian,ross_fabian}. This reflection spectrum contains a number of fluorescent emission lines, most notably the iron K lines appearing in the spectrum around 6.4\keV. These emission lines are broadened by the combination of Doppler shifts, due to the orbital motion of the material in the accretion disc, and gravitational redshifts where photons are emitted in the strong gravitational field just outside the black hole, producing a distinctive line profile with a blueshifted peak and extended redshifted wing comprising the emission from the innermost regions of the disc \citep{fabian+89,martocchia_matt}. The shape of the relativistically broadened emission lines, in particular the iron K$\alpha$ line, encodes a wealth of information, including the spin of the black hole \citep{brenneman_reynolds, reynolds_spin_review} via the extremal redshift, with the accretion discs around more rapidly spinning black holes extending closer to the event horizon, producing a broader line. Also encoded is the geometry of the inner accretion flow and the corona, via the \textit{emissivity profile}, the relative flux of the reflected emission received from each radius on the disc \citep{1h0707_emis_paper, understanding_emis_paper}.

The \textit{X-ray Imaging and Spectroscopy Mission}, hereafter \textit{XRISM} \citep{xrism}, launched September 6 2023, enables revolutionary, high-resolution X-ray spectroscopy in the iron K band. \textit{XRISM} carries the \textit{Resolve} microcalorimeter spectrometer \citep{xrism_resolve}, providing imaging spectroscopy with resolution around 4.5\eV\ at 6\keV\ ($R = \lambda / \Delta\lambda \sim 1000$). This represents a significant advancement over CCD imaging spectroscopy with energy resolution around 120\eV\ ($R\sim 50$) in this band. \textit{Resolve} also provides significant advances over the \textit{High Energy Transmission Grating} (HETG) on board \textit{Chandra} with significantly greater effective area, increased bandpass (up to 12\keV, while HETG provides collecting area only up to $8\sim9$\keV\ in the brightest sources), and increased resolution (HETG provides a resolution of only 27\eV\ at 6\keV).

Accurate measurements of the properties of black holes and the structure and geometry of the corona and inner accretion flow rely not just on the ability to measure the shape of relativistically broadened emission lines, but also the ability to separate the broad components of the lines from the underlying continuum and from narrow emission and absorption lines in the same energy band. These may include narrow emission components of the iron K fluorescent lines, produced by X-ray reflection from more distant, cold/low-ionisation material (\textit{e.g.} the putative torus or clouds within the optical broad line region or BLR), and absorption lines from ionised species, including Fe\,\textsc{xxv} and Fe\,\textsc{xxvi}, imprinted upon the spectrum by a range of outflows, ranging from moderately ionised and relatively low velocity warm absorbers \citep{halpern-84,blustin+2005} to highly ionised ultrafast outflows (UFOs) launched from the inner accretion flow at velocities up to 30 per cent of the speed of light \citep{tombesi+2010,laha+2021}. The \textit{XRISM Resolve} spectrometer can readily resolve these absorption and emission line components \citep{ngc4151_xrism,xiang+2025,xu+2025,mehdipour+2025} and separate them from the continuum and relativistic broad line seen from the inner accretion disc \citep{brenneman+2025}.

The detection of a relativistically broadened iron K line from the inner accretion disc was first reported by \citet{tanaka+95} in observations of the AGN MCG--6-30-15 using \textit{ASCA}. MCG--6-30-15, hereafter MCG--6, is a nearby Seyfert 1.2 galaxy at redshift $z = 0.007749$. Since then, MCG--6 has been one of the most extensively studied AGN in the X-ray band, with extensive datasets having been collected by \textit{BeppoSAX}, \textit{RXTE}, \textit{Chandra}, \textit{XMM-Newton}, \textit{Suzaku} and \textit{NuSTAR}. Each of these observations has reported a broad iron K$\alpha$ emission feature consistent with a relativistically broadened line in the reflection spectrum from the inner accretion disc \citep{fabian+2002,reynolds+2004,marinucci+2014}. In addition, the X-ray spectrum of MCG--6 exhibits absorption lines from a complex, multi-zone system of outflowing winds \citep{reynolds+1995,young+2005,holczer+2010}, which must be properly modelled in order to accurately derive the parameters of the black hole and inner accretion flow from the relativistic broad line and reflection spectrum.

\textit{XRISM} observed MCG--6 in February 2024 as part of a deep campaign coordinated with \textit{NuSTAR}, \textit{XMM-Newton}, \textit{Chandra}, \textit{NICER} and the \textit{Neil Gehrels Swift Observatory}. The initial results from this campaign were presented by \citet{brenneman+2025}. The \textit{XRISM Resolve} spectrometer reveals a wealth of spectral features in the iron K band, including narrow components of the iron K$\alpha$ and K$\beta$ emission lines, Fe\,\textsc{xxv} and Fe\,\textsc{xxvi} absorption lines from two highly-ionised outflowing winds, and corresponding emission features from these winds. Beneath these narrow spectral lines, reflection from the inner regions of the accretion disc and relativistically broadened iron K$\alpha$ line are significantly detected in both the \textit{Resolve} spectrum, and the broadband spectrum obtained when \textit{NuSTAR} and \textit{XMM-Newton} data are combined with the \textit{XRISM} data.

Like many Seyfert 1 AGN, in particular the narrow line Seyfert 1 (NLS1) AGN \citep{gallo_nls1}, MCG--6 displays extreme variability in its X-ray emission, with flux variations by factors of a few on timescales of just hours and X-ray flares emitted from close proximity to the black hole \citep{iwasawa+99}. This variability can be traced to changes in the location of the corona close to the black hole, with strong light bending enhancing the reflection from the inner accretion disk during periods of lower flux as the corona moves closer to the black hole \citep{miniutti+03,miniutti+04}. Measurements of the X-ray variability and timing properties provide powerful constraints alongside measurements of the time-averaged spectrum. X-ray reverberation \citep{reverb_review} has been detected in MCG--6 \citep{kara_mcg6, wavelet_paper} where variations in energy bands dominated by reflection from the accretion disc lag behind the correlated variations in bands dominated by directly observed continuum emission from the corona, confirming the detection of X-ray reflection from the inner disc and further constraining the geometry of the accretion flow via the light travel time between the corona and inner accretion disc. Measurements of the variability allow different spectral components to be separated based upon the timescales upon which they vary \citep[\textit{e.g.}][]{parker_pca}, and any model used to explain or interpret the data must not only describe the shape of the spectrum, but must provide a physically consistent picture of the parameters that vary on different timescales.

Following the time-averaged analysis of the data from the \textit{XRISM} campaign presented by \citet{brenneman+2025}, in this paper, we study the variability of the X-ray spectrum of MCG--6 that was observed during the 2024 \textit{XRISM} campaign. By modelling the variation of the spectrum over time, we are able to measure how the reflection from the inner accretion disc varies in order to understand the time evolution of the corona that gives rise to the X-ray variability that we observe. In addition to the reflection from the accretion disc, variability was also observed in the absorption lines imprinted by the multi-component outflows in MCG--6, which will be discussed in a subsequent paper.

\section{Observations and data reduction}
MCG--6-30-15 was observed by \textit{XRISM}, \textit{NuSTAR} and \textit{XMM-Newton} during a coordinated campaign between February 5 and February 8, 2024. The observations are summarised in Table~\ref{tab:obs}. Spectral analysis was performed using \textsc{xspec} version 12.15.0c \citep{xspec}, fitting models to the data by minimising the modified version of the Cash statistic \citep{cash}, often referred as the $C$-statistic\footnote{\url{https://heasarc.gsfc.nasa.gov/docs/xanadu/xspec/manual/node336.html\#AppendixStatistics}}.

\begin{deluxetable*}{llllllll}
\tablewidth{0pt}
\tablecaption{Details of the coordinated \textit{XRISM}, \textit{NuSTAR} and \textit{XMM-Newton} observation obtained of MCG--6-30-15 in February 2024 that are used in this analysis. For each instrument, the count rate and signal-to-noise ratio refers to that over the full bandpass of the instrument.\label{tab:obs}}
\tablehead{
\colhead{Observatory} & \colhead{OBSID} & \colhead{Start Time (UTC)} & \colhead{Instrument} & \colhead{Energy Band} & \colhead{Net Exposure} & \colhead{Count Rate} & \colhead{S/N}
}
\startdata
XRISM & 000161000 & 2024-02-05 23:13 & Resolve & $2-12$\keV & 142\ks & 0.704\ctsps & 28 \\
& & & Xtend & $0.3-12$\keV & 112\ks & 6.562\ctsps & 88 \\
NuSTAR & 60902004002 & 2024-02-06 00:31 & FPMA & $3-55$\keV & 101\ks & 1.193\ctsps & 39 \\
& & & FPMB & $3-55$\keV & 99.6\ks & 1.101\ctsps & 41 \\
XMM-Newton & 0921420101 & 2024-02-06 00:56 & EPIC pn & $0.5-10$\keV & 44.7\ks & 27.24\ctsps & 699 \\
\enddata
\end{deluxetable*}

\subsection{XRISM}
\textit{XRISM} observed MCG--6 simultaneously with the \textit{Resolve} microcalorimeter spectrometer and the \textit{Xtend} CCD imaging spectrometer with a total exposure time of 213\ks\ once Earth occultations are taken into account. The data from both \textit{Resolve} and \textit{Xtend} were reduced using the first public release of the \textit{XRISM} mission-specific tools in \textsc{heasoft} version 6.34, using version 11 of the \textit{XRISM} calibration database, or \textsc{caldb}. Data reduction followed the steps outlined in the \textit{XRISM ABC Guide}\footnote{\url{https://heasarc.gsfc.nasa.gov/docs/xrism/analysis/abc guide/Contents.html}}.

The \textit{Resolve} observations were conducted with no filter. Screening of the event list was performed to ensure pulse shape validity and to avoid events for which the pulse rise was coincident with the frame time. Time filtering was also performed to exclude events recorded during passages of the South Atlantic Anomaly (SAA), periods where observations pass through the Earth's sunlit limb, and the recycling intervals of the instrument's adiabatic demagnetisation refrigerator (ADR).

The count rate recorded by the \textit{Resolve} pixel array from MCG--6 is sufficiently low that 95 per cent of events were classified as high spectral resolution (\textit{i.e.} `Hp' events) where pulse profiles from each pixel do not overlap one another in time and can be cleanly separated, producing the highest resolution photon energy measurements. We therefore restrict our analysis to just the high resolution events. The $^{55}$Fe calibration pixel shows a FWHM of $4.50\pm0.03$\eV\ for the Mn K$\alpha$ emission line, with a gain uncertainty less than 1\eV\ across the bandpass used in our analysis. The \textit{Resolve} spectra cover the 1.7-12\keV\ energy range. Data below 1.7\keV\ were not available from \textit{Resolve} as these observations were conducted with the gate valve closed.

Spectra and light curves were extracted using \textsc{xselect} from the entire $6\times 6$ pixel array, excluding pixel 12 (the calibration pixel), and pixel 27, which shows anomalous gain variations. The response matrix (RMF) and ancillary response, encoding the effective area as a function of energy (ARF) were generated using the tools \textsc{rslmkrmf} and \textsc{xaarfgen} respectively, assuming a point-like source for calculation of the point spread function (PSF) and effective area as a function of energy in the ARF. Time-resolved spectral analysis was performed by further filtering the time ranges from which spectra were extracted in \textsc{xselect} and analysed using the time-averaged RMF and ARF response matrices computed for the entire observation.

Since the PSF fills the \textit{Resolve} $6\times 6$ pixel array, it is not possible to extract and subtract a background spectrum from a region of the field of view with no source photons. It is therefore necessary to incorporate the instrumental non-X-ray background (NXB) into the spectral model that is used to analyse the data. A background model was constructed from the \textit{XRISM Resolve} \textsc{nxb database} v1.0 \footnote{\url{https://heasarc.gsfc.nasa.gov/docs/xrism/analysis/nxb/index.html}} for the observation in question, and was added to the spectral model through a diagonal response matrix, following the standard procedure in \textsc{xspec}. The NXB model includes the background continuum via a power law, in addition to instrumental emission lines from Al, Cr, Mn, Fe, Ni, and Cu, each of which is included with a Gaussian profile. The continuum level of the background is a factor of 100 below the emission from MCG--6 in these observations in the centre of the bandpass and only the strongest instrumental lines are detectable in the spectrum, though with the high spectral resolution, these are cleanly separated from the source spectrum. We find that including the NXB in the model decreases the best-fitting $C$-statistic by 28 (with no additional free parameters when the background components are frozen at the levels computed using the \textsc{nxb database}), however we find that including the NXB model does not significantly alter the values inferred for the source parameters.

We fit models to the \textit{XRISM Resolve} at the native instrumental energy binning of 0.5\eV\ per channel. Since the background spectrum is included in the model and not subtracted, the $C$-statistic represents the true Poisson likelihood and can be applied to finely-binned data even when the number of counts per bin is small, so as to maintain the full spectral resolution of the instrument.

Due to uncertainties in the cross-calibration with other instruments, including \textit{Resolve} and \textit{XMM-Newton} EPIC pn, we do not use spectral data from the \textit{XRISM Xtend} CCDs in this analysis, however we do examine the light curve recorded by \textit{Xtend} to assess the variability of the X-ray flux over the 0.3-12\keV\ band. The \textit{Xtend} CCDs were operated in 1/8 window mode to observe MCG--6 in order to avoid photon pile-up. Source photons were extracted from a rectangular region on the strip of the CCD that is read out in the window mode, 5' long and 3.65' wide, centred on the point source and binned by arrival time using \textsc{xselect} to create the light curves. A background region of the same size was selected as far as possible from the point source to create background light curves that can be subtracted from the source light curves.

\subsection{NuSTAR}
The \textit{NuSTAR} observations were reduced using the \textit{NuSTAR} data-analysis system, \textsc{nustardas}, v2.1.4. The event lists from each of the focal plane module (FPM) detectors were reprocessed and filtered using the \textsc{nupipeline} task, applying the most recent calibration available at the time of writing. We extracted the source photons from a circular region, 90\,arcsec in diameter, centred on the point source. Since MCG--6 is a nearby, bright X-ray source, we find that this large extraction region provides ample signal-to-noise above the instrumental background. The background was extracted from a region, the same size, away from the point source, on each detector. Source and background spectra, along with their corresponding RMF and ARF were extracted using the \textsc{nuproducts} tasks. In addition, \textsc{nuproducts} was used to extract source and background light curves with all appropriate dead time and exposure corrections applied.

The modified $C$-statistic that is employed when an on-chip background spectrum is subtracted requires that there be at least one count per spectral bin, and is a more reliable approximation of the Poisson likelihood when the source counts are maximised. In order to maximise the counts per bin, we therefore sum the FPMA and FPMB spectra to produce a single \textit{NuSTAR} spectrum from each time interval (after having inspected the FPMA and FPMB spectra to ensure their consistency), and use response matrices averaged between the two detectors. We then apply the optimal binning algorithm of \citet{kaastra_bleeker} using the tool \textsc{ftgrouppha} to obtain the optimal trade-off between maintaining spectral resolution and maximising the number of counts per bin.

\subsection{XMM-Newton}
We include in our analysis data collected by the EPIC pn camera on board \textit{XMM-Newton}, due to the instrument's superior sensitivity, particularly when analysing the variability of the X-ray emission. During the observations, the pn camera was operated in small window mode so as not to be impacted by photon pile-up from a source as bright as MCG--6. The \textit{XMM-Newton} observations were reduced using the \textsc{xmm science analysis system (sas)} v21.0.0. The event lists were reprocessed using the \textsc{epproc} task, applying the latest available version of the calibration. Source photons were extracted from a circular region, 35\,arcsec in diameter, and the background was extracted from a circular region of the same size, located on the same detector chip. The source and background spectra were extracted using the \textsc{evselect} task and the corresponding response and ancillary response matrices were generated using \textsc{rmfgen} and \textsc{arfgen}. \textit{XMM-Newton} spectra were binned using the same optimal binning algorithm applied to the \textit{NuSTAR} spectra. Light curves were extracted from the \textit{XMM-Newton} observations, also using \textsc{evselect}, and were corrected to account for dead time and exposure variations using the \textsc{epiclccorr} task.

\section{The time-averaged spectrum}
\citet{brenneman+2025} present analysis of the time-averaged spectrum including the high-resolution data from \textit{Resolve} over the 2-11\keV\ energy range, and broadband data including spectra from \textit{XMM-Newton} EPIC pn (0.3-12\keV) and \textit{NuSTAR} (3-55\keV). We here summarise the results of this analysis that will be the basis for the spectral model adopted here.

The full, broadband spectra from the coordinated \textit{XRISM}, \textit{NuSTAR} and \textit{XMM-Newton} campaign are well-described by a model in which the primary X-ray continuum is produced by a compact corona close to the black hole. The primary X-ray continuum illuminates the inner accretion disc, producing the characteristic reflection spectrum, including a relativistically broadened iron K line, detected both in the high-resolution \textit{XRISM Resolve} spectrum and the lower-resolution \textit{NuSTAR} and \textit{XMM-Newton} spectra, in addition to the Compton hump, which is significantly detected by \textit{NuSTAR}. Both the primary X-ray continuum and the reflection from the accretion disc are modelled using \textsc{relxilllpCp} version 2.3 \citep{garcia+2014}, which includes the continuum produced by the Comptonisation of thermal seed photons in the corona, the reprocessing of the X-ray continuum by the accretion disc plasma using \textsc{xillver} \citep{garcia+2013}, and the relativistic broadening of the line including both Doppler shifts and gravitational redshifts via the \textsc{relline} kernel \citep{dauser+10}. In the `lp' variant of the model, the emissivity profile of the accretion disc (that is the variation of reflected flux as a function of radius, which governs the shape of the redshifted wing of the broad emission lines) is calculated under the simplifying assumption that the disc is illuminated by a point source (also known as a `lamppost'). Such a model reduces the geometry of the corona to just one free parameter: the height, $h$, of the corona above the black hole which can be interpreted as a proxy for the compactness of the corona and its proximity to the event horizon.\footnote{The height or extent of the corona is measured in units of the gravitational radius of the black hole: $r_\mathrm{g} = GM/c^2$ for the mass, $M$, of the black hole in question.} We shall return to the geometry of the corona and the emissivity profile of the accretion disc in \S\ref{sec:emis}.

From analysis of the full, broadband spectrum, it is clear that relativistic reflection from the inner regions of the accretion disc is necessary to simultaneously model the entire broadband spectrum from 0.3 to 55\keV\ and the high-resolution \textit{XRISM Resolve} spectra in the 2-12\keV\ band, including the broad features within the iron K band and the Compton hump. The detection of reflection from the inner accretion disc is further supported by the detection of reverberation time lags in MCG--6 \citep{kara_mcg6, wavelet_paper}.

In addition to the relativistic reflection spectrum and the broad component of the iron K line, the spectral model includes narrow components of the iron K$\alpha$ and K$\beta$ lines produced by the reflection or reprocessing of the X-ray continuum by more distant material, both of which are detected in the \textit{Resolve} spectrum. The narrow components of emission lines throughout the broadband X-ray spectrum are included via the line component of the \textsc{MYTorus} model \citep{mytorus}. An updated version of the model is used to predict the line emission at the high resolution necessary to describe the \textit{XRISM} spectrum. We note that the inclusion of the scattered continuum component of \textsc{MYTorus} is disfavoured by the data, suggesting that the narrow components of the X-ray emission lines do not emerge from an optically thick torus surrounding the central engine, but instead may originate in more tenuous clouds, perhaps associated with the optical broad line region (BLR).

A total of four photoionised outflow components are required to explain the absorption lines seen in the spectrum. Two high-ionisation components are detected in the \textit{XRISM Resolve} spectrum. One close to rest with respect to MCG--6 producing prominent, narrow Fe\,\textsc{xxv} and Fe\,\textsc{xxvi} absorption lines at 6.67 and 6.97\keV\ respectively. The other produces broader blueshifted Fe\,\textsc{xxv} and Fe\,\textsc{xxvi} corresponding to an outflow velocity of $0.06c$. In addition, two lower-ionisation warm absorbers are required to model the soft ($< 1$\keV) X-ray spectrum, as reported in previous analyses \citep{reynolds+1995,lee+2001,young+2005}. These photoionised outflows are modelled using the \textsc{cloudy} spectral synthesis code v23.01 \citep{cloudy,cloudy2023} and incorporated into the \textsc{xspec} model via pre-calculated tables following the methods outlined in \citet{ogorzalek+2022}. Details of the photoionisation modelling of the outflows and absorption lines will be presented in subsequent papers. Alongside the absorption lines in the iron K band, corresponding emission features from the slower wind components are tentatively detected, and are included as slightly-redshifted Gaussian emission lines corresponding to Fe\,\textsc{xxv} and Fe\,\textsc{xxvi}. Finally, the iron K photoelectric absorption edge is detected in the \textit{Resolve} spectra from the dusty ISM in the host galaxy that had been previously detected in \textit{Chandra} HETG spectra \citep{lee+2001},  and is included here using the \textsc{tbvarabs} model.

\section{Time variability}
\begin{figure*}[ht!]
\plotone{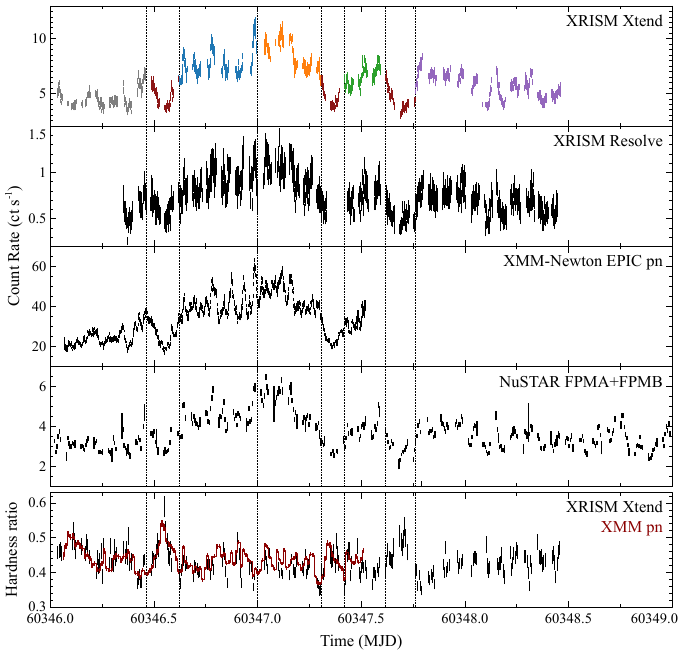}
\caption{Light curves showing the observed X-ray flux from MCG--6-30-15 during the coordinated campaign with \textit{XRISM Xtend} (0.3-12\keV), \textit{XRISM Resolve} (2-12\keV), \textit{XMM-Newton} EPIC pn (0.3-10\keV) and \textit{NuSTAR} (3-55\keV). The count rate from the two \textit{NuSTAR} focal plane module detectors (FPMA and FPMB) are summed for the purposes of creating the light curve. The bottom panel shows the hardness ratio, $H/S$, between the 2-10\keV\ hard X-ray band and 0.3-2\keV\ soft X-ray band, measured with both \textit{XRISM Xtend} and \textit{XMM-Newton} EPIC pn. The \textit{XMM-Newton} hardness ratio is multiplied by a constant factor of 1.8 for display purposes to match the \textit{XRISM Xtend} hardness ratio, accounting for the energy-dependent difference in effective area between the two instruments. We identify eight time intervals across which we perform the time-resolved spectral analysis, denoted by vertical lines as well as colours in the top panel. These intervals consist of the rise (blue) and fall (orange) of the flare, in addition to three flux dips (red) and the period at the end of the observation after the flare and dips (purple).
\label{fig:lc}}
\end{figure*}

Light curves showing the observed X-ray flux from MCG--6-30-15 during the  campaign are shown in Figure~\ref{fig:lc}. Significant variability was observed with flux variations by more than a factor of three between the dimmest and brightest states that were observed, and periods during which the flux rapidly varies by around 50 per cent on timescales of just 30 minutes.

A high flux period, or `flare', was observed in the X-ray light curve between MJD\,60346 and 60347, where the count rate increases by a factor of three from the baseline level at the start of the observation, in addition to three short dips where the X-ray flux drops and the hardness ratio peaks over a period of approximately three hours, at MJD\,60346.5, 60347.5 and 60347.75. As noted by \citet{brenneman+2025}, the peaks in the hardness ratio appear in both the soft and hard X-ray bands, as measured by \textit{XRISM} and \textit{XMM-Newton}, and \textit{NuSTAR}, suggesting that these variations are a broadband phenomenon, arising in the primary continuum emission from the corona, and cannot be attributed to changes in absorption that would only be manifested in the soft X-ray band.

To understand the mechanisms underlying this variability, we identify eight principal time periods within which we perform spectral analysis: the start of the observation, before the dips (MJD\,60346.0-60346.5), each of the three dips, the rise of the high flux period (MJD\,60346.6-60347.0), the fall of the high flux period (MJD\,60347.0-60347.3), the period between the dips (MJD\,60347.4-60347.6) and the low flux period after the dips (MJD\,60347.8-60348.5).

\section{Modelling the variable X-ray spectrum}
\label{sec:model}
To understand the mechanisms underlying this variability, we extract spectra from the \textit{XRISM}, \textit{NuSTAR} and \textit{XMM-Newton} observations from each of the time intervals and apply a model based upon the best-fitting model to the time-averaged spectra.

\begin{figure*}[ht!]
    \gridline{
        \fig{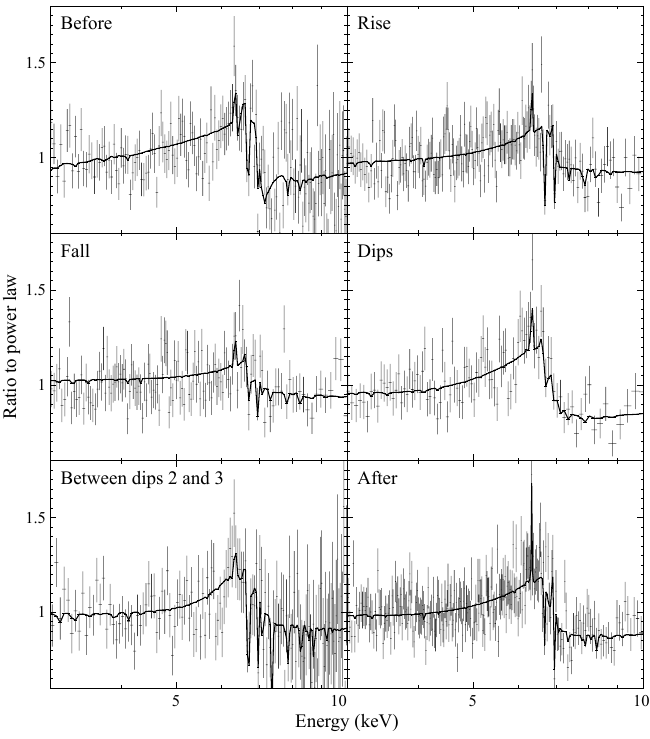}{0.47\textwidth}{(a)}
        \fig{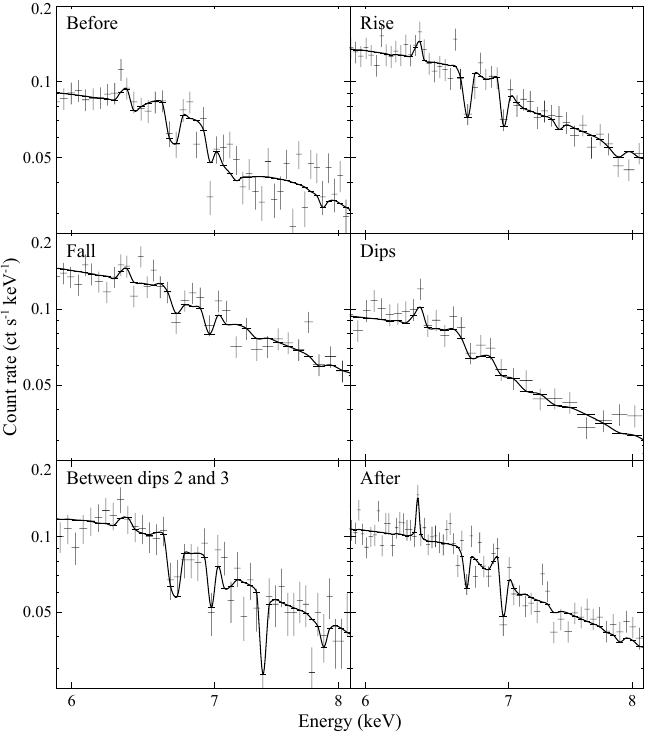}{0.47\textwidth}{(b)}
    }
    \caption{The variation of the high-resolution X-ray spectrum of MCG--6, measured by \textit{XRISM Resolve}, between the observed time intervals. (a) Shows the ratio of the observed spectrum to the best-fitting power law, to reveal the broad and narrow absorption and emission lines overlaying the continuum spectrum. The profile of the relativistically broadened iron K line, in the reflection spectrum from the inner disc, can clearly be seen with a blueshifted peak and extended redshifted wing. The solid line shows the best-fitting spectral model, described in \S\ref{sec:model}. (b) Shows a zoom-in of the X-ray spectrum over the 6-8\keV\ band, showing the multitude of narrow absorption and emission line features in the iron K band. The spectra from the three dips are summed together to increase the signal-to-noise for visualisation purposes only, however in the spectral analysis these three dip time intervals are treated independently. 
        \label{fig:spectra}}
\end{figure*}

We see the variation in the observed spectra in Fig~\ref{fig:spectra}. The ratio of the high-resolution \textit{XRISM Resolve} spectra of MCG--6 to the best-fitting power law reveals the broad and narrow emission and absorption lines that are superimposed on the underlying continuum. The zoom-in of the 6-8\keV\ spectrum shows the variation of the narrow absorption lines between the time intervals. Notably, the ratio of each spectrum to the underlying power law reveals the characteristic shape of the relativistically broadened iron K$\alpha$ line, with a blueshifted peak just above the rest frame energy of 6.4\keV, and a broad redshifted wing, extending to low energies. This line profile is characteristic of X-ray reflection from the inner accretion disc, and the extended wing is comprised of gravitationally-redshifted emission from the innermost radii of the disc.

In Figure~\ref{fig:spectra}(a) we see how the strength of the reflection and broad iron K$\alpha$ line relative to the underlying continuum varies over the course of the observation, with the reflection weakest during the fall of the flare or high flux period, and strongest during the short dips that are observed in the light curve. In Figure~\ref{fig:spectra}(b) we see variation in the narrow Fe\,\textsc{xxv} and Fe\,\textsc{xxvi} absorption lines, at 6.67 and 6.97\keV\ respectively, that are attributed to the ionised outflowing wind. In this paper, we shall concentrate on the variation of the X-ray emitting corona and reflection from the inner disc. A detailed study of the variation of the outflows will be presented in subsequent papers.

For each time interval, we simultaneously apply the model to the spectra obtained by \textit{XRISM Resolve}, \textit{NuSTAR} and \textit{XMM-Newton} EPIC pn, providing broadband coverage from 0.3-55\keV\ and high-resolution spectroscopy with \textit{Resolve} over the 2-12\keV\ band. We model the eight time intervals simultaneously, allowing only those parameters that should be able to vary on short timescales vary between the intervals, while tying the values of parameters that should not vary. Such a simultaneous fitting scheme enables us to employ all of the data to obtain the tightest possible constraints on the tied parameters, while marginalising over any uncertainty in these parameters when measuring the variation of the other parameters over the course of the observations.

The following parameters are not expected to vary on the timescale of the observations, thus are tied between the intervals:
\begin{itemize}
    \setlength\itemsep{0em}
    \item The spin parameter of the black hole, $a = J/Mc$, and the dimensionless spin parameter $a_\star = a / GMc^{-2}$
    \item The inclination at which the accretion disc is viewed, $i$
    \item The density of the accretion disc plasma, $n_\mathrm{e}$
    \item The metallicity (\textit{i.e.} iron abundance, $A_\mathrm{Fe}$) within the accretion disc plasma
    \item The normalisation of the narrow components of the emission lines, in the reflection spectrum from material at large distances from the black hole, which due to the light travel time across the emitting region will respond to variability on timescales much longer than those considered here
\end{itemize}
On the other hand, the X-ray emission from the corona can vary on short timescales. Indeed, the typical cooling time of the X-ray emitting corona is sufficiently short that it requires constant injection of energy \citep{fabian+2015}, thus the structure and internal properties of the corona may also be expected to vary on short timescales, in addition to the reprocessing of the coronal X-ray emission by any components of the system that can respond rapidly. We therefore allow the following parameters to vary between each of the time intervals:
\begin{itemize}
    \setlength\itemsep{0em}
    \item The flux (normalisation) of the primary X-ray continuum, emitted by the corona
    \item The photon index, $\Gamma$ of the X-ray continuum spectrum (dictated by the characteristic temperature and optical depth through the corona)
    \item The temperature of the corona, $kT_\mathrm{e}$, can, in principle, vary between the time intervals, however we find that the best-fitting coronal temperature in MCG--6 that the fit is not improved when the temperature is allowed to vary between time intervals, thus we tie the value.
    \item The reflection fraction ($R$, the ratio of the reflected to continuum flux, determined by the geometry and motion of the corona)
    \item The emissivity profile of the reflection from the accretion disc (\textit{i.e.} the height of the point source, $h$, in the simplified `lamppost' model)
    \item The ionisation parameter of the plasma in the accretion disc ($\xi = 4\pi F/n$)
    \item The column densities, ionisation parameters and velocities of the absorbers (note that only the velocities of the two high-ionisation components that are detected via absorption lines in \textit{Resolve} are allowed to vary, the bulk velocities and velocity broadening of the two lower-ionisation components are frozen at the values measured from lines identified in the \textit{XMM-Newton} RGS spectra, which are not included in the simultaneous fit)
    \item The normalisations of the Gaussian emission components identified from the outflows
\end{itemize}
An initial estimate of the best-fitting parameter values was estimated by minimising the modified $C$-statistic using \textsc{xspec}. The model was found to provide a good fit to the time-resolved spectral data, with a best-fitting $C$-statistic of 128,379 for 161,413 degrees of freedom and 183 free parameters.

\begin{figure}[ht!]
    \plotone{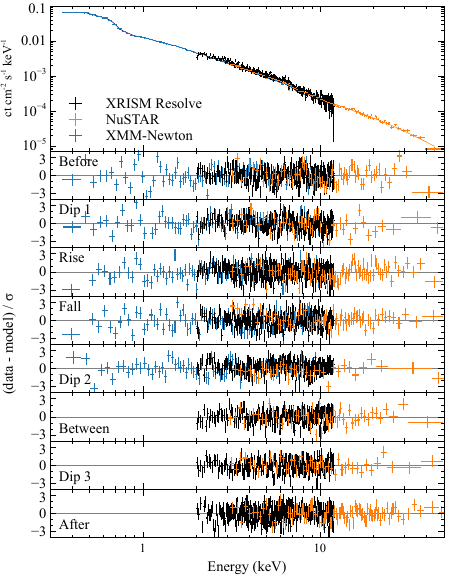}
    \caption{\textit{Top panel:} The X-ray spectra of MCG--6-30-15 obtained with \textit{XRISM} Resolve (2-12\keV), \textit{NuSTAR} (3-55\keV) and \textit{XMM-Newton} EPIC pn (0.3-10\keV). The spectra are shown from the time interval corresponding to the rise of the high flux period, and are divided by the effective area of each instrument to obtain a continuous spectrum across the entire bandpass. The lines show the best-fitting model. \textit{Lower panels:} The residual between the observed spectrum and the best-fitting model in each of the eight time intervals, scaled by the error bar in each spectral bin.
        \label{fig:spec_fit}}
\end{figure}

The parameter space was then explored via Markov Chain Monte Carlo (MCMC) to estimate the statistical uncertainty on each parameter. Chains were run using the algorithm of \citet{goodman_weare}, also implemented in \textsc{xspec}, with 150 walkers for 8,000,000 steps, after discarding (or `burning') the first 500,000 steps of the chains so as to remove any memory of the starting points of the chains. The best-fitting parameter values from each time interval are shown in Table~\ref{tab:param} and the spectra as well as the ratio between the data and the best-fitting model are shown in Fig.~\ref{fig:spec_fit}.

During the final dip and the time period after the dips and flare, no soft-band (0.3-2\keV) spectra are available, since \textit{XMM-Newton} was not observing for the latter part of the observation, and we do not use the \textit{XRISM Xtend} spectra due to the calibration inconsistency between it and \textit{XMM} EPIC pn. This means that parameters relying on the soft X-ray data, including the disc ionisation parameter and the properties of the lower ionisation warm absorbers, are less well constrained during these periods. We, however, marginalise over the uncertainty in these parameters in the MCMC analysis.

\begin{figure*}[ht!]
    \gridline{
        \fig{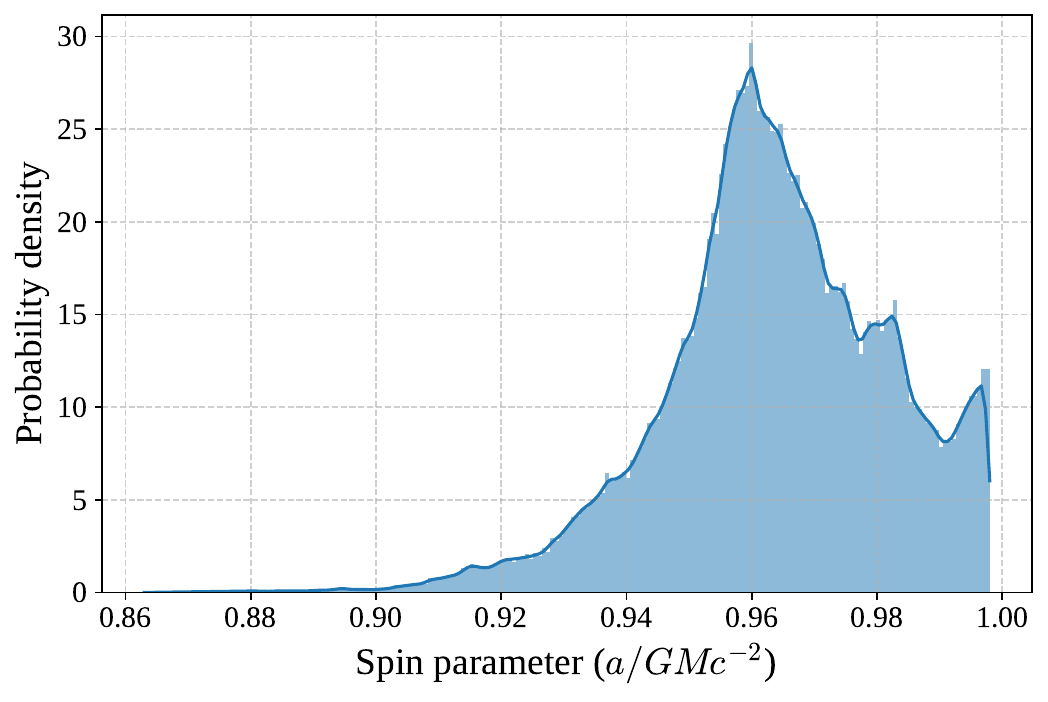}{0.47\textwidth}{(a)}
        \fig{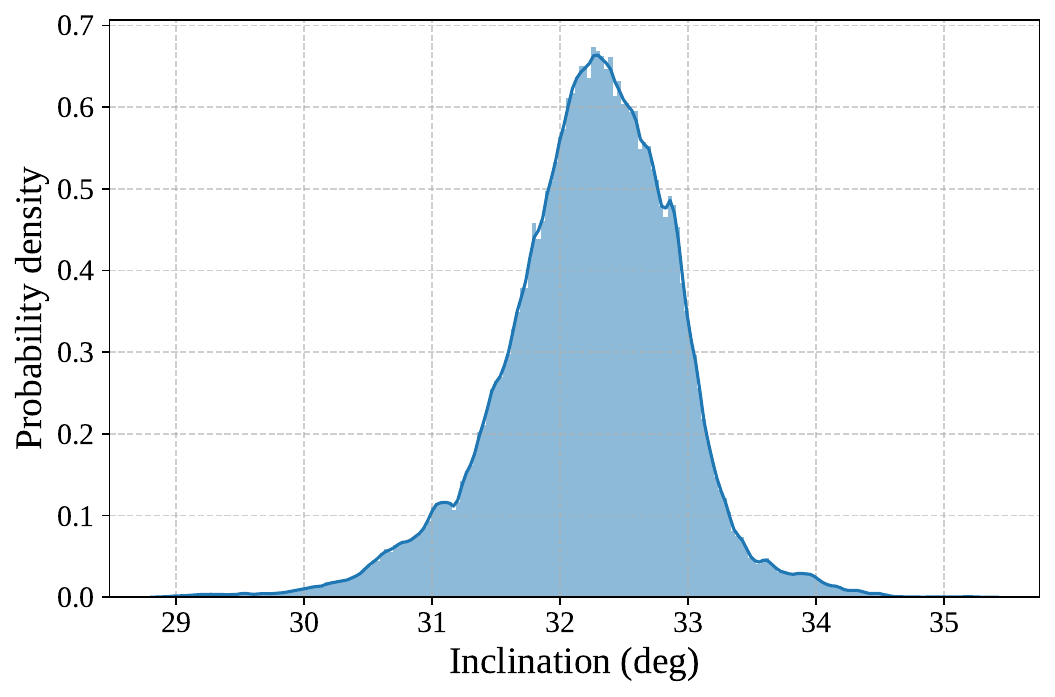}{0.47\textwidth}{(b)}
    }
    \gridline{
        \fig{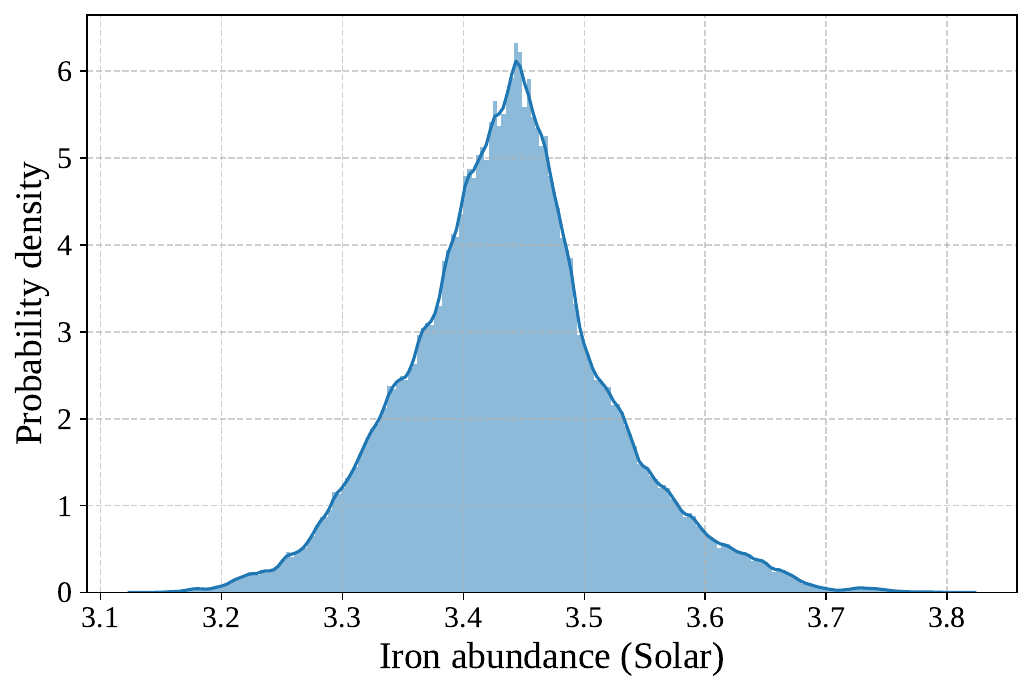}{0.47\textwidth}{(c)}
        \fig{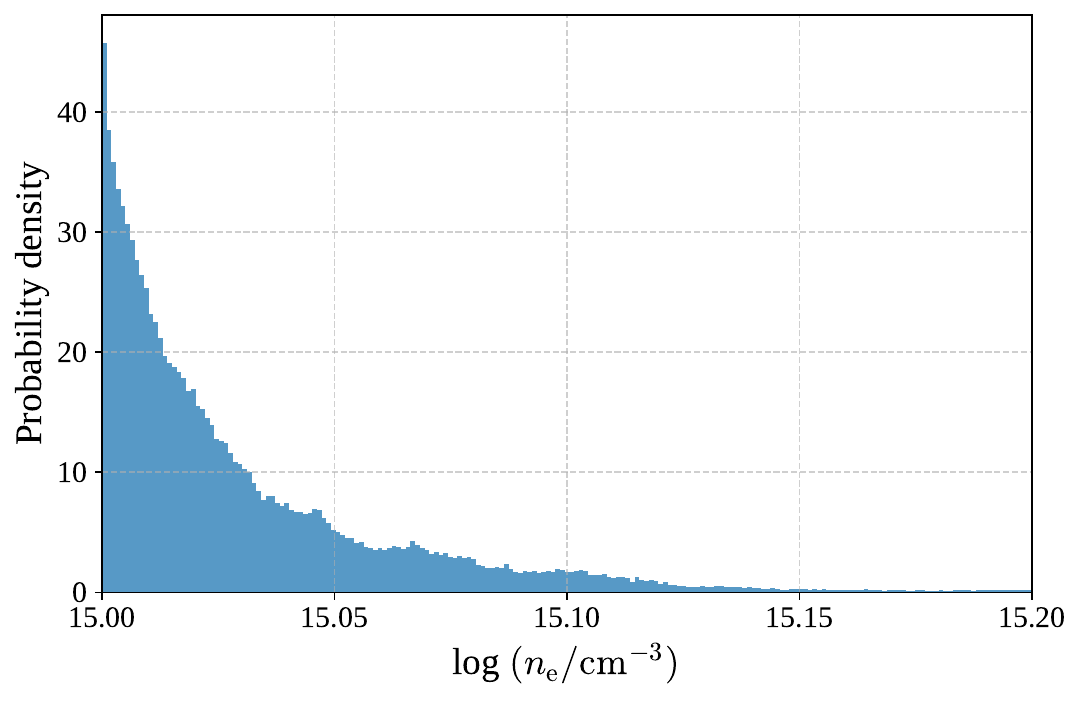}{0.48\textwidth}{(d)}
    }
    \caption{Posterior distributions for key parameters of the relativistic reflection model from the inner accretion disc (\textsc{relxilllpCp}) that are not expected to vary on the timescale of the observations, thus are tied between the time intervals: (a) The dimensionless spin parameter of the black hole, $a_\star = cJ/GM^{2}$, (b) the inclination of the normal to the accretion disc to the line of sight, (c) the iron abundance within the accretion disc plasma, and (d) the electron density within the disc plasma. Posteriors are derived from MCMC exploration of the parameter space of the model, fit simultaneously to the eight time intervals. The histograms are derived from the MCMC chain steps and the solid lines show the kernel density estimate, derived by smoothing the raw histogram using a Gaussian kernel, implemented in \textsc{seaborn}.
        \label{fig:tied_params}}
\end{figure*}

Fitting the time-resolved spectra, we obtain tight constraints on the spin of the black hole, the inclination of the accretion disc, the iron abundance in the accretion disc plasma, and density of the disc. The posterior distributions for these parameters, derived from MCMC exploration of the parameter space, are shown in Figure~\ref{fig:tied_params}. The black hole is found to be rapidly spinning, with dimensionless spin parameter $a_\star > 0.93$ (where the lower limit is defined by the 95 per cent confidence interval).

The accretion disc in MCG--6 is observed at an inclination of $32.3_{-0.7}^{+0.6}$\,deg between the disc normal and the line of sight. The iron abundance was found to be $3.44_{-0.15}^{+0.07}$ times the Solar value, and the accretion disc density was constrained be relatively low with $\log(n_\mathrm{e} / \mathrm{cm}^{-3}) < 15.1$, with no evidence for elevation above the canonical value ($10^{15}\,\mathrm{cm}^{3}$), as has been found in a number of AGN with black hole masses close to that of MCG--6 \citep{jiang_highden}.

 We note that by analysing the time-resolved spectra, we are able to obtain tighter constraints on the spin of the black hole than was possible via analysis of the time-averaged spectrum, and we shall return to this point in \S\ref{sec:spin}.

\begin{figure*}[ht!]
\plotone{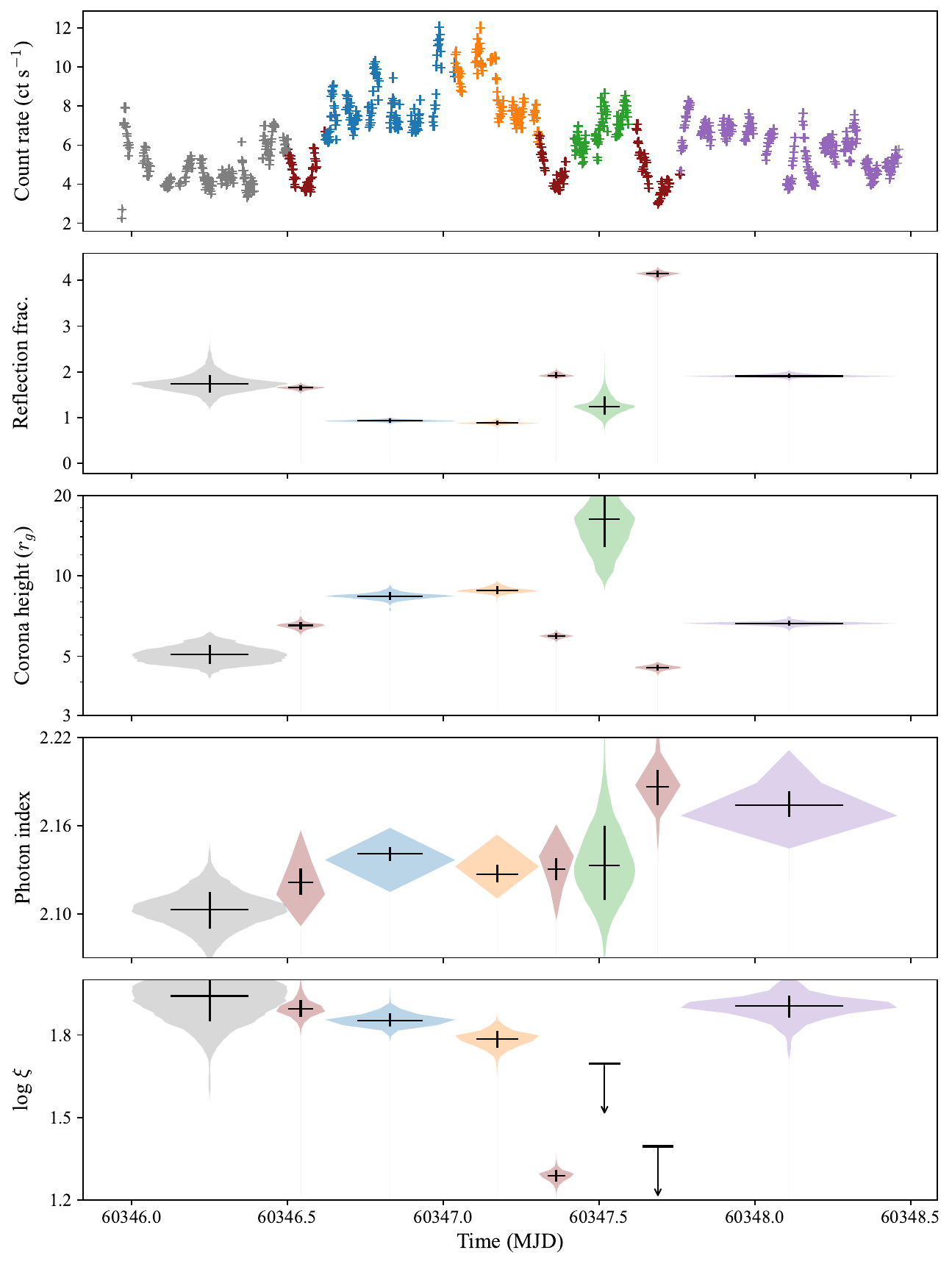}
\caption{The variation of parameters of the relativistic reflection model from the inner accretion disc (\textsc{relxilllpCp}) between the time intervals. The top panel shows the \textit{XRISM Xtend} light curve in which the colours represent the time intervals into which the observations were divided for the purposes of spectral analysis. The lower panels show the variation of the reflection fraction, the height of the corona (derived from the emissivity profile of the accretion disc assuming it to be illuminated by a point source on the spin axis of the black hole), the photon index of the X-ray continuum emitted by the corona, and the ionisation parameter of the accretion disc, $\log\xi$. The widths of the shaded regions represent the posterior probability densities at each parameter value, derived from MCMC exploration of the parameter space, while the error bars represent the $1\sigma$ confidence limits. The upper limit of $\log\xi$ shown for the third dip represents the 95 per cent confidence limit.
\label{fig:violin}}
\end{figure*}

We see significant variation in the model parameters related to the geometry of the corona and the coronal X-ray emission over the course of the observations, shown in Figure~\ref{fig:violin}. The reflection fraction (defined as the ratio of the flux reflected from the accretion disc to the continuum flux observed directly from the corona, over the full bandpass) drops substantially from $R = 1.74_{-0.16}^{+0.17}$ at the beginning of the observation to $R = 0.93_{-0.02}^{+0.02}$ during the rise, falling again to $R = 0.88_{-0.02}^{+0.03}$ on the drop from the high to low flux state. The second and third dips are once again marked by a significant increase in the reflection fraction, reaching $R = 1.91_{-0.01}^{+0.07}$ during the second dip and $R = 4.15_{-0.06}^{+0.03}$ during the third. Between the dips, the reflection fraction drops again to $R=1.23_{-0.14}^{+0.2}$, and settles to a value of $R = 1.90_{-0.01}^{+0.04}$ as the light curve stabilises in the latter part of the observation.

These variations in reflection fraction can be directly mapped to changes in the size-scale of the X-ray emitting corona. In the \textsc{relxilllpCp} model, the scale of the corona is represented by a single parameter, the height, $h$, under the simplifying assumption that the corona can be represented by a point source of X-ray emission located upon the spin axis above the black hole, though we shall return to this assumption in the next section. We see that the drops in reflection fraction during the rise and fall of the high flux state correspond to an increase in the scale-height of the corona, while the increased reflection fraction during the dips correspond to time periods during which the corona has collapsed to a more confined region around the black hole, with a smaller scale height, when gravitational light bending focuses a greater fraction of the emission onto the inner accretion disc. During the entire observations, however, the reflection spectrum is consistent with the accretion disc being illuminated by a relatively compact corona, with height ranging from $4.53_{-0.07}^{+0.09}$\rg\ during the most extreme dip to $8.8_{-0.1}^{+0.3}$\rg\ during the fall from the high flux state. Between the second and third dips, the corona briefly expands again, with the best fitting corona height found to be $17_{-3}^{+4}$\rg. The geometry of the corona will be discussed in more detail in \S\ref{sec:emis}.

We see some evidence for variation of the photon index, $\Gamma$, of the continuum spectrum (where the slope of the power law spectrum is defined in terms of count rate by $dN_E = E^{-\Gamma}\,dE$, which is in turn related to the energetics of the corona, and the optical depth through which photons Compton scatter within the corona to produce the X-ray continuum). The spectrum softens from $\Gamma = 2.121_{-0.007}^{+0.009}$ to $\Gamma = 2.141_{-0.004}^{+0.004} $ as the flux rises, then hardens to $\Gamma = 2.125_{-0.003}^{+0.008}$ as it falls. Following the third dip, the spectrum softens quite substantially to $\Gamma = 2.174_{-0.008}^{+0.008}$.

We also see some slight variation in the ionisation parameter of the accretion disc. Where the plasma in the accretion disc of density $n$ is photoionised by irradiation from the corona with X-ray flux $F$, the ionisation parameter is defined as $\xi = 4\pi F / n$. The ionisation of the disc drops slightly between the first dip, the rise and the fall of the high flux state, from $\log(\xi / \mathrm{erg}\,\mathrm{cm}\,\mathrm{s}^{-1}$) = $ 1.89_{-0.02}^{+0.03}$ to $\log\xi = 1.79_{-0.04}^{+0.02}$, before quite a significant drop to $\log\xi\sim 1.3$ during the second and third dips.

Irradiation of the accretion disc by a compact corona close to the black hole is expected to produce a gradient in the ionisation parameter across the disc \citep[see \textit{e.g.}][]{svoboda+12,plunging_region_paper}. Detailed discussion and modelling of accretion disc ionisation gradients is beyond the scope of this work, however it is possible to include radial variation of the ionisation parameter in the \textsc{relxilllpCp} variant of the reflection model. The ionisation profile of the disc will depend upon both the irradiation profile (\textit{i.e.} the emissivity profile of the reflection spectrum) and the density profile, and in \textsc{relxilllpCp}, the ionisation profile can be modelled as either a simple phenomenological power law, or the expected profile from a point source corona irradiating a disc with density profile according to the standard $\alpha$-disc prescription. We find that neither prescription significantly improves the fit to the data and that the observed spectrum is consistent with a constant value at all radius (the best fitting power law index was found to be $-0.3\pm 0.3$). We note, however, that some of the steepness of the emissivity profile on the innermost radii of the disc can be produced by unmodeled gradients in the disc ionisation \citep{svoboda+12}.

\section{The geometry of the corona: is it really a lamppost?}
\label{sec:emis}
The geometry of the corona is constrained by the emissivity profile of the accretion disc, that is the variation in reflected flux as a function of radius across the disc. Due to the variation in Doppler shifts and gravitational redshift as a function of distance from the black hole, changes in the emissivity profile are manifested as changes in the detailed shape of the relativistically broadened emission lines in the reflection spectrum \citep{1h0707_emis_paper}. If the emissivity profile of the disc can be measured or constrained from the model fit to the X-ray spectrum, it is possible to infer the location, size and geometry of the corona that is illuminating the disc \citep{understanding_emis_paper,gonzalez+2017}.

The \textsc{relxilllpCp} model employed for the initial analysis calculates the emissivity profile of the disc under the simplifying assumption that the corona can be represented by a point source (or `lamppost') located on the spin axis of the black hole at height $h$. The variable parameters are the height of the point source, and its velocity, which can lead to relativistic beaming of its emission away from the disc. This model also includes the enhancement to the inner disc emissivity profile from returning radiation; \textit{i.e.} the reflected photons that are returned to the disc by light bending in the strong gravitational field to produce higher-order reflection components \citep{ross_fabian_ballantyne,return_radiation_paper}.

Of course any corona that produces X-ray emission via the inverse-Compton scattering of thermal photons from the accretion disc must have finite surface area and cannot truly be an infinitesimal point source, however the lamppost model is a useful approximation for the purposes of modelling, that has proven to be adequate in many cases where the signal-to-noise within the data is limited.

We may test for evidence of deviation from a point-like or lamppost corona by instead fitting a reflection model that incorporates a generalised prescription for the accretion disc emissivity profile. Emissivity profiles that arise from most realistic configurations of the corona approximately take the form of a once- or twice-broken power law \citep{understanding_emis_paper}. The emissivity falls off steeply over the innermost radii of the disc, where relativistic effects enhance the observed reflection as strong light bending focuses the emission from the corona towards the black hole and the inner disc. The profile then flattens over intermediate radii, before tending toward the classical solution $\epsilon\propto r^{-3}$ over the outer disc. The break radius between the flattened part of the emissivity profile and the $r^{-3}$ decay over the outer disc roughly corresponds to the radial extent of the corona, and an extended corona can be distinguished from a point source where both strong light bending from the innermost radii and an extended flattened section due to the spatial extent of the corona are required simultaneously.

We replace the \textsc{relxilllpCp} component of the model, describing the continuum and its reflection from the inner accretion disc, with the \textsc{relxillCp3} variant of the model\footnote{\textsc{relxillCp3} is included in the \textsc{relxill++} model package, which provides extended functionality to the original \textsc{relxill} models: \url{https://github.com/wilkinsdr/relxillpp}}, which parametrises the emissivity profile as a twice broken power law, in place of the point source approximation. The free parameters are the inner, middle and outer power law indices ($q_1$, $q_2$ and $q_3$) and the two break radii ($r_{\mathrm{b},1}$ and $r_{\mathrm{b},2}$ at which these transitions in power law index occur). Since the configuration of the corona can change on relatively short timescales, all five of these parameters are permitted to vary between the eight time intervals.

We find that allowing the emissivity profile of the disc to deviate from the lamppost model provides a significantly improved fit to the data. We compare the \textsc{relxilllpCp} and \textsc{relxillCp3} models using the deviance information criterion or DIC (\citealt{dic}, see also \citealt{ogorzalek+2022, 1zw1_flare_paper}). The DIC is given by
\begin{equation}
    \mathrm{DIC} = D(\hat{\theta}) + 2p_D
\end{equation}
Where the deviance at a given set of parameter values, $\theta$, the deviance is defined as $D(\theta) = -2\ln\mathcal{L}$, for likelihood $\mathcal{L}$ (\textit{i.e.} the $C$-statistic for those parameter values, where the uncertainties on the data points follow a Poisson distribution), and $\hat\theta$ denotes the best-fitting parameters for which the likelihood is maximised.

Similar to the Bayesian Information Criterion (BIC) and Akaike Information Criterion (AIC), the DIC draws on information theory to formalise the relationship between the reduction in fit statistic and the number of free parameters in determining the statistical significance to which one model may be preferred over another. The DIC can be directly computed from the chains resulting from MCMC calculations, and conveys two principal advantages in comparing models. Firstly, DIC takes into account the \textit{volume} of parameter space in which the models provide a good description of the data, rather than focusing entirely on single point in parameter space at which the maximum likelihood is found. Secondly, in considering the number of free parameters within a model, the DIC defines an \textit{effective} number of free parameters, thereby not penalising a model for parameters that are not constrained by the data. The effective number of free parameters is defined as:
\begin{equation}
    p_D = \overline{D(\theta)} - D(\overline{\theta}) 
\end{equation}
Where bars denote the mean over the sample of values derived from the chains. A model yielding a lower DIC is preferred by the data, and the $\Delta\mathrm{DIC}$ between two models quantifies the evidence in favour of one model over another. $\Delta\mathrm{DIC}$ values between zero and two suggest only marginal evidence, while $\Delta\mathrm{DIC}$ greater than 6 shows strong evidence for one model in favour of another \citep{kass+1995}.

\begin{figure}[ht!]
\includegraphics[width=0.98\textwidth]{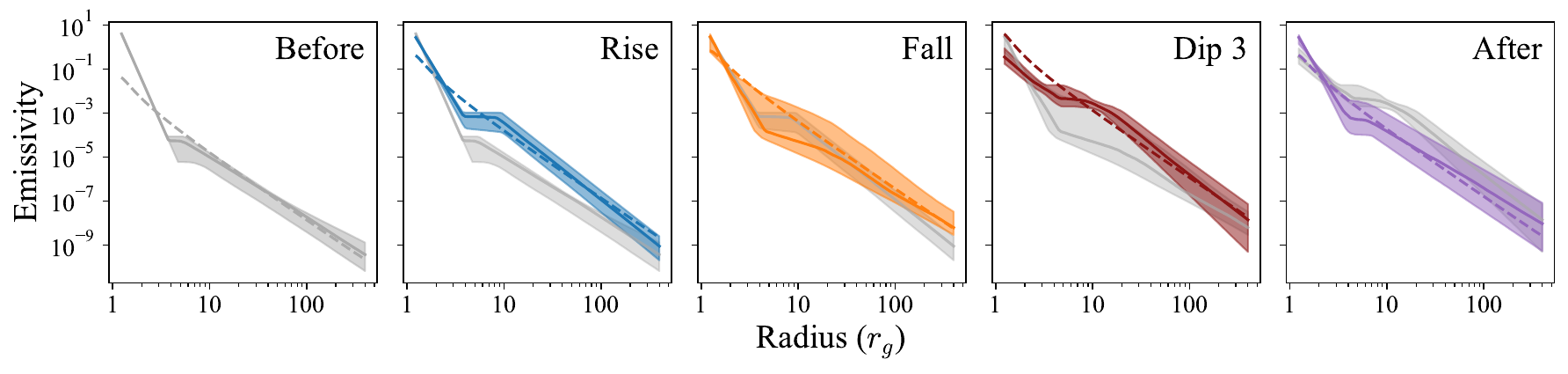}
\caption{The best-fitting emissivity profiles from selected time intervals, \textit{i.e.} the variation in reflected flux originating as a function of radius on the accretion disc, derived from fitting the \textsc{relxillCp3} model to the time-resolved spectra, in which the emissivity profile is parametrised as a twice-broken power law. The dashed lines, for comparison, show the emissivity profiles corresponding to the best fitting lamppost or point source models used in \textsc{relxilllpCp}. The grey contours in the latter panels show the emissivity profile from the previous panel, showing the evolution between the time intervals. The emissivity is in arbitrary units and represents the relative flux emitted from each radius (defined in the rest frame of the emitting material).
\label{fig:emis}}
\end{figure}

The twice-broken power law model for the emissivity profile yields a reduction in DIC of 21 relative to the lamppost model (which corresponds to $\Delta C = -56$ for 18 additional free parameters), showing that there is significant evidence for extension of the corona beyond a simple point source in the time-resolved spectra of MCG--6, with the \textsc{relxillCp3} model strongly preferred by the data. The best-fitting emissivity profiles are shown in Figure~\ref{fig:emis}, and are compared to the profiles corresponding to the best fitting lamppost or point source models. The parameters of the derived emissivity profiles are listed in Table~\ref{tab:relxill3}.

When modelling the accretion disc emissivity profile with a twice-broken power law, we find that the best-fitting values for many of the parameters of interest are consistent (within uncertainty) with those values inferred using the lamppost model for the accretion disc emissivity. Fitting the \textsc{relxillCp3} to the eight time intervals simultaneously, we find that the spin parameter $a_\star = 0.95\pm 0.01$, the inclination of the accretion disc to the line of sight, $i = 32.3_{-0.1}^{+0.9}$. and the iron abundance in the accretion disc, $A_\mathrm{Fe} = 3.39_{-0.01}^{+0.05}$.

In each of these time intervals, we see the characteristic form of the emissivity profile expected for a compact X-ray source close to the black hole. We see the emissivity fall steeply over the innermost $3\sim 4$\rg (where the gravitational radius, $r_\mathrm{g} = GM/c^2$, is the characteristic scale length in the gravitational field around the black hole, and the event horizon of a maximally spinning black hole is located at $\sim 1\,r_\mathrm{g}$), where strong light bending and relativistic effects enhance the reflected flux seen from the inner radii. The best-fitting emissivity profiles are steeper than the best-fitting point source models, showing that part of the corona extends closer to the black hole to produce the observed enhancement in the reflection seen from the innermost radii.

The emissivity profile flattens over intermediate radii, suggesting that there is some extension of the corona radially over the surface of the inner accretion disc, and we can identify the outer break radius with the radial extent of the corona over the disc. At the beginning of the observations, the outer break radius is detected at $(5\pm 1)$\rg, indicating a compact corona, in a confined region close to the black hole. As the flux increases, the corona expands, with the outer break radius reaching $(9.4\pm 1.7)$\rg. The corona is most extended during the fall from the high flux state, where the outer break radius is detected at $15_{-0.14}^{+18}$\rg. In the low flux state at the end of the observation, the corona has one again collapsed to a confined region within $\sim 5$\rg\ of the black hole. During the dips, the corona is especially compact, and we note the outer power law index of the emissivity profile has steepened from $\sim 3$ to $(4.5\pm 1.7)$. Light bending from a corona located close to the event horizon of the black hole results in more of the emission from the corona being focused onto the inner disc, steeply falling at large radii.

We note that during the fall, the emissivity profile does not flatten to the same degree over the middle section, but reaches a power law index $q_2 = (2.0\pm 0.5)$. Such a profile can arise from a vertically extended corona \citep{mrk335_corona_paper}, as the emission region expands upwards, away from the disc, as the high flux period begins, and, in this case, the outer break radius relates to whichever is the largest of the vertical and radial extent of the corona.

Comparing the twice-broken power law emissivity profiles to the best-fitting point source or lamppost emissivity profiles, we see that in all time intervals except the rise, the twice-broken power laws closely resemble the best-fitting point source profiles, showing that while there is some physical extent to the corona, it is still relatively compact, extending no more than $5\sim 10$\rg\ over the inner disc.

\subsection{Variation of the inner radius of the accretion disc}
We also tested allowing the inner radius of the accretion disc to vary freely between the time intervals, to determine whether the spectral variability could be explained by changes to the inner accretion disc rather than to the corona. Initially, allowing the inner radius to vary within the simplified point source model yielded an improvement to the fit, with the DIC statistic decreasing by $\Delta \mathrm{DIC}=-18$ (corresponding to $\Delta C=-19$ for 6 additional free parameters). Within this model, the black hole spin remains close to maximal ($a_\star > 0.96$) and during the rise and fall time periods of the high flux state, the inner radius of the disc was most tightly constrained, lying within $r_\mathrm{in} < 1.1\,r_\mathrm{ISCO}$ (where the innermost stable circular orbit for a maximally spinning black hole lies at $r_\mathrm{ISCO} = 1.235\,r_\mathrm{g}$). We note that this means that the drop in reflection fraction during the high flux state cannot be attributed to truncation of the inner accretion disc. When a point source corona is assumed, there is some evidence for slight variation of the inner accretion disc. During the periods of lower flux, notably the first dip and the period at the end of the observation, the constraint on the inner radius of the accretion disc loosens, with $r_\mathrm{in} < 3.5\,r_\mathrm{ISCO}$ and $r_\mathrm{in} < 3\,r_\mathrm{ISCO}$ for these periods respectively. This model, however, still requires variation of the height of the corona between the time intervals, yielding similar values to those found when the inner disc radius is fixed.

However, we find that when we relax the assumption of a point source corona, and allow the inner radius of the disc to vary within the reflection model in which the disc emissivity profile is described by a twice-broken power law, the data no longer prefer the model in which the inner radius of the disc may vary between time intervals. The \textsc{relxillcp3} model with variable inner disc radius yields an increase in DIC statistic of $\Delta \mathrm{DIC} = +3$ ($\Delta C = -12$ but for 8 additional free parameters) relative to the equivalent \textsc{relxillcp3} model in which the inner disc radius does not vary, and the inner disc radius is constrained tightly to the ISCO at all times. It thus may be that the slight variation observed in the inner disc radius when fitting a simplified point source model is simply compensating for variation of the inner disc emissivity in a way that cannot be accounted for within the model. We thus conclude that the observed variability can be attributed to changes in the luminosity and geometry of the corona alone, with no significant variation of the inner disc over the course of the observations.

\subsection{Inferring the location and motion of the corona from the reflection fraction}

We may gain further information about the extent and motion of the corona from the measured reflection fractions, defined as the ratio of the reflected to continuum flux in the observed spectrum \citep{1h0707_jan11,mrk335_corona_paper}. In the low flux period at the end of the observations, after the dips, the reflection fraction is measured to be $R = 1.90_{-0.01}^{+0.04}$. In the classical case, the reflection fraction is expected to be unity, since one half of the flux emitted from an isotropic source is emitted downwards to be reflected off the disc, and one half is emitted upwards, escaping to be observed as the continuum. For a compact corona close to the black hole, light bending focuses more of the emission towards the black hole and hence onto the inner disc. Using Equation 11 of \citet{gonzalez+2017}, we can estimate the expected reflection fraction as a function of corona height and velocity (for a point source), accounting for the solid angle aberration between the source and observer frame due to special relativistic beaming and general relativistic light bending. For a compact corona at rest at the inferred height of around 6\rg, the expected reflection fraction due to light bending is 1.85, close to be observed value (note that the precise value will depend upon the spatial extent of the corona and the distribution of luminosity throughout the volume of the corona).

When modelling the reflection from the inner accretion disc under the assumption of a point source corona using the \textsc{relxilllpCp} model, we can self-consistently link the emissivity profile for a given coronal height to the reflection fraction, or we may fit the emissivity profile/coronal height independently to the data. We find that linking the emissivity profile and reflection fraction produces a worse fit to the data, with $\Delta C = +57$ relative to the case in which the two parameters are allowed to vary freely. We find that the best-fitting reflection fraction varies between a minimum of 0.88-0.94 times that expected for the best-fitting point source height during the rise and fall of the high-flux period, and 1.33 times the expected value during the low flux period at the end of the observation. It should be noted, however, that the predicted reflection fraction for a given coronal height depends strictly on the assumption of an infinitesimal point source, and will be sensitive to the spatial extent of the corona, the luminosity profile as a function of position within the corona, and any motion.

During the short high flux period, the reflection fraction drops below unity, to $R = 0.93_{-0.02}^{+0.02}$ as the flux rises, and $R = 0.88_{-0.02}^{+0.03}$ as it falls. Such a drop in reflection fraction can be understood if, during the short, flare-like high flux state, the corona is accelerated away from the disc, reaching a mildly relativistic velocity. Special relativistic beaming will direct a greater fraction of the emission away from the disc to be observed as part of the continuum \citep{beloborodov,mrk335_flare_paper,1zw1_flare_paper}. Under the simplifying assumption of a point-like X-ray source at a height of 9\rg, reflection fractions of 0.93 and 0.88 correspond to the corona accelerating to $0.24c$ and $0.27c$ on the rise and fall of the flare. 

In addition to the reflection fraction, relativistic beaming of the continuum emission due to the motion of the corona is also manifested in the emissivity profile of the accretion disc. The upward beaming of emission reduces the emissivity of the innermost radii of the disc relative to the outer radii, producing subtle variations in the observed line profiles \citep{dauser+13,gonzalez+2017}. Interestingly, we find that in MCG--6, while the measured reflection fractions imply motion of corona away from the disc, when fitting the emissivity profile of the disc via the \textsc{relxilllpCp} reflection model, the best-fitting velocity of the point source is consistent with zero (note that the reflection fraction is not tied to the height or velocity of the point source; the height and velocity are constrained by the shape of the relativistically broadened emission lines, while the reflection fraction is determined by the strength of the reflection relative to the continuum). This apparent discrepancy can be understood if (as inferred above) the corona is spatially extended. The part of the corona closest to the black hole strongly irradiates the inner disc, while the upper parts of the corona are accelerated away from the black hole, beaming their emission and increasing the flux of the observed continuum with respect to the reflection component of the spectrum.

\subsection{The origin of the flux dips}

The rise in reflection fraction during the dips is consistent with the drop in the scale height or spatial extent of the corona that is inferred from the variation of the accretion disc emissivity profile. During the first and second dips, the height of the corona (in the simplified lamppost model) is $h = 6.6_{-0.2}^{+0.1}$\rg\ with a reflection fraction of $R = 1.65_{-0.04}^{+0.03}$ (compared to the predicted value of $R = 1.7$ for a corona at rest at this height), during the second dip, $h = 5.9_{-0.1}^{+0.1}$\rg\ with $ R = 1.91_{-0.01}^{+0.07}$ (predicted $R = 1.9$), and the effect is most extreme during the third dip when $h = 4.53_{-0.07}^{+0.09}$\rg\ and $R = 4.15_{-0.06}^{+0.03}$.

During the third dip, the predicted reflection fraction for a point source at a height of 4.5\rg\ is only $R = 2.2$, however the larger observed reflection is likely indicative of the corona having finite spatial extent with a significant fraction of the emission originating closer to the black hole (for a point source, a reflection fraction of $R = 4.15$ is obtained when the corona is located 2.5\rg\ from the black hole).

From these findings, we can conclude that the dips in the light curve correspond to time periods in which the corona has collapsed to a confined region around the black hole, enhancing the reflection observed from the innermost regions of the accretion disc. We also see evidence of the corona accelerating away from the disc again during the brief period between the second and third dips, when the reflection fraction briefly drops again to $1.23_{-0.14}^{+0.2}$, with a best-fitting point source height of $17_{-4}^{+3}$\rg\ likely as continued energy dissipation results in further reconfiguration of the magnetic fields that are accelerating the corona.

\section{Discussion}
The coordinated MCG--6-30-15 observing campaign between \textit{XRISM}, \textit{NuSTAR} and \textit{XMM-Newton} (in addition to \textit{Chandra} and \textit{NICER} whose data will be discussed in a subsequent publication) has yielded a rich data set, revealing not just the spectrum, but the variability of the reflection from the innermost accretion disc, multi-component, multi-phase winds and outflows launched by the central engine, and the structure of more distant reflecting material. The combination of the high spectral resolution provided by \textit{XRISM Resolve} (in addition to the \textit{XMM-Newton} Reflection Grating Spectrometer and \textit{Chandra} grating spectrometers) enables detailed measurements of narrow absorption lines, constraining the physical properties of the outflows and enabling these narrow spectral features to be unambiguously separated from the broad iron K line in the reflection spectrum from the inner disc (see \citealt{brenneman+2025} for more details). Moreover, the simultaneous broadband coverage of \textit{XRISM}, \textit{NuSTAR} and \textit{XMM-Newton} enables the full reflection spectrum to be separated from the continuum. The reflection spectrum exhibits Doppler shifts and gravitational redshifts resulting from the orbit of the accretion disc plasma in the strong gravitational field around the black hole, enabling measurements of the spin of the black hole \citep{brenneman_reynolds,reynolds_spin_review} and the structure and geometry of the corona and inner accretion flow \citep{understanding_emis_paper, propagating_lag_paper, plunging_region_paper}.

When considering the reflection from the innermost regions of the accretion disc, the principal advantage of the high resolution spectra provided in the iron K band by \textit{XRISM Resolve} comes from the ability to accurately characterise the narrow Fe\,\textsc{xxv} and Fe\,\textsc{xxvi} absorption lines produced by the ionising outflow components, in addition to the narrow component of the Fe\,K emission lines, produced by reflection from distant material. In CCD-resolution spectra, these narrow spectral features are blended with the underlying broad component of the emission line, either increasing uncertainties in parameters derived from the reflection spectrum (assuming uncertainties in the absorption and distant reflection parameters can be marginalised over), or biasing parameter estimates if these components are not fully included in the spectral model. At CCD resolution, only a single unresolved absorption feature is seen in the spectrum from the outflows, making it almost impossible to measure the ionisation states and column densities of the high ionisation outflow components that do not produce strong spectral features in the soft X-ray band that is covered by the grating spectrometers. With the ability to resolve these components, we find that the underlying broad emission feature is consistent with the relativistic broadened emission line that is expected from the relatively simple model of X-ray reflection from the inner accretion disc, while tightening constraints on parameters derived from this model relative to values previously reported in the literature from analysis of CCD-resolution spectra \citep{brenneman_reynolds,marinucci+2014}. With a complete characterisation of the absorption lines, we find that alternative models in which the broad spectral features in the iron K bands result from spectral curvature produced by a complex structure of low ionisation absorbers \citep[\textit{e.g.}][]{miller+08} are strongly disfavoured.

We find that both the spectrum and its variability can be self-consistently explained by a model in which the X-ray continuum, emitted from the corona of accelerated particles close to the black hole, illuminates the inner accretion disc, producing the characteristic reflection spectrum. The accretion disc extends inwards to the innermost stable orbit around a rapidly-spinning black hole ($a_\star > 0.93$) and the variability observed in the model parameters can be explained by variation in location and velocity of the X-ray emitting corona. No additional component is required to explain the soft X-ray excess (for example Comptonisation of photons in a `warm corona' or warm atmosphere above the accretion disc) and we find that the soft X-ray spectrum can be entirely explained by the combination of Bremsstrahlung emission and relativistically broadened emission lines (notably Fe\,L, O and N) in the reflection spectrum, which are included in the \textsc{relxill} models. Indeed, adding an additional soft excess component to the spectral model representing the Comptonisation of photons within a warm corona, similar to the models described in \citet{done_jin}, is significantly disfavored by the data, increasing the DIC statistic by $\Delta \mathrm{DIC}=+33$ ($\Delta C = -11$ but with 18 additional free parameters). In addition to variation in the corona, we find that the column densities and ionisation states of the outflows vary over the course of the observations. Detailed photoionisation modelling of the outflows and their variability will be presented in a subsequent paper.

\subsection{Geometry, structure and evolution of the corona}
The location, geometry and structure of the corona (\textit{i.e.} the particle acceleration region close to the black hole in which the X-ray continuum is produced) may be constrained via the emissivity profile of the accretion disc (the variation in reflected flux as a function of radius). The emissivity profile corresponds to the illumination pattern of the disc by the corona, and is hence dependent upon the coronal geometry.

A simplified model in which the corona is represented by a point source (a `lamppost') located at variable height, $h$ on the spin axis above the black hole, provides a good description of the time-resolved spectra of MCG--6. A model in which the emissivity profile is parametrised as a twice-broken power law, however, is preferred by the data, providing evidence for the corona having finite spatial extent. A corona that produces X-ray emission through the Comptonisation of UV seed photons emitted by the accretion disc must have finite spatial extent so as to have non-zero cross section for scattering these photons, however the precise size, location and production mechanism of the corona remains unknown. Comparing the twice-broken power law emissivity profiles fit to the data to the predictions of General Relativistic ray tracing simulations \citep[\textit{e.g.}][]{understanding_emis_paper}, and to the equivalent best-fitting point source emissivity profiles, reveals that while some spatial extent of the corona is required, the corona remains compact, and is contained within $\sim 10$\rg\ of the black hole over the majority of the observations.

Such a corona, extended over the surface of the inner accretion disc, has previously been inferred from measurements of the X-ray reflection emissivity profile \citep{1h0707_emis_paper} and the extent of the corona had been found to increase in the brighter flux states \citep{1h0707_var_paper, mrk335_corona_paper}. Similarly, recent analysis of X-ray polarisation measurements in a handful of AGN observed with the \textit{Imaging X-ray Polarization Explorer (IXPE)} provide evidence for a corona extended over the accretion disc in a `slab' or `wedge' geometry, as opposed to a spherical corona located above the black hole that would be akin to the `lamppost' geometry \citep{ixpe_ngc4151,ixpe_mcg5,ixpe_ic4329a}. It is important to stress, however, that such polarisation measurements are made integrating over long exposure times, thus represent the time-averaged polarisation signature dominated by slowly-varying components of the corona and do not account for the evolution of the corona on short timescales, or any structures associated with rapid flaring behaviour.

A short, flare-like high flux period was observed in the X-ray emission from MCG--6 during the campaign, during which the count rate increased by a factor of three over the course of one day. We are able to infer the time evolution of the corona from variations in the disc emissivity and reflection fraction that are fit to the spectra extracted from the different time intervals. In the simplified point source model of the corona, we find a self-consistent evolution between the inferred scale height of the corona and the reflection fraction. The scale height of the corona increases during both the rise and fall of the X-ray flux, while the reflection fraction drops to values below unity, as expected due to relativistic beaming of the emission as the corona accelerates away from the disc. Similar acceleration of the corona away from the black hole and accretion disc has been observed during X-ray flares from the AGN Markarian~335 \citep{mrk335_flare_paper} and I\,Zwicky\,1 \citep{1zw1_nature,1zw1_flare_paper}.

From the drop in reflection fraction during the high-flux period, we infer that the corona accelerates to a velocity of $0.27c$. Similar velocities around $0.3c$ have been associated with ultrafast outflows of ionised gas detected via blueshifted absorption lines. \citet{gu+2025} note that assuming a magnetic flux density of $10^4$\,Gauss within the corona, and a mass density of $10^{-15}$\gpcm, Alfv\'en speed is approximately $0.3c$. It is thus plausible that the material is accelerated by rapid magnetic reconnection in a process analogous to a Solar coronal mass ejection (CME). A similar process could be operating during the X-ray flare in MCG--6, accelerating the X-ray emitting plasma within the corona away from the black hole and accretion disc.

Similarly, allowing for an extended corona via the twice-broken power law model for the disc emissivity profile, we see a decrease in the steepness of the emissivity profile over the innermost radii, and flattening of the middle section of the emissivity profile over a broader range of radii during the high flux state. Moreover, the gradient of the middle section of the emissivity profile during the rise is indicative of the corona extending vertically showing that the corona expands to fill a larger volume as it accelerates away from the disc. Employing an extended corona model, we infer that the corona becomes most extended during the rising state, reaches an extent of $\sim 25$\rg\ (vertical extension of the corona is most likely here, noting the gradient of the middle portion of the emissivity profile). As the X-ray flux drops, the corona extends to $\sim 9$\rg, but has accelerated away from the disc, leading to the reduction in the observed reflection fraction. We note that the extended corona model yields a slightly larger estimate for the size of the corona than the point source model, since a point source model must place the corona closer to the black hole in order to produce the observed steepening of the emissivity profile across the innermost radii of the disc. An extended corona will naturally place emission sites both closer and further from the black hole, so can reproduce both the inner steepening of the profile and the emission from the corona at larger radius. 

The drop in reflection fraction during the high flux states could also be attributed to Comptonisation of the X-ray reflection from the accretion disc by a more spatially extended corona. \citet{comptonisation_paper} show that when the corona is extended over the surface of the accretion disc, a fraction of the reflected photons (including the fluorescent line emission) are expected to be Compton up-scattered by the same population of accelerated particles that produce the X-ray continuum. This results in a drop in the apparent reflection fraction as Compton scattering smooths the reflection spectrum into a continuum that is indistinguishable from the primary power law spectrum. The detection of highly redshifted line photons from an accretion disc extending to the ISCO beneath an extended corona, however, requires that the corona be patchy with a covering factor less than approximately 0.8, such that a sufficient number of redshifted photons can escape to produce the small radius of the ISCO that is inferred in MCG--6. From the calculations presented by \citet{comptonisation_paper}, an increase the covering fraction of a patchy corona from approximately 0.25 to 0.85 could explain the drop in reflection fraction from 1.9 in the low flux state to 0.8 in the high flux state. We however note that the variation in covering fraction of a patchy Comptonising corona does not naturally explain the observed anti-correlation between the reflection fraction and the height/extent of the corona that is derived from the emissivity profile, and does not explain the extremely high reflection fraction measured during the third dip, thus we conclude that in this instance, motion and variation in the spatial extent of the corona between the low and high flux states is the more natural explanation.

It is also interesting to note that when fitting the \textsc{relxilllpCp} point source model with free source velocity parameter to account for relativistic beaming in the computation of the emissivity profile, that the best-fitting source velocity during all time intervals is close to zero. The apparent discrepancy between the source velocity implied by the reflection fraction and the source velocity implied by the inner disc emissivity profile can be resolved if the corona is not only extended, but if the plasma accelerates as it moves away from the base. The slowly-moving base of the corona illuminates the inner disc, while the more rapidly moving upper parts of the corona dominate the directly-observed continuum emission, producing the observed drop in the reflection fraction, as discussed by \citet{understanding_emis_paper}.

Both before and after the high flux state or flare, a series of short dips are observed in the X-ray light curve, which correspond to peaks in the hardness ratio (approximating the overall slope of the spectrum). The best-fitting emissivity profile of the accretion disc (using both the point source and extended corona models) and the enhanced reflection fraction observed during these dips reveal that the corona had collapsed into a confined region, with the bulk of emission arising within just $\sim2.5$\rg\ of the black hole, with the corona accelerating away from the disc again during the brief periods between the dips. During these dips, the intrinsic luminosity of the corona drops, and during the third dip (the most pronounced), we see the spectrum soften from a photon index of $\Gamma = 2.12$ as the flux falls to $\Gamma = 2.19$.  Assuming the temperature of the corona remains constant at the inferred value of 43\keV\ and using the relationship between the photon index, $\Gamma$ and the temperature, $T$, and optical depth to electron scattering, $\tau$, within the corona (\citealt{sunyaev_trumper}, see also \citealt{1zw1_flare_paper}), we find that this softening of the spectrum corresponds to a reduction in optical depth through the corona of 20 per cent as the corona collapses.

\begin{figure*}[ht!]
\includegraphics[width=\textwidth]{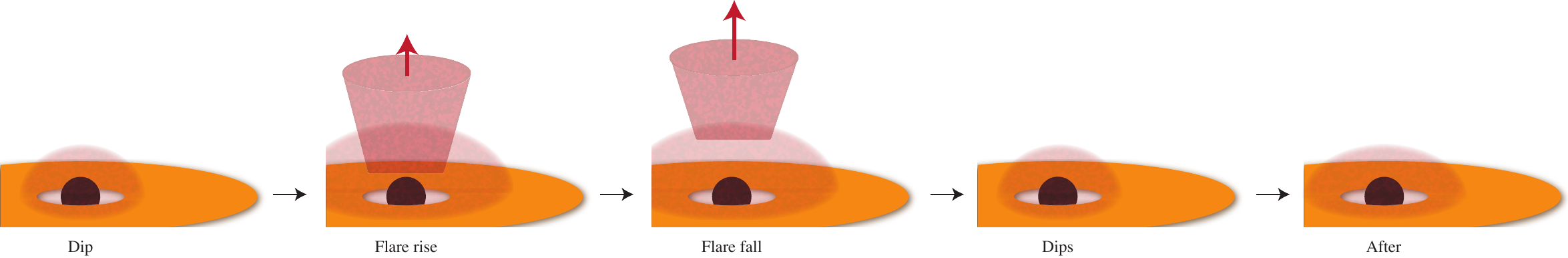}
\caption{The evolution of the corona over the course of the flare-like high flux state and flux dips in MCG--6, inferred from variations in the emissivity profile of the accretion disc and reflection fraction measured from the time-resolved analysis of the time-resolved spectral analysis of the X-rays reflected from the inner accretion disc. During the flux dips, the corona collapsed to a confined region, just a few gravitational radii from the black hole. Strong gravitational light bending from such a compact corona focuses the X-ray emission onto the innermost radii of the disc. As the flare rose, the corona expanded away from the accretion disc, and was accelerated away from the disc on fall of the flare. Relativistic beaming of the X-ray emission away from the disc leads to a drop in the detected reflection fraction.
\label{fig:corona_evo}}
\end{figure*}

A schematic of the evolution of the corona inferred over the course of the flare-like high flux state and the dips in MCG--6 is shown in Figure~\ref{fig:corona_evo}.

The flux dips appear analogous to the extended low flux states that have been observed in 1H\,0707$-$495 \citep{1h0707_jan11} and Mrk~335 \citep{parker_mrk335}. The collapse of the corona to a confined region around the black hole results in the concentration of the coronal X-ray emission onto the innermost regions of the disc due to the strong light bending experienced in close proximity to the black hole. Such low flux states provide a unique view of the strong relativistic effects (light bending, Doppler and gravitational redshifts) at play around a black hole, and enable some of the clearest measurements of the inner accretion disc and the spin of the black hole.

We note also variability in the X-ray flux on shorter timescales, with short peaks appearing on timescales of approximately hours, although due to limited signal-to-noise, we are not able to infer the evolution of the corona on such short timescales from analysis of the reflection spectrum. The corona, however, is likely a highly dynamic system requiring constant injection of energy by, \textit{e.g.} magnetic reconnection to offset the short cooling time via the radiation it emits \citep{fabian+2015,fabian+2017}. The corona we see likely comprises an ensemble of short timescale flares from distinct reconnection events \citep{merloni_fabian}, and while on longer timescales we are likely seeing the evolution of the ensemble as we observe changes in the location, geometry and spatial extent of the corona, on short timescales the variability will start to be dominated by individual flaring events.

\subsection{Impact of variability on black hole spin measurements and reflection parameters}
\label{sec:spin}
\begin{figure*}[p]
    \gridline{
        \fig{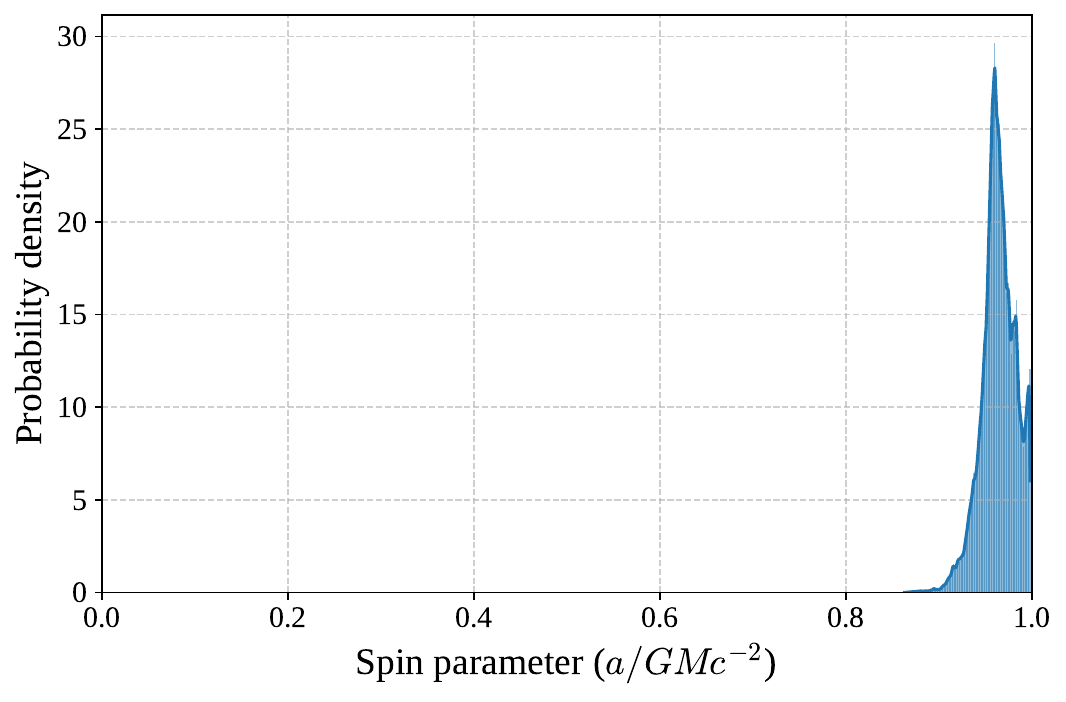}{0.3\textwidth}{}
        \fig{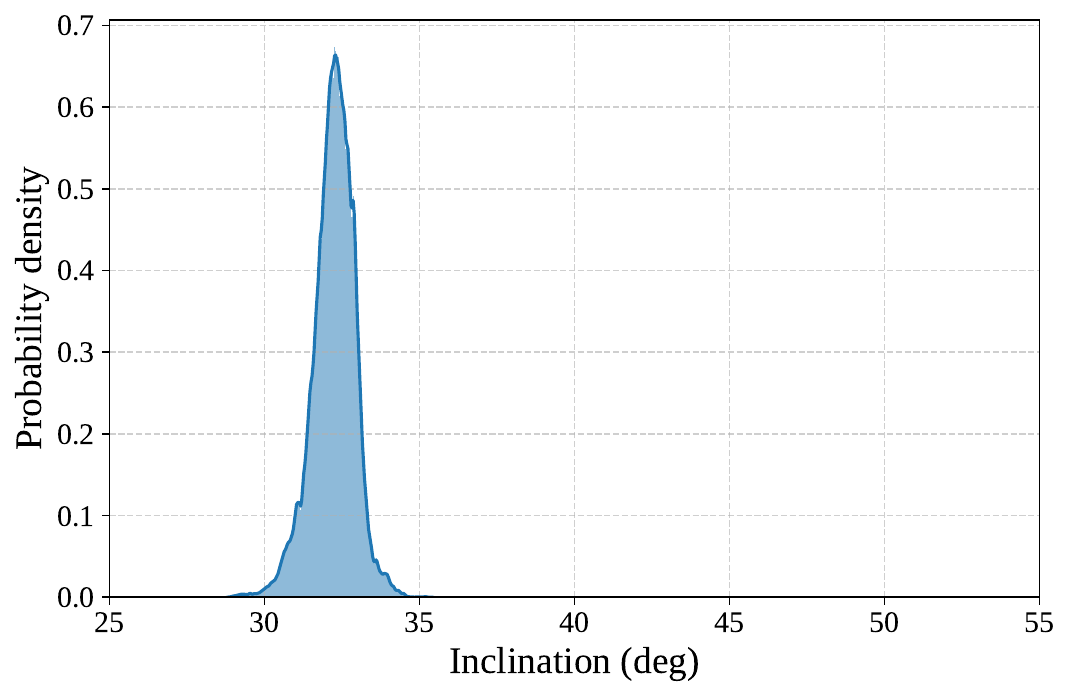}{0.3\textwidth}{(a) Time-resolved, $0.3-55$\keV}
        \fig{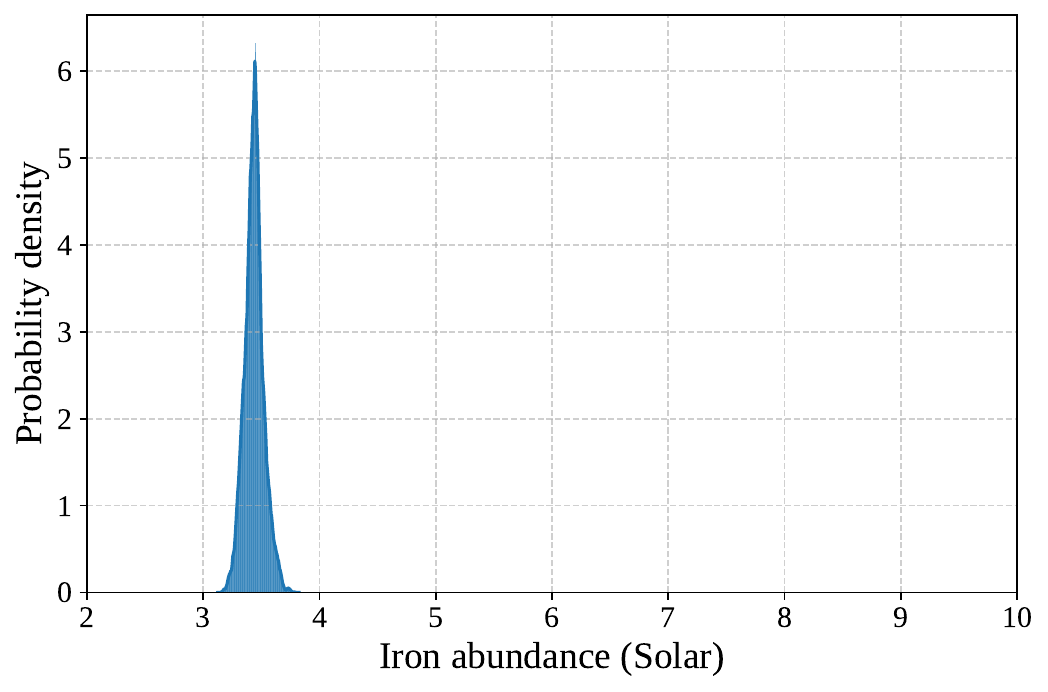}{0.3\textwidth}{}
    }
    \gridline{
        \fig{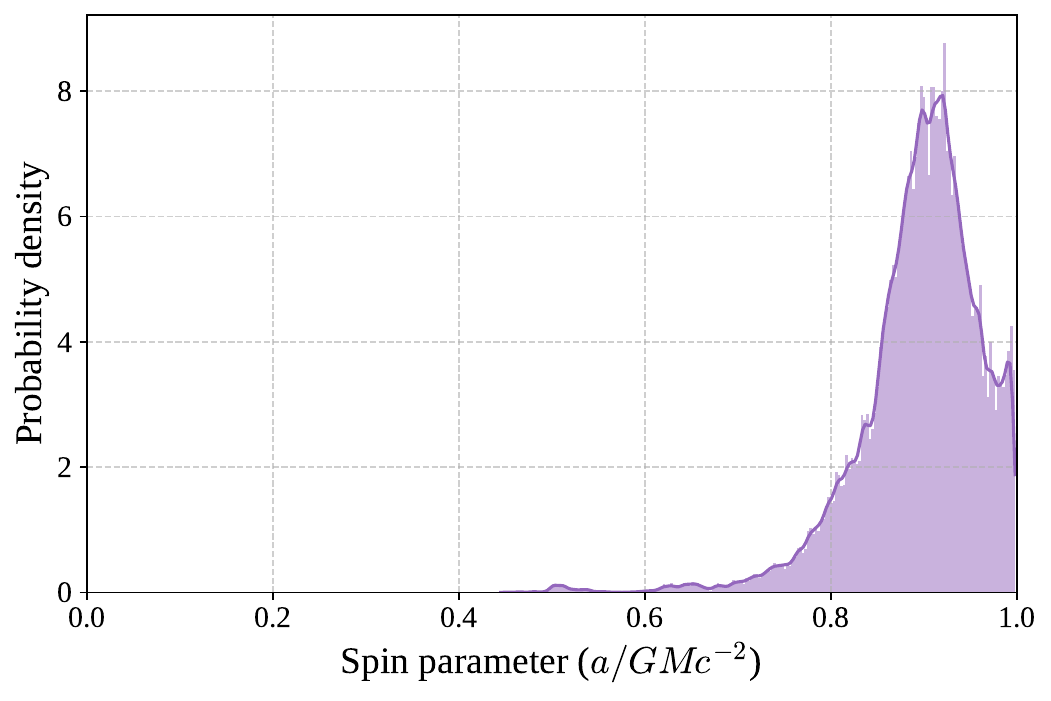}{0.3\textwidth}{}
        \fig{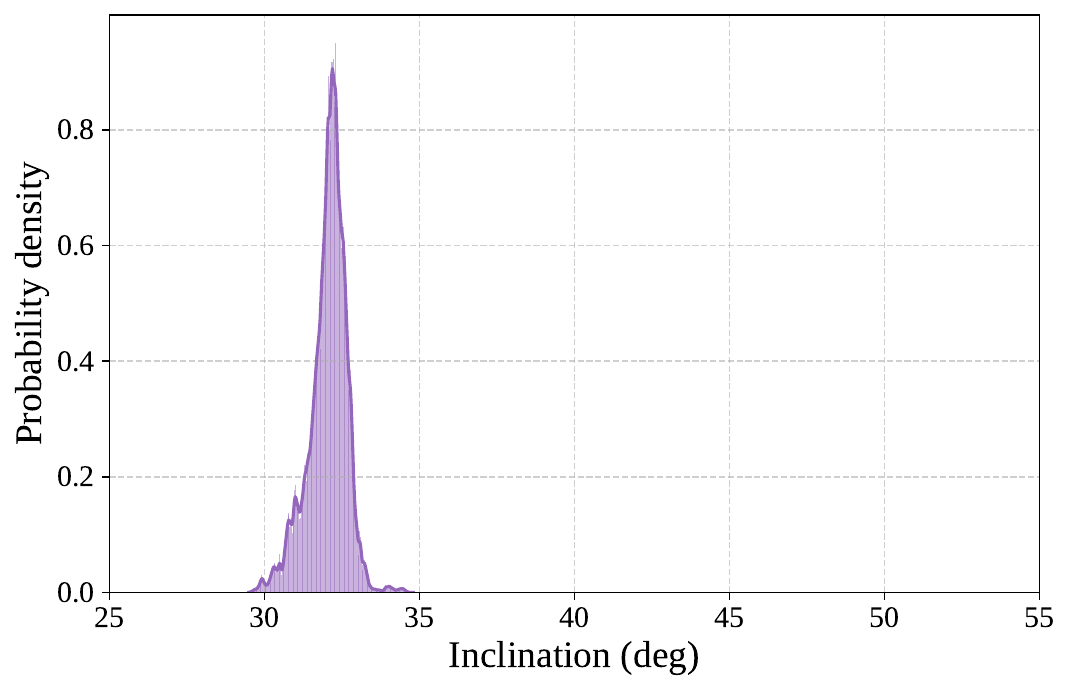}{0.3\textwidth}{(b) Time-resolved, $2-55$\keV}
        \fig{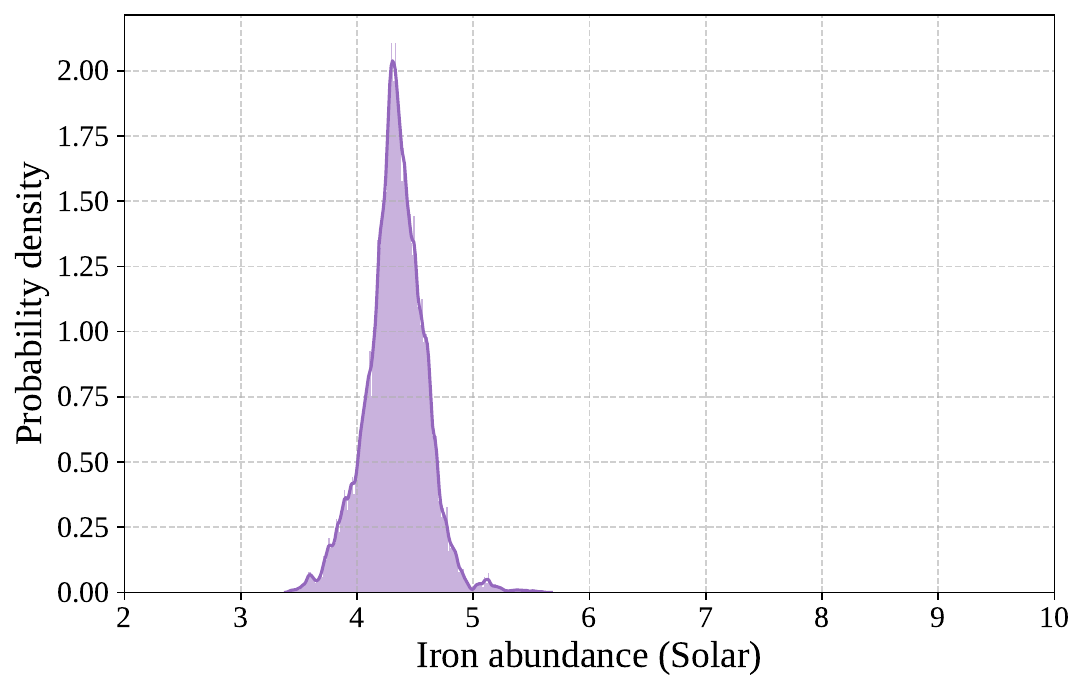}{0.3\textwidth}{}
    }
    \gridline{
        \fig{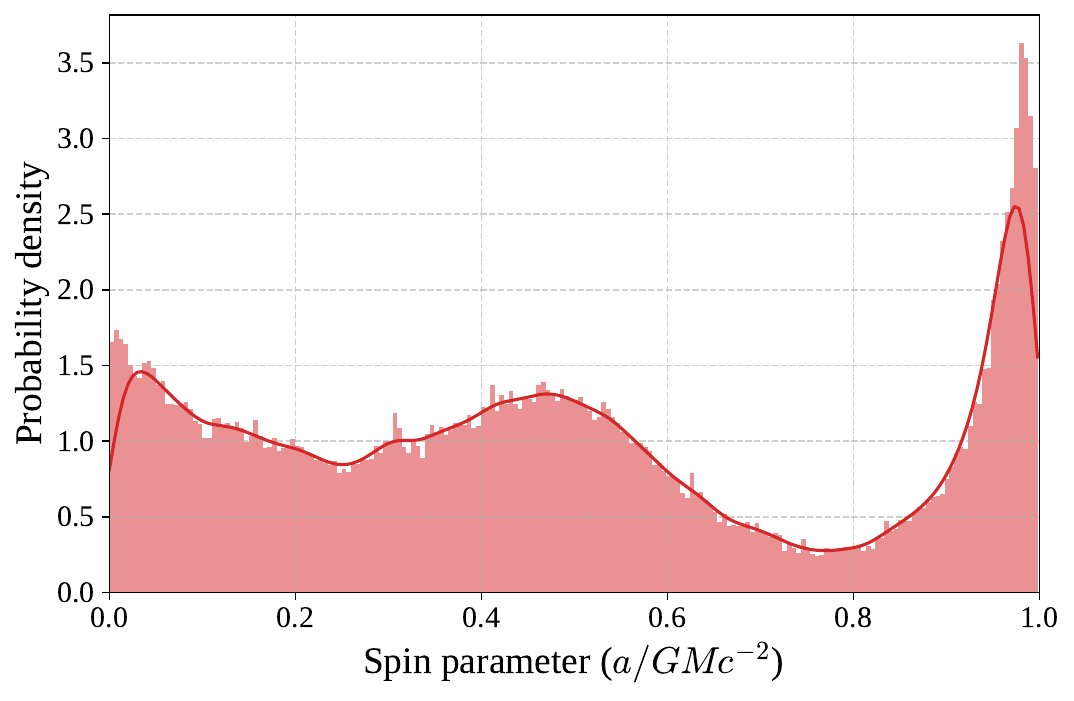}{0.3\textwidth}{}
        \fig{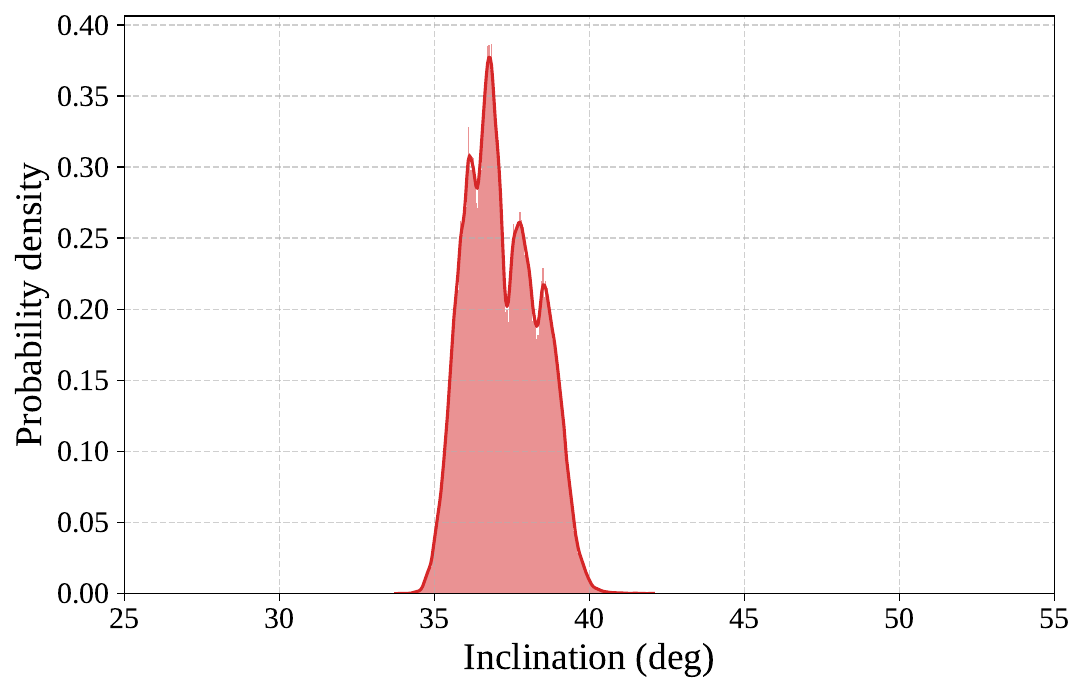}{0.3\textwidth}{(c) Time-averaged, $0.3-55$\keV}
        \fig{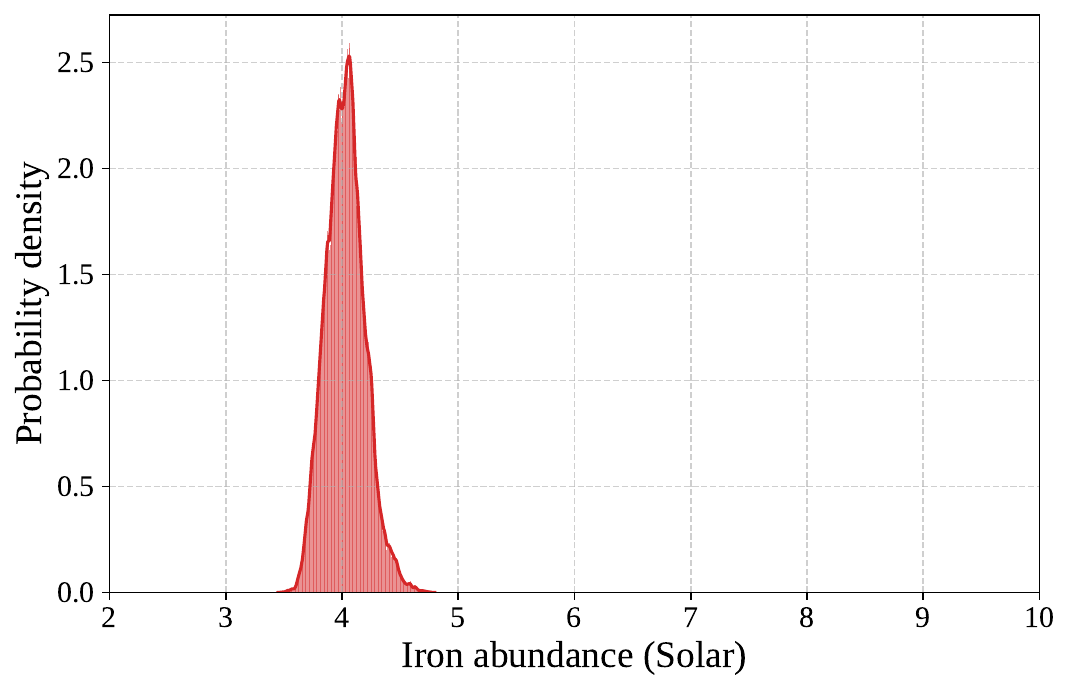}{0.3\textwidth}{}
    }
    \gridline{
        \fig{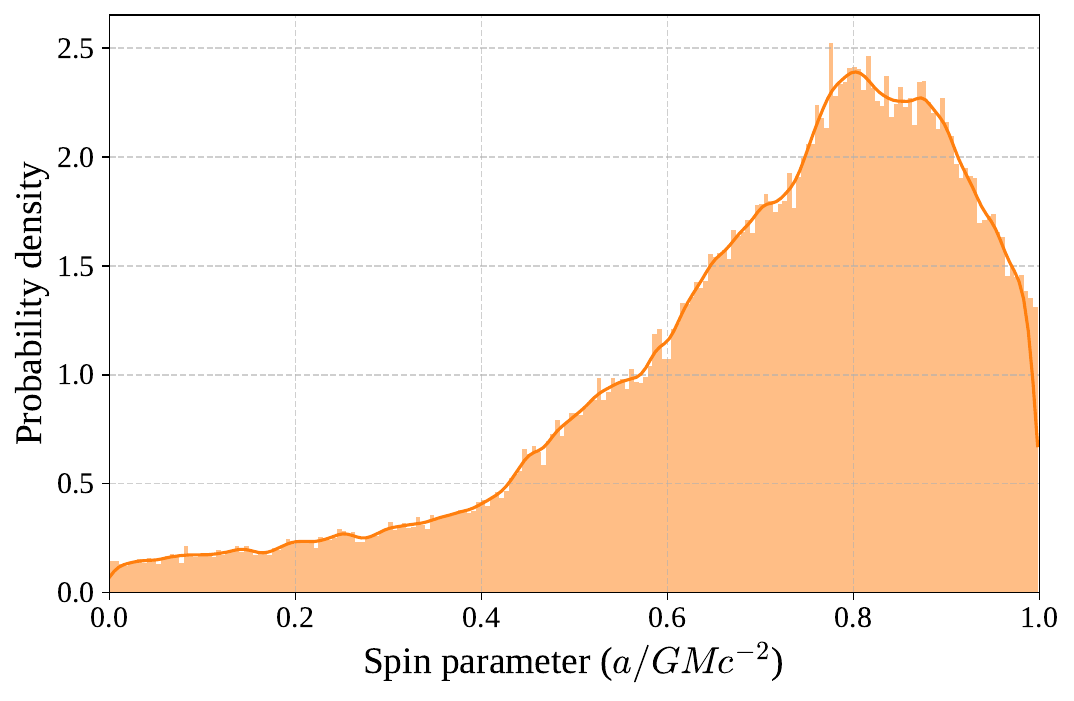}{0.3\textwidth}{}
        \fig{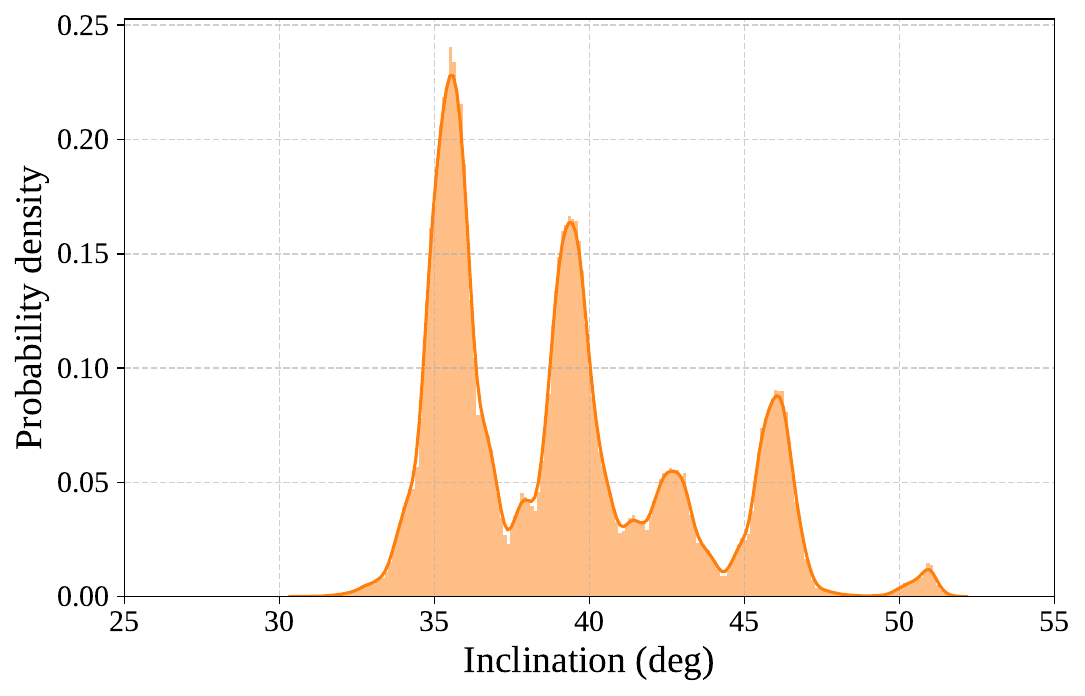}{0.3\textwidth}{(d) Time-averaged, $2-55$\keV}
        \fig{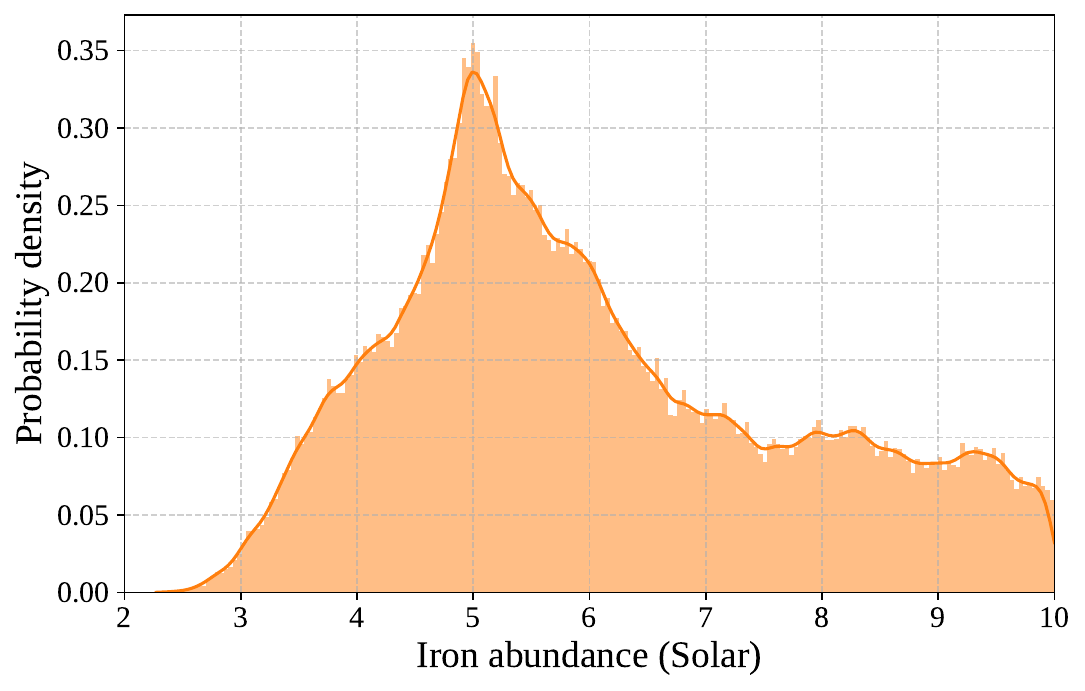}{0.3\textwidth}{}
    }
    \caption{Posterior distributions for the black hole spin, the inclination of the accretion disc, and the iron abundance in the accretion disc plasma in MCG--6, inferred from fitting the reflection model to the (a) time-resolved \textit{XRISM Resolve}, \textit{NuSTAR} and \textit{XMM-Newton} data spanning the 0.3-55\keV\ band, (b) time-resolved \textit{XRISM Resolve} and \textit{NuSTAR} data spanning the more limited 2-55\keV\ band, (c) time-averaged data spanning the 0.3-55\keV\ band, and (d) time-averaged data, spanning the limited 2-55\keV\ band.
        \label{fig:time_var_avg_params}}
\end{figure*}

We note that when fitting the time-resolved spectra, we obtain tighter constraints on the spin of the black hole, in addition to other parameters of the inner disc reflection model, than were obtained when fitting the spectra averaged over the full duration of these same observations \citep{brenneman+2025}. To investigate this effect further, we fit the same model to both the time-resolved and time-averaged spectra both over the full 0.3-55\keV\ energy band (combining \textit{XRISM Resolve}, \textit{NuSTAR} and \textit{XMM-Newton} data), and the restricted 2-55\keV\ band, which includes the two most prominent features of the reflection spectrum: the broad iron K line and the Compton hump (utilising only the \textit{XRISM Resolve} and \textit{NuSTAR} data). In each case, the parameter space was explored using MCMC and the posterior distributions derived for the black hole spin, disc inclination and iron abundance are shown in Fig~\ref{fig:time_var_avg_params}.

When fitting the time-resolved spectra, the black hole is tightly constrained close to maximal spin, with a dimensionless spin parameter $a_\star > 0.93$. On the other hand, when the time-averaged spectra are fit, a maximal value of the spin parameter provides the best fit, the constraint is significantly looser and lower spin values are not formally excluded by the data. We can understand these findings if the location and/or geometry of the corona is varying over the course of the observations, changing the pattern of illumination of the inner accretion disc (\textit{i.e.} the emissivity profile). 

Inference of the black hole spin relies upon identifying the inner edge of the accretion disc (assuming that the disc extends inwards to the innermost stable orbit, or ISCO, which is itself a function of the spin of the black hole) from the maximal redshift detected in the wing of the broad iron K line (and other spectral features when fitting the full reflection spectrum, although it is the iron K line in which gravitational redshifts are manifested most clearly). The broad line sits atop the power law continuum and accurate determination of the extremal redshift relies upon separating the decaying line flux from the underlying continuum. This means that the ability to accurately determine the spin is, in part, dependent upon the emissivity profile of the disc. It has previously been noted that if the innermost accretion disc is weakly illuminated by a corona at a relatively large scale height, there exists a degeneracy in which a model can underestimate the spin of the black hole if the model assumes too much emission from the inner disc relative to the outer disc. In order to fit the data, in this case, the inner disc can be truncated, fitting a lower value of the spin, in order to remove the excess inner disc emission \citep{fabian+2012_cygx1,fabian+2014}.

If the location or geometry of the corona were changing over the course of the observations of MCG--6, the emissivity profile of the accretion disc that describes the time-averaged spectrum would be a superposition of the emissivity profiles that arise from the configuration of the corona at different times. Fitting a single point source model or single smooth power law model would therefore not be able to completely describe the spectrum. Times in which the corona is more extended produce more emission from the outer disc relative to the inner disc, while the flux dips in which the corona were more compact would produce more emission from the inner disc. It is therefore likely that lower spin values are inferred in order to `correct' the relative emissivity between the inner and outer disc when the coronal geometry and emissivity profile are not allowed to vary in time.

We also note that while the values inferred for the disc inclination are consistent between the time-resolved and time-averaged spectra when the full 0.3-55\keV\ bandpass is incorporated, when the time-averaged spectrum is fit over the limited 2-55\keV\ band, the posterior distribution is multi-modal. When fitting the time-averaged broadband data, a higher inclination of $38\pm 1$\,deg is obtained, compared to $32_{-0.7}^{+0.5}$\,deg obtained from the time-resolved fit. Fitting the time-averaged spectrum over the restricted 2-55\keV\ band, however, produces additional peaks in the posterior distribution (corresponding to local minima of the fit statistic). While the best-fit is consistent with the inclination derived from the broadband spectra, inclination values to up 51\,deg may be inferred. The disc inclination is primarily constrained by the blueshifted edge of the broad iron K line, comprised of emission from the rapidly-orbiting material close to the ISCO. As with the black hole spin, significant variation of the emissivity profile over the course of the observation can blur this edge of the line in the time-averaged spectrum, biasing the inferred inclination.

A long-standing puzzle in the analysis of X-ray reflection from the accretion discs in AGN has been extremely high iron abundances required to model the observed spectra, sometimes in excess of ten times the Solar value \citep{garcia+2018}. These findings prompt the question of whether AGN accretion discs really possess such enhanced metallicities (for example if the host galaxy had gone through a significant phase of star formation in the nucleus), or whether there exists a systematic error in the reflection models and analysis. Accounting for densities of the accretion disc plasma above the canonically assumed value of $10^{15}\,\mathrm{cm}^{-3}$ addresses this issue to some degree \citep{xillver_density,jiang_highden}, however we here find that underlying spectral variability can also lead to over-estimation of the iron abundance when the reflection is modelled over only the 2-55\keV\ band. While fitting the time-resolved spectra, or the time-averaged spectra over the full 0.3-55\keV\ band yields iron abundances between three and four times the Solar value, when the time-averaged spectrum is fit over the 2-55\keV\ band, the inferred iron abundance is significantly higher, with the best fit around five times the Solar value, but the derived posterior distribution unbounded to higher values. The iron abundance in the reflection model determines the flux of the iron K and L emission lines, as well as the iron K absorption edge, relative to the continuum, however if the ratio of the reflected to continuum flux (\textit{i.e.} the reflection fraction) is varying due to changes in the coronal geometry, the inferred iron abundance may be biased to account for the relative strength of the emission lines and absorption edges. Including the soft X-ray band mitigates this problem since the iron L line is included in addition to the iron K line, providing additional constraints on the iron abundance, however we note that in the best-fitting model to MCG--6, the soft X-ray band is dominated by reflection from the inner accretion disc. This additional constraint may not be possible if other spectral components dominate the soft X-ray band.

We conclude that in order to obtain the most precise estimates of the black hole spin and other parameters of the reflection spectrum arising from the inner accretion disc, it is necessary to account for any variation over the period over which the spectral data is obtained. This issue is particularly acute for the lower mass supermassive black holes, with masses below $10^7$\Msun, most notably the `complex' subset of narrow line Seyfert 1 AGN \citep{gallo_nls1}, in which spectral variability on short timescales is commonly observed.

Where available, broadband spectral data can also be greatly beneficial in constraining the parameters of the reflection model. The hard X-ray band, from 10-50\keV\ contains the Compton hump, which aids in separating the reflection component from the underlying continuum. In the case that the soft X-ray band is dominated by reflection from the inner disc, and no other emission components contribute to the soft excess, the soft band (containing the iron L line in addition to relativistically broadened lines from elements including nitrogen and oxygen) can help to break a number of degeneracies in the accretion disc density, ionisation parameter and iron abundance.

\section{Conclusions}
Analysing time-resolved, high-resolution X-ray spectra of MCG--6-30-15 obtained by the \textit{XRISM Resolve} microcalorimeter spectrometer, alongside broadband spectra across the 0.3-55\keV\ band obtained with \textit{NuSTAR} and \textit{XMM-Newton}, we find that the X-ray emission from MCG--6 can be well described by a model in which X-ray emission from a corona of accelerated particles illuminates the inner regions of the accretion disc around a rapidly spinning ($a_\star > 0.93$) supermassive black hole. 

Reprocessing of the X-ray continuum by the accretion disc plasma produces a characteristic reflection spectrum, including a broadened iron K emission line whose profile is derived from relativistic effects, including strong light bending, Doppler shifts and gravitational redshifts in the extreme gravitational field just outside the black hole.

The spectral resolution of \textit{XRISM Resolve} unambiguously separates the relativistically broadened iron K line from narrow absorption lines produced by photoionised outflows, and from narrow components of the iron K emission line produced as X-rays are reflected from more distant material.

The corona is compact and resides within $\sim 10$\rg\ over the majority of the observations, although to fully describe the profile of the relativistically broadened reflection features in the spectrum, a model in which the corona has slight spatial extent is preferred over the simplified `lamppost' model in which the corona is represented by a point source.

The X-ray emission from the corona in MCG--6 exhibits rapid variability, and during the coordinated \textit{XRISM}, \textit{NuSTAR} and \textit{XMM-Newton} campaign, a short, flare-like high flux period was observed in the X-ray emission, during which the count rate increased by a factor of three over the course of one day, in addition to a series of short dips, each lasting around three hours, which correspond to peaks in the hardness of the X-ray spectrum. 

Performing a time-resolved analysis of the spectrum over the course of the observations, we are able to infer the evolution and motion of the corona that underlies the observed variability in the X-ray emission. We find that both the X-ray spectrum and its variability can be described by a self-consistent model in which the time evolution of the observed X-ray emission can be described by underlying changes in the luminosity, spatial extent and acceleration of the corona.

We infer that during the rise of the flare-like high flux state, the corona expands and accelerates away from the disc, extending to $\sim 15$\rg\ over the disc, and as the flux falls, is ejected away from the disc at a velocity of approximately $0.27c$, producing an observable reduction in the reflection fraction (the ratio of the reflected flux to the directly observed continuum flux) below unity. During each of the dips, we found that the corona had collapsed to a confined region around the black hole, with an extent of just 2.5\rg\ during the third dip in which the reflection fraction reached the highest value. During these flux dips, the coronal X-ray emission is focused onto the inner regions of the accretion disc, enabling some of the clearest measurements of the relativistic effects imprinted on emission from the innermost radii of the disc.

We find that in order to obtain the most precise estimates of the black hole spin and other parameters of the reflection spectrum arising from the inner accretion disc, it is necessary to account for any spectral variation over the period over which data is obtained. Averaging spectra over periods of significant spectral variability can bias the inferred parameter values, including underestimating the spin of the black hole, and overestimating of the iron abundance within the accretion disc. We therefore recommend that light curves and spectral hardness ratios are examined before modelling the reflection spectrum, and a time-resolved spectral analysis is performed, simultaneously fitting the model to spectra extracted from time intervals identified from the light curves. Such detailed analysis will be important in making the most accurate inferences from high-resolution spectra obtained with \textit{XRISM} and future high resolution X-ray spectrometers.

\begin{deluxetable*}{lllllll}
\digitalasset
\tablecaption{Best-fitting model parameters derived from the joint, time-resolved fit to the \textit{XRISM}, \textit{NuSTAR} and \textit{XMM-Newton} observations of MCG--6. This model consists of \textsc{relxilllpCp}, describing both the directly-observed X-ray continuum emission from the corona, and its reflection off the inner regions of the accretion disc, in addition to absorption from four photoionised outflow components modelled using the spectral synthesis code \textsc{cloudy}. The first two, high-ionisation outflows are detected via Fe\,\textsc{xxv} and Fe\,\textsc{xxvi} absorption lines in the \textit{XRISM} Resolve spectrum, and their velocities are fit to the observed spectrum. The second two, lower-ionisation outflows represent warm absorbers detected in the soft X-ray band whose velocities are fixed at values derived from absorption lines in the \textit{XMM-Newton} RGS spectrum. \textsc{mytorus} describes the narrow components of emission lines in the reflection spectra arising from distant material. \textsc{tbvarabs} models the neutral iron K absorption edge from the dusty ISM in the host galaxy, and \textsc{tbabs} models the Galactic absorption through which MCG--6 is observed. The eight values for variable parameters represent each time interval, while the `tied' value is shown for parameters that are fit simultaneously to the spectra but not allowed to vary between intervals. $^f$ denotes parameters whose values are fixed.\label{tab:param}}
\tablehead{
\colhead{Component} & \colhead{Parameter} & \colhead{Tied} & \colhead{Before} & \colhead{Dip 1} & \colhead{Rise} & \colhead{Fall}
}
\startdata
\textsc{relxilllpCp} & Redshift, $z$                                                                & 0.007749$^f$                   &                                  &                                 &                                   &                                \\
                                  & Black hole spin, $a_\star = cJ/GM^{2}$                                       & $ > 0.93 $                     &                                  &                                 &                                   &                                \\
                                  & Disc inclination, $i$ / deg                                                  & $ 32.3_{-0.7}^{+0.6} $         &                                  &                                 &                                   &                                \\
                                  & Disc density, $\log (n_\mathrm{e} / \mathrm{cm}^{-3})$                       & $ < 15.1 $                     &                                  &                                 &                                   &                                \\
                                  & Iron abundance, $A_\mathrm{Fe} / \mathrm{Solar}$                             & $ 3.44_{-0.15}^{+0.07} $       &                                  &                                 &                                   &                                \\
                                  & Continuum photon index, $\Gamma$                                             &                                & $ 2.102_{-0.011}^{+0.013} $      & $ 2.121_{-0.007}^{+0.009} $     & $ 2.141_{-0.004}^{+0.004} $       & $ 2.125_{-0.003}^{+0.008} $    \\
                                  & Corona temperature, $kT_\mathrm{e} / \mathrm{keV}$                           & $ 43.1_{-0.4}^{+0.7} $         &                                  &                                 &                                   &                                \\
                                  & Corona height, $h / r_\mathrm{g}$                                            &                                & $ 5.1_{-0.4}^{+0.4} $            & $ 6.6_{-0.2}^{+0.1} $           & $ 8.4_{-0.2}^{+0.2} $             & $ 8.8_{-0.1}^{+0.3} $          \\
                                  & Ionisation parameter, $\log(\xi / \mathrm{erg},\mathrm{cm},\mathrm{s}^{-1}$) &                                & $ 1.96_{-0.12}^{+0.06} $         & $ 1.89_{-0.02}^{+0.03} $        & $ 1.85_{-0.02}^{+0.03} $          & $ 1.79_{-0.04}^{+0.02} $       \\
                                  & Reflection fraction, $R$                                                     &                                & $ 1.74_{-0.16}^{+0.17} $         & $ 1.65_{-0.04}^{+0.03} $        & $ 0.93_{-0.02}^{+0.02} $          & $ 0.88_{-0.02}^{+0.03} $       \\
                                  & Normalisation $/10^{-4}$                                                     &                                & $ 5.0_{-0.3}^{+0.5} $            & $ 4.53_{-0.08}^{+0.09} $        & $ 5.82_{-0.09}^{+0.08} $          & $ 6.04_{-0.1}^{+0.08} $        \\ \hline 
\textsc{cloudy}      & Ionisation parameter, $\log(\xi / \mathrm{erg},\mathrm{cm},\mathrm{s}^{-1})$ &                                & $> 4.5$                          & $> 4.5 $                        & $> 4.5 $                          & $ 4.74_{-0.05}^{+0.08} $       \\
                                  & Column density, $\log(n_\mathrm{H} / \mathrm{cm}^{-2})$                      &                                & $ 22.6_{-0.18}^{+0.10} $         & $ 22.7_{-0.1}^{+0.05} $         & $ 22.5_{-0.2}^{+0.04} $           & $ 22.5_{-0.2}^{+0.04} $        \\
                                  & Velocity broadening, $\log(v / \mathrm{km},\mathrm{s}^{-1})$                 &                                & $ 3.01_{-0.09}^{+0.2} $          & $ 3.09_{-0.06}^{+0.04} $        & $ 2.88_{-0.09}^{+0.04} $          & $ 2.81_{-0.07}^{+0.04} $       \\
                                  & Outflow velocity, $v / c$                                                    &                                & $ 0.00752_{-0.0001}^{+0.00003} $ & $ 0.00699_{-0.0002}^{+0.0001} $ & $ 0.00766_{-0.00002}^{+0.00002} $ & $ 0.0067_{-0.0001}^{+0.0002} $ \\ \hline 
\textsc{cloudy}      & Ionisation parameter, $\log(\xi / \mathrm{erg},\mathrm{cm},\mathrm{s}^{-1})$ &                                & $> 4.5 $                         & $> 4.5 $                        & $> 5.5 $                          & $> 5.5 $                       \\
                                  & Column density, $\log(n_\mathrm{H} / \mathrm{cm}^{-2})$                      &                                & $ 23.0_{-0.2}^{+0.1} $           & $ 22.3_{-0.4}^{+0.2} $          & $ 23.1_{-0.5}^{+0.1} $            & $ 22.5_{-0.3}^{+0.2} $         \\
                                  & Velocity broadening, $\log(v / \mathrm{km},\mathrm{s}^{-1})$                 &                                & $ 4.19_{-0.25}^{+0.01} $         & $ 2.22_{-0.03}^{+0.08} $        & $< 1.58 $                         & $ 2.25_{-0.02}^{+0.05} $       \\
                                  & Outflow velocity, $v / c$                                                    &                                & $ 0.0427_{-0.012}^{+0.003} $     & $ 0.0635_{-0.001}^{+0.0004} $   & $ 0.0618_{-0.0008}^{+0.002} $     & $ 0.0621_{-0.0005}^{+0.0006} $ \\ \hline 
\textsc{cloudy}      & Ionisation parameter, $\log(\xi / \mathrm{erg},\mathrm{cm},\mathrm{s}^{-1})$ &                                & $ 2.72_{-0.04}^{+0.05} $         & $ 2.75_{-0.02}^{+0.02} $        & $ 2.74_{-0.02}^{+0.05} $          & $ 2.74_{-0.03}^{+0.05} $       \\
                                  & Column density, $\log(n_\mathrm{H} / \mathrm{cm}^{-2})$                      &                                & $ 22.28_{-0.03}^{+0.03} $        & $ 22.3_{-0.02}^{+0.02} $        & $ 22.1_{-0.007}^{+0.02} $         & $ 22_{-0.1}^{+0.01} $          \\ \hline 
\textsc{cloudy}      & Ionisation parameter, $\log(\xi / \mathrm{erg},\mathrm{cm},\mathrm{s}^{-1})$ &                                & $ 1.8_{-0.3}^{+0.3} $            & $ 1.85_{-0.04}^{+0.03} $        & $ 1.9_{-0.02}^{+0.08} $           & $ 1.83_{-0.05}^{+0.04} $       \\
                                  & Column density, $\log(n_\mathrm{H} / \mathrm{cm}^{-2})$                      &                                & $< 20.7 $                        & $< 20.7 $                       & $ 20.5_{-0.1}^{+0.2} $            & $ 20.5_{-0.2}^{+0.1} $         \\ \hline 
\textsc{mytorus}     & Line component normalisation                                                 & $ 0.0103_{-0.0002}^{+0.0002} $ &                                  &                                 &                                   &                                \\ \hline 
\textsc{tbvarabs}    & Dusty ISM iron column, $10^{17}~n_\mathrm{Fe} / \mathrm{cm}^{-2}$            & $ 3.73_{-0.08}^{+0.03} $       &                                  &                                 &                                   &                                \\ \hline 
\textsc{tbabs}       & Galactic absorption, $n_\mathrm{H}$ / $10^{20}~\mathrm{cm}^{-2}$             & $ 3.59^f $                     &                                  &                                 &                                   &                     \\          
\enddata
\end{deluxetable*}

\begin{deluxetable*}{lllllll}
\rotate
\digitalasset
\tablecaption{Continuation of Table~\ref{tab:param}}
\tablehead{
\colhead{Component} & \colhead{Parameter} & \colhead{Dip 2} & \colhead{Between} & \colhead{Dip 3} & \colhead{After}
}
\startdata
\textsc{relxilllpCp} & Redshift, $z$                                                                &                              &                                 &                                 &                                   \\
                                  & Black hole spin, $a_\star = cJ/GM^{2}$                                       &                              &                                 &                                 &                                   \\
                                  & Disc inclination, $i$ / deg                                                  &                              &                                 &                                 &                                   \\
                                  & Disc density, $\log (n_\mathrm{e} / \mathrm{cm}^{-3})$                       &                              &                                 &                                 &                                   \\
                                  & Iron abundance, $A_\mathrm{Fe} / \mathrm{Solar}$                             &                              &                                 &                                 &                                   \\
                                  & Continuum photon index, $\Gamma$                                             & $ 2.131_{-0.007}^{+0.007} $  & $ 2.12_{-0.02}^{+0.03} $        & $ 2.189_{-0.014}^{+0.008} $     & $ 2.174_{-0.008}^{+0.008} $       \\
                                  & Corona temperature, $kT_\mathrm{e} / \mathrm{keV}$                           &                              &                                 &                                 &                                   \\
                                  & Corona height, $h / r_\mathrm{g}$                                            & $ 5.9_{-0.1}^{+0.1} $        & $ 17_{-3}^{+4} $                & $ 4.53_{-0.07}^{+0.09} $        & $ 6.65_{-0.1}^{+0.07} $           \\
                                  & Ionisation parameter, $\log(\xi / \mathrm{erg},\mathrm{cm},\mathrm{s}^{-1}$) & $ 1.29_{-0.02}^{+0.01} $     & $< 1.7$                         & $ < 1.2 $                       & $ 1.90_{-0.04}^{+0.03} $          \\
                                  & Reflection fraction, $R$                                                     & $ 1.91_{-0.01}^{+0.07} $     & $ 1.23_{-0.14}^{+0.2} $         & $ 4.15_{-0.06}^{+0.03} $        & $ 1.90_{-0.01}^{+0.04} $          \\
                                  & Normalisation $/10^{-4}$                                                     & $ 4.49_{-0.09}^{+0.07} $     & $ 3.3_{-0.2}^{+0.2} $           & $ 5.45_{-0.04}^{+0.07} $        & $ 4.99_{-0.02}^{+0.02} $          \\ \hline
\textsc{cloudy}      & Ionisation parameter, $\log(\xi / \mathrm{erg},\mathrm{cm},\mathrm{s}^{-1})$ & $ 4.63_{-0.07}^{+0.07} $     & $ 4.63_{-0.07}^{+0.2} $         & $ 4.74_{-0.09}^{+0.1} $         & $ 5.04_{-0.07}^{+0.06} $          \\
                                  & Column density, $\log(n_\mathrm{H} / \mathrm{cm}^{-2})$                      & $ 22.5_{-0.3}^{+0.4} $       & $ 22.8_{-0.1}^{+0.2} $          & $ 22.4_{-0.4}^{+0.2} $          & $ 22.7_{-0.2}^{+0.08} $           \\
                                  & Velocity broadening, $\log(v / \mathrm{km},\mathrm{s}^{-1})$                 & $ 1.99_{-0.03}^{+0.06} $     & $ 2.8_{-0.2}^{+0.09} $          & $ 2.4_{-0.05}^{+0.03} $         & $ 2.7_{-0.05}^{+0.02} $           \\
                                  & Outflow velocity, $v / c$                                                    & $ < 0.3$                     & $ 0.00603_{-0.0002}^{+0.0004} $ & $ 0.00634_{-0.0002}^{+0.0007} $ & $ 0.00758_{-0.00002}^{+0.00002} $ \\ \hline
\textsc{cloudy}      & Ionisation parameter, $\log(\xi / \mathrm{erg},\mathrm{cm},\mathrm{s}^{-1})$ & $ 4.94_{-0.1}^{+0.07} $      & $> 5.5$                         & $< 4.76 $                       & $ 4.91_{-0.06}^{+0.04} $          \\
                                  & Column density, $\log(n_\mathrm{H} / \mathrm{cm}^{-2})$                      & $> 22.9 $                    & $ 23.2_{-0.2}^{+0.1} $          & $ 22.7_{-0.3}^{+0.1} $          & $ 22.6_{-0.2}^{+0.08} $           \\
                                  & Velocity broadening, $\log(v / \mathrm{km},\mathrm{s}^{-1})$                 & $ 1.54_{-0.02}^{+0.04} $     & $ 3.1_{-0.3}^{+0.2} $           & $ 3.91_{-0.1}^{+0.1} $          & $ 3.52_{-0.08}^{+0.06} $          \\
                                  & Outflow velocity, $v / c$                                                    & $ 0.0642_{-0.001}^{+0.002} $ & $ 0.0487_{-0.0007}^{+0.003} $   & $ 0.0463_{-0.0008}^{+0.0004} $  & $ 0.0688_{-0.002}^{+0.0005} $     \\ \hline
\textsc{cloudy}      & Ionisation parameter, $\log(\xi / \mathrm{erg},\mathrm{cm},\mathrm{s}^{-1})$ & $ 2.71_{-0.04}^{+0.03} $     & $ 3.6_{-1.1}^{+1.3} $           & $ 3.22_{-0.07}^{+0.07} $        & $> 5.35 $                         \\
                                  & Column density, $\log(n_\mathrm{H} / \mathrm{cm}^{-2})$                      & $ 22.1_{-0.04}^{+0.02} $     & $ 21.9_{-1.3}^{+0.08} $         & $ 21.9_{-0.3}^{+0.2} $          & $< 20 $                           \\ \hline
\textsc{cloudy}      & Ionisation parameter, $\log(\xi / \mathrm{erg},\mathrm{cm},\mathrm{s}^{-1})$ & $ 4.16_{-0.02}^{+0.05} $     & $ < 2.3 $                       & $ 1.01_{-0.005}^{+0.007} $      & $ 1.38_{-0.04}^{+0.04} $          \\
                                  & Column density, $\log(n_\mathrm{H} / \mathrm{cm}^{-2})$                      & $ 22.3_{-0.1}^{+0.1} $       & $< 20.6$                        & $< 20 $                         & $ 21.7_{-0.1}^{+0.1} $            \\ \hline
\textsc{mytorus}     & Line component normalisation                                                 &                              &                                 &                                 &                                   \\ \hline
\textsc{tbvarabs}    & Dusty ISM iron column, $10^{17}~n_\mathrm{Fe} / \mathrm{cm}^{-2}$            &                              &                                 &                                 &                                   \\ \hline
\textsc{tbabs}       & Galactic absorption, $n_\mathrm{H}$ / $10^{20}~\mathrm{cm}^{-2}$             &         \\     
\enddata
\end{deluxetable*}

\begin{table*}
\caption{The parameters of the accretion disc emissivity profile, fit to the observed spectrum in time intervals, using the \textsc{relxillCp3} model. The emissivity profile is parametrised as a twice-broken power law ($\epsilon \propto r^{-q}$), with indices $q_\mathrm{in}$, $q_\mathrm{mid}$ and $q_\mathrm{out}$, and break radii $r_{\mathrm{br},1}$ and $r_{\mathrm{br},2}$. A steep inner index ($q_\mathrm{in} > 3$) is indicative of gravitational light bending and the associated blueshift enhancing the irradiation of the innermost radii of the accretion disc. When the corona is radially extended over the surface of the accretion disc, $q_\mathrm{mid}\sim 0$ and the outer break radius, $r_{\mathrm{br},2}$ indicates the radial extent of the corona over the disc. During the dip, the emissivity profile is consistent with just a single break in the power law, as the two break radii are seen to converge, as would be produced by a compact corona with little extension over the disc.\label{tab:relxill3}}
\begin{center}
\begin{tabular}{lccccc}
\hline
\colhead{Parameter} &  \colhead{Before} & \colhead{Rise} & \colhead{Fall} & \colhead{Dip 3} & \colhead{After} \\
\hline
$q_\mathrm{in}$ & $>9$ & $7.0\pm 1.4$ & $9\pm 4$ & $4.1\pm 0.8$ & $5.2_{-0.4}^{+4}$\\
$r_{\mathrm{br},1}$ & $3.7\pm0.3$ & $3.8\pm0.4$ & $3.6\pm1.0$ & $5\pm 2$ & $3.4\pm 0.9$\\
$q_\mathrm{mid}$ & $<2$ & $<0.6$ & $2.0\pm 0.5$ & -- & $0.16\pm 0.09$\\
$r_{\mathrm{br},2}$ & $5\pm 1$ & $9.4\pm 1.7$ & $15.0_{-0.4}^{+18}$ & $5.8_{-0.6}^{+6}$ & $4.8\pm 1.9$\\
$q_\mathrm{out}$ & $2.9\pm 0.4$ & $3.5\pm 0.3$ & $2.1\pm 1.3$ & $4.5\pm 1.7$ & $2.8\pm 0.3$\\
\hline
\end{tabular}
\end{center}
\end{table*}

\begin{acknowledgments}
This work was supported by the NASA \textit{XRISM} Guest Scientist program under grant 80NSSC23K0631. The material is based upon work supported by NASA under award number 80GSFC24M0006. AJ acknowledges support from NASA Grant 80NSSC25K0082. We thank the directors and the scheduling and operations teams of \textit{NuSTAR} and \textit{XMM-Newton} for the coordination of the observations that greatly enhanced the scientific return of the {\it XRISM} data. DRW thanks Stanford University, the Stanford Research Computing Center and The Ohio State University College of Arts and Sciences for providing computational resources and support. We thank the anonymous referee for their valuable feedback on the original version of this manuscript.
\end{acknowledgments}

\begin{contribution}
DRW performed the time-resolved spectral analysis and authored this manuscript. LWB and DRW performed the time-resolved spectral analysis, deriving the baseline spectral model. AO developed and computed absorption models for the ionised outflows using \textsc{cloudy}. All co-authors contributed to discussions on the data reduction, analysis and interpretation, and contributed to the authoring and proof-reading of this article.


\end{contribution}

%
\facilities{\textit{XRISM}, \textit{NuSTAR}, \textit{XMM-Newton}}

\software{XSPEC \citep{xspec}, Heasoft, XMM-Newton Science Analysis System (SAS), Relxill \citep{garcia+2014,dauser+2014}, Cloudy \citep{cloudy,cloudy2023}, Veusz, Seaborn, Matplotlib}

\bibliography{agn}{}

@ARTICLE{fabian+2017,
       author = {{Fabian}, A.~C. and {Lohfink}, A. and {Belmont}, R. and {Malzac}, J. and {Coppi}, P.},
        title = "{Properties of AGN coronae in the NuSTAR era - II. Hybrid plasma}",
      journal = {\mnras},
         year = 2017,
        month = may,
       volume = {467},
       number = {3},
        pages = {2566-2570},
          doi = {10.1093/mnras/stx221},
archivePrefix = {arXiv},
       eprint = {1701.06774},
 primaryClass = {astro-ph.HE},
       adsurl = {https://ui.adsabs.harvard.edu/abs/2017MNRAS.467.2566F}
}

@ARTICLE{comptonisation_paper,
       author = {{Wilkins}, D.~R. and {Gallo}, L.~C.},
        title = "{The Comptonization of accretion disc X-ray emission: consequences for X-ray reflection and the geometry of AGN coronae}",
      journal = {\mnras},
         year = 2015,
        month = mar,
       volume = {448},
       number = {1},
        pages = {703-712},
          doi = {10.1093/mnras/stu2524},
archivePrefix = {arXiv},
       eprint = {1412.0015},
 primaryClass = {astro-ph.HE},
       adsurl = {https://ui.adsabs.harvard.edu/abs/2015MNRAS.448..703W}
}

@INPROCEEDINGS{xrism_resolve,
       author = {{Ishisaki}, Yoshitaka and {Kelley}, Richard L. and {Awaki}, Hisamitsu and {Balleza}, Jesus C. and {Barnstable}, Kim R. and {Bialas}, Thomas G. and {Boissay-Malaquin}, Rozenn and {Brown}, Gregory V. and {Canavan}, Edgar R. and {Cumbee}, Renata S. and {Carnahan}, Timothy M. and {Chiao}, Meng P. and {Comber}, Brian J. and {Costantini}, Elisa and {den Herder}, Jan-Willem and {Dercksen}, Johannes and {de Vries}, Cor P. and {DiPirro}, Michael J. and {Eckart}, Megan E. and {Ezoe}, Yuichiro and {Ferrigno}, Carlo and {Fujimoto}, Ryuichi and {Gorter}, Nathalie and {Graham}, Steven M. and {Grim}, Martin and {Hartz}, Leslie S. and {Hayakawa}, Ryota and {Hayashi}, Takayuki and {Hell}, Natalie and {Hoshino}, Akio and {Ichinohe}, Yuto and {Ishida}, Manabu and {Ishikawa}, Kumi and {James}, Bryan L. and {Kenyon}, Steven J. and {Kilbourne}, Caroline A. and {Kimball}, Mark O. and {Kitamoto}, Shunji and {Leutenegger}, Maurice A. and {Maeda}, Yoshitomo and {McCammon}, Dan and {Miko}, Joseph J. and {Mizumoto}, Misaki and {Okajima}, Takashi and {Okamoto}, Atsushi and {Paltani}, Stephane and {Porter}, Frederick S. and {Sato}, Kosuke and {Sato}, Toshiki and {Sawada}, Makoto and {Shinozaki}, Keisuke and {Shipman}, Russell and {Shirron}, Peter J. and {Sneiderman}, Gary A. and {Soong}, Yang and {Szymkiewicz}, Richard and {Szymkowiak}, Andrew E. and {Takei}, Yoh and {Tamura}, Keisuke and {Tsujimoto}, Masahiro and {Uchida}, Yuusuke and {Wasserzug}, Stephen and {Witthoeft}, Michael C. and {Wolfs}, Rob and {Yamada}, Shinya and {Yasuda}, Susumu},
        title = "{Status of resolve instrument onboard X-Ray Imaging and Spectroscopy Mission (XRISM)}",
    booktitle = {Space Telescopes and Instrumentation 2022: Ultraviolet to Gamma Ray},
         year = 2022,
       editor = {{den Herder}, Jan-Willem A. and {Nikzad}, Shouleh and {Nakazawa}, Kazuhiro},
       series = {Society of Photo-Optical Instrumentation Engineers (SPIE) Conference Series},
       volume = {12181},
        month = aug,
          eid = {121811S},
        pages = {121811S},
          doi = {10.1117/12.2630654},
       adsurl = {https://ui.adsabs.harvard.edu/abs/2022SPIE12181E..1SI}
}

@ARTICLE{xrism,
       author = {{Tashiro}, Makoto S.},
        title = "{XRISM: X-ray imaging and spectroscopy mission}",
      journal = {International Journal of Modern Physics D},
         year = 2022,
        month = jan,
       volume = {31},
       number = {2},
          eid = {2230001},
        pages = {2230001},
          doi = {10.1142/S0218271822300014},
       adsurl = {https://ui.adsabs.harvard.edu/abs/2022IJMPD..3130001T}
}

@ARTICLE{gu+2025,
       author = {{Gu}, Liyi and {Fukumura}, Keigo and {Kaastra}, Jelle and {Eckart}, Megan and {Ballhausen}, Ralph and {Behar}, Ehud},
        title = "{Delving into the depths of NGC 3783 withj XRISM: III. Birth of an ultrafast outflow during a soft flare}",
      journal = {\aap~submitted},
         year = 2025,
}

@ARTICLE{ixpe_ic4329a,
       author = {{Ingram}, A. and {Ewing}, M. and {Marinucci}, A. and {Tagliacozzo}, D. and {Rosario}, D.~J. and {Veledina}, A. and {Kim}, D.~E. and {Marin}, F. and {Bianchi}, S. and {Poutanen}, J. and {Matt}, G. and {Marshall}, H.~L. and {Ursini}, F. and {De Rosa}, A. and {Petrucci}, P.-O. and {Madejski}, G. and {Barnouin}, T. and {Gesu}, L. Di and {Dov{\v{c}}iak}, M. and {Gianolli}, V.~E. and {Krawczynski}, H. and {Loktev}, V. and {Middei}, R. and {Podgorny}, J. and {Puccetti}, S. and {Ratheesh}, A. and {Soffitta}, P. and {Tombesi}, F. and {Ehlert}, S.~R. and {Massaro}, F. and {Agudo}, I. and {Antonelli}, L.~A. and {Bachetti}, M. and {Baldini}, L. and {Baumgartner}, W.~H. and {Bellazzini}, R. and {Bongiorno}, S.~D. and {Bonino}, R. and {Brez}, A. and {Bucciantini}, N. and {Capitanio}, F. and {Castellano}, S. and {Cavazzuti}, E. and {Chen}, C.-T. and {Ciprini}, S. and {Costa}, E. and {Del Monte}, E. and {Lalla}, N. Di and {Marco}, A. Di and {Donnarumma}, I. and {Doroshenko}, V. and {Enoto}, T. and {Evangelista}, Y. and {Fabiani}, S. and {Ferrazzoli}, R. and {Garc{\'\i}a}, J.~A. and {Gunji}, S. and {Heyl}, J. and {Iwakiri}, W. and {Jorstad}, S.~G. and {Kaaret}, P. and {Karas}, V. and {Kislat}, F. and {Kitaguchi}, T. and {Kolodziejczak}, J.~J. and {Monaca}, F. La and {Latronico}, L. and {Liodakis}, I. and {Maldera}, S. and {Manfreda}, A. and {Marscher}, A.~P. and {Mitsuishi}, I. and {Mizuno}, T. and {Muleri}, F. and {Negro}, M. and {Ng}, C.-Y. and {O'Dell}, S.~L. and {Omodei}, N. and {Oppedisano}, C. and {Papitto}, A. and {Pavlov}, G.~G. and {Peirson}, A.~L. and {Perri}, M. and {Pesce-Rollins}, M. and {Pilia}, M. and {Possenti}, A. and {Ramsey}, B.~D. and {Rankin}, J. and {Roberts}, O.~J. and {Romani}, R.~W. and {Sgr{\`o}}, C. and {Slane}, P. and {Spandre}, G. and {Swartz}, D.~A. and {Tamagawa}, T. and {Tavecchio}, F. and {Taverna}, R. and {Tawara}, Y. and {Tennant}, A.~F. and {Thomas}, N.~E. and {Trois}, A. and {Tsygankov}, S.~S. and {Turolla}, R. and {Vink}, J. and {Weisskopf}, M.~C. and {Wu}, K. and {Xie}, F. and {Zane}, S.},
        title = "{The X-ray polarization of the Seyfert 1 galaxy IC 4329A}",
      journal = {\mnras},
         year = 2023,
        month = nov,
       volume = {525},
       number = {4},
        pages = {5437-5449},
          doi = {10.1093/mnras/stad2625},
archivePrefix = {arXiv},
       eprint = {2305.13028},
 primaryClass = {astro-ph.HE},
       adsurl = {https://ui.adsabs.harvard.edu/abs/2023MNRAS.525.5437I}
}

@ARTICLE{ixpe_mcg5,
       author = {{Tagliacozzo}, D. and {Marinucci}, A. and {Ursini}, F. and {Matt}, G. and {Bianchi}, S. and {Baldini}, L. and {Barnouin}, T. and {Cavero Rodriguez}, N. and {De Rosa}, A. and {Di Gesu}, L. and {Dov{\v{c}}iak}, M. and {Harper}, D. and {Ingram}, A. and {Karas}, V. and {Kim}, D.~E. and {Krawczynski}, H. and {Madejski}, G. and {Marin}, F. and {Middei}, R. and {Marshall}, H.~L. and {Muleri}, F. and {Panagiotou}, C. and {Petrucci}, P.-O. and {Podgorny}, J. and {Poutanen}, J. and {Puccetti}, S. and {Soffitta}, P. and {Tombesi}, F. and {Veledina}, A. and {Zhang}, W. and {Agudo}, I. and {Antonelli}, L.~A. and {Bachetti}, M. and {Baumgartner}, W.~H. and {Bellazzini}, R. and {Bongiorno}, S.~D. and {Bonino}, R. and {Brez}, A. and {Bucciantini}, N. and {Capitanio}, F. and {Castellano}, S. and {Cavazzuti}, E. and {Chen}, C.-T. and {Ciprini}, S. and {Costa}, E. and {Del Monte}, E. and {Di Lalla}, N. and {Di Marco}, A. and {Donnarumma}, I. and {Doroshenko}, V. and {Ehlert}, S.~R. and {Enoto}, T. and {Evangelista}, Y. and {Fabiani}, S. and {Ferrazzoli}, R. and {Garcia}, J.~A. and {Gunji}, S. and {Heyl}, J. and {Iwakiri}, W. and {Jorstad}, S.~G. and {Kaaret}, P. and {Kislat}, F. and {Kitaguchi}, T. and {Kolodziejczak}, J.~J. and {La Monaca}, F. and {Latronico}, L. and {Liodakis}, I. and {Maldera}, S. and {Manfreda}, A. and {Marscher}, A.~P. and {Massaro}, F. and {Mitsuishi}, I. and {Mizuno}, T. and {Negro}, M. and {Ng}, C.-Y. and {O'Dell}, S.~L. and {Omodei}, N. and {Oppedisano}, C. and {Papitto}, A. and {Pavlov}, G.~G. and {Peirson}, A.~L. and {Perri}, M. and {Pesce-Rollins}, M. and {Pilia}, M. and {Possenti}, A. and {Ramsey}, B.~D. and {Rankin}, J. and {Ratheesh}, A. and {Roberts}, O.~J. and {Romani}, R.~W. and {Sgr{\`o}}, C. and {Slane}, P. and {Spandre}, G. and {Swartz}, D.~A. and {Tamagawa}, T. and {Tavecchio}, F. and {Taverna}, R. and {Tawara}, Y. and {Tennant}, A.~F. and {Thomas}, N.~E. and {Trois}, A. and {Tsygankov}, S.~S. and {Turolla}, R. and {Vink}, J. and {Weisskopf}, M.~C. and {Wu}, K. and {Xie}, F. and {Zane}, S.},
        title = "{The geometry of the hot corona in MCG-05-23-16 constrained by X-ray polarimetry}",
      journal = {\mnras},
         year = 2023,
        month = nov,
       volume = {525},
       number = {3},
        pages = {4735-4743},
          doi = {10.1093/mnras/stad2627},
archivePrefix = {arXiv},
       eprint = {2305.10213},
 primaryClass = {astro-ph.HE},
       adsurl = {https://ui.adsabs.harvard.edu/abs/2023MNRAS.525.4735T}
}

@ARTICLE{ixpe_ngc4151,
       author = {{Gianolli}, V.~E. and {Kim}, D.~E. and {Bianchi}, S. and {Ag{\'\i}s-Gonz{\'a}lez}, B. and {Madejski}, G. and {Marin}, F. and {Marinucci}, A. and {Matt}, G. and {Middei}, R. and {Petrucci}, P.-O. and {Soffitta}, P. and {Tagliacozzo}, D. and {Tombesi}, F. and {Ursini}, F. and {Barnouin}, T. and {De Rosa}, A. and {Di Gesu}, L. and {Ingram}, A. and {Loktev}, V. and {Panagiotou}, C. and {Podgorny}, J. and {Poutanen}, J. and {Puccetti}, S. and {Ratheesh}, A. and {Veledina}, A. and {Zhang}, W. and {Agudo}, I. and {Antonelli}, L.~A. and {Bachetti}, M. and {Baldini}, L. and {Baumgartner}, W.~H. and {Bellazzini}, R. and {Bongiorno}, S.~D. and {Bonino}, R. and {Brez}, A. and {Bucciantini}, N. and {Capitanio}, F. and {Castellano}, S. and {Cavazzuti}, E. and {Chen}, C.-T. and {Ciprini}, S. and {Costa}, E. and {Del Monte}, E. and {Di Lalla}, N. and {Di Marco}, A. and {Donnarumma}, I. and {Doroshenko}, V. and {Dov{\v{c}}iak}, M. and {Ehlert}, S.~R. and {Enoto}, T. and {Evangelista}, Y. and {Fabiani}, S. and {Ferrazzoli}, R. and {Garc{\'\i}a}, J.~A. and {Gunji}, S. and {Heyl}, J. and {Iwakiri}, W. and {Jorstad}, S.~G. and {Kaaret}, P. and {Karas}, V. and {Kislat}, F. and {Kitaguchi}, T. and {Kolodziejczak}, J.~J. and {Krawczynski}, H. and {La Monaca}, F. and {Latronico}, L. and {Liodakis}, I. and {Maldera}, S. and {Manfreda}, A. and {Marscher}, A.~P. and {Marshall}, H.~L. and {Massaro}, F. and {Mitsuishi}, I. and {Mizuno}, T. and {Muleri}, F. and {Negro}, M. and {Ng}, C.-Y. and {O'Dell}, S.~L. and {Omodei}, N. and {Oppedisano}, C. and {Papitto}, A. and {Pavlov}, G.~G. and {Peirson}, A.~L. and {Perri}, M. and {Pesce-Rollins}, M. and {Pilia}, M. and {Possenti}, A. and {Ramsey}, B.~D. and {Rankin}, J. and {Roberts}, O.~J. and {Romani}, R.~W. and {Sgr{\`o}}, C. and {Slane}, P. and {Spandre}, G. and {Swartz}, D.~A. and {Tamagawa}, T. and {Tavecchio}, F. and {Taverna}, R. and {Tawara}, Y. and {Tennant}, A.~F. and {Thomas}, N.~E. and {Trois}, A. and {Tsygankov}, S.~S. and {Turolla}, R. and {Vink}, J. and {Weisskopf}, M.~C. and {Wu}, K. and {Xie}, F. and {Zane}, S.},
        title = "{Uncovering the geometry of the hot X-ray corona in the Seyfert galaxy NGC 4151 with IXPE}",
      journal = {\mnras},
         year = 2023,
        month = aug,
       volume = {523},
       number = {3},
        pages = {4468-4476},
          doi = {10.1093/mnras/stad1697},
archivePrefix = {arXiv},
       eprint = {2303.12541},
 primaryClass = {astro-ph.GA},
       adsurl = {https://ui.adsabs.harvard.edu/abs/2023MNRAS.523.4468G}
}

@ARTICLE{iwasawa+99,
       author = {{Iwasawa}, K. and {Fabian}, A.~C. and {Young}, A.~J. and {Inoue}, H. and {Matsumoto}, C.},
        title = "{Variation of the broad X-ray iron line inMCG-6-30-15 during a flare}",
      journal = {\mnras},
         year = 1999,
        month = jun,
       volume = {306},
       number = {1},
        pages = {L19-L24},
          doi = {10.1046/j.1365-8711.1999.02671.x},
archivePrefix = {arXiv},
       eprint = {astro-ph/9904078},
 primaryClass = {astro-ph},
       adsurl = {https://ui.adsabs.harvard.edu/abs/1999MNRAS.306L..19I}
}

@ARTICLE{martocchia_matt,
       author = {{Martocchia}, Andrea and {Matt}, Giorgio},
        title = "{Iron Kalpha line intensity from accretion discs around rotating black holes}",
      journal = {\mnras},
         year = 1996,
        month = oct,
       volume = {282},
       number = {4},
        pages = {L53-L57},
          doi = {10.1093/mnras/282.4.L53},
       adsurl = {https://ui.adsabs.harvard.edu/abs/1996MNRAS.282L..53M}
}

@INPROCEEDINGS{garcia+2018,
       author = {{Garc{\'\i}a}, J.~A. and {Kallman}, T.~R. and {Bautista}, M. and {Mendoza}, C. and {Deprince}, J. and {Palmeri}, P. and {Quinet}, P.},
        title = "{The Problem of the High Iron Abundance in Accretion Disks around Black Holes}",
    booktitle = {Workshop on Astrophysical Opacities},
         year = 2018,
       series = {Astronomical Society of the Pacific Conference Series},
       volume = {515},
        month = aug,
        pages = {282},
          doi = {10.48550/arXiv.1805.00581},
archivePrefix = {arXiv},
       eprint = {1805.00581},
 primaryClass = {astro-ph.HE},
       adsurl = {https://ui.adsabs.harvard.edu/abs/2018ASPC..515..282G}
}

@ARTICLE{lee+2001,
       author = {{Lee}, Julia C. and {Ogle}, Patrick M. and {Canizares}, Claude R. and {Marshall}, Herman L. and {Schulz}, Norbert S. and {Morales}, Raquel and {Fabian}, Andrew C. and {Iwasawa}, Kazushi},
        title = "{Revealing the Dusty Warm Absorber in MCG -6-30-15 with the Chandra High-Energy Transmission Grating}",
      journal = {\apjl},
         year = 2001,
        month = jun,
       volume = {554},
       number = {1},
        pages = {L13-L17},
          doi = {10.1086/320912},
archivePrefix = {arXiv},
       eprint = {astro-ph/0101065},
 primaryClass = {astro-ph},
       adsurl = {https://ui.adsabs.harvard.edu/abs/2001ApJ...554L..13L}
}

@ARTICLE{blustin+2005,
       author = {{Blustin}, A.~J. and {Page}, M.~J. and {Fuerst}, S.~V. and {Branduardi-Raymont}, G. and {Ashton}, C.~E.},
        title = "{The nature and origin of Seyfert warm absorbers}",
      journal = {\aap},
         year = 2005,
        month = feb,
       volume = {431},
        pages = {111-125},
          doi = {10.1051/0004-6361:20041775},
archivePrefix = {arXiv},
       eprint = {astro-ph/0411297},
 primaryClass = {astro-ph},
       adsurl = {https://ui.adsabs.harvard.edu/abs/2005A&A...431..111B}
}

@ARTICLE{reynolds+1995,
       author = {{Reynolds}, C.~S. and {Fabian}, A.~C. and {Nandra}, K. and {Inoue}, H. and {Kunieda}, H. and {Iwasawa}, K.},
        title = "{ASCA PV observations of the Seyfert 1 galaxy MCG-6-30-15: rapid variability of the warm absorber}",
      journal = {\mnras},
         year = 1995,
        month = dec,
       volume = {277},
       number = {3},
        pages = {901-912},
          doi = {10.1093/mnras/277.3.901},
archivePrefix = {arXiv},
       eprint = {astro-ph/9506086},
 primaryClass = {astro-ph},
       adsurl = {https://ui.adsabs.harvard.edu/abs/1995MNRAS.277..901R}
}

@ARTICLE{tombesi+2010,
       author = {{Tombesi}, F. and {Cappi}, M. and {Reeves}, J.~N. and {Palumbo}, G.~G.~C. and {Yaqoob}, T. and {Braito}, V. and {Dadina}, M.},
        title = "{Evidence for ultra-fast outflows in radio-quiet AGNs. I. Detection and statistical incidence of Fe K-shell absorption lines}",
      journal = {\aap},
         year = 2010,
        month = oct,
       volume = {521},
          eid = {A57},
        pages = {A57},
          doi = {10.1051/0004-6361/200913440},
archivePrefix = {arXiv},
       eprint = {1006.2858},
 primaryClass = {astro-ph.HE},
       adsurl = {https://ui.adsabs.harvard.edu/abs/2010A&A...521A..57T}
}

@ARTICLE{laha+2021,
       author = {{Laha}, Sibasish and {Reynolds}, Christopher S. and {Reeves}, James and {Kriss}, Gerard and {Guainazzi}, Matteo and {Smith}, Randall and {Veilleux}, Sylvain and {Proga}, Daniel},
        title = "{Ionized outflows from active galactic nuclei as the essential elements of feedback}",
      journal = {Nature Astronomy},
         year = 2021,
        month = jan,
       volume = {5},
        pages = {13-24},
          doi = {10.1038/s41550-020-01255-2},
archivePrefix = {arXiv},
       eprint = {2012.06945},
 primaryClass = {astro-ph.GA},
       adsurl = {https://ui.adsabs.harvard.edu/abs/2021NatAs...5...13L}
}

@ARTICLE{holczer+2010,
       author = {{Holczer}, Tomer and {Behar}, Ehud and {Arav}, Nahum},
        title = "{X-ray Absorption Analysis of MCG -6-30-15: Discerning Three Kinematic Systems}",
      journal = {\apj},
         year = 2010,
        month = jan,
       volume = {708},
       number = {2},
        pages = {981-994},
          doi = {10.1088/0004-637X/708/2/981},
archivePrefix = {arXiv},
       eprint = {0910.2402},
 primaryClass = {astro-ph.GA},
       adsurl = {https://ui.adsabs.harvard.edu/abs/2010ApJ...708..981H}
}

@ARTICLE{brenneman+2025,
       author = {{Brenneman}, Laura W. and {Wilkins}, Daniel R. and {Ogorzalek}, Anna and {Rogantini}, Daniele and {Fabian}, Andrew C. and {Garcia}, Javier A.},
        title = "{A Sharper View of the X-ray Spectrum of MCG--6-30-15 with
  XRISM, XMM-Newton and NuSTAR}",
      journal = {\apj~submitted},
         year = 2025,
}

@ARTICLE{xiang+2025,
       author = {{Xiang}, Xin and {Miller}, Jon M. and {Behar}, Ehud and {Boissay-Malaquin}, Rozenn and {Brenneman}, Laura and {Buhariwalla}, Margaret and {Byun}, Doyee and {Done}, Chris and {Gallo}, Luigi and {Gerolymatou}, Dimitra and {Hagen}, Scott and {Kaastra}, Jelle and {Paltani}, Stephane and {Porter}, Frederick S. and {Mushotzky}, Richard and {Noda}, Hirofumi and {Mehdipour}, Missagh and {Minezaki}, Takeo and {Tashiro}, Makoto and {Zoghbi}, Abderahmen},
        title = "{XRISM Spectroscopy of Accretion-driven Wind Feedback in NGC 4151}",
      journal = {\apjl},
         year = 2025,
        month = aug,
       volume = {988},
       number = {2},
          eid = {L54},
        pages = {L54},
          doi = {10.3847/2041-8213/adee9b},
       adsurl = {https://ui.adsabs.harvard.edu/abs/2025ApJ...988L..54X}
}

@ARTICLE{xu+2025,
       author = {{Xu}, Yerong and {Gallo}, Luigi C and {Hagino}, Kouichi and {Reeves}, James N and {Tombesi}, Francesco and {Mizumoto}, Misaki and {Luminari}, Alfredo and {Gonzalez}, Adam G and {Behar}, Ehud and {Boissay-Malaquin}, Rozenn and {Braito}, Valentina and {Cond{\'o}}, Pierpaolo and {Done}, Chris and {Miyamoto}, Aiko and {Mizukawa}, Ryuki and {Odaka}, Hirokazu and {Sato}, Riki and {Tanimoto}, Atsushi and {Tashiro}, Makoto and {Yaqoob}, Tahir and {Yamada}, Satoshi},
        title = "{Unraveling the structure of the stratified ultra-fast outflows in PDS 456 with XRISM}",
      journal = {\pasj},
         year = 2025,
        month = jul,
          doi = {10.1093/pasj/psaf070},
archivePrefix = {arXiv},
       eprint = {2506.05273},
 primaryClass = {astro-ph.HE},
       adsurl = {https://ui.adsabs.harvard.edu/abs/2025PASJ..tmp...73X}
}

@ARTICLE{mehdipour+2025,
       author = {{Mehdipour}, Missagh and {Kaastra}, Jelle S. and {Eckart}, Megan E. and {Gu}, Liyi and {Ballhausen}, Ralf and {Behar}, Ehud and {Diez}, Camille M. and {Fukumura}, Keigo and {Guainazzi}, Matteo and {Hagino}, Kouichi and {Kallman}, Timothy R. and {Kara}, Erin and {Li}, Chen and {Miller}, Jon M. and {Mizumoto}, Misaki and {Noda}, Hirofumi and {Ogawa}, Shoji and {Panagiotou}, Christos and {Tanimoto}, Atsushi and {Zhao}, Keqin},
        title = "{Delving into the depths of NGC 3783 with XRISM: I. Kinematic and ionization structure of the highly ionized outflows}",
      journal = {\aap},
         year = 2025,
        month = jul,
       volume = {699},
          eid = {A228},
        pages = {A228},
          doi = {10.1051/0004-6361/202555623},
archivePrefix = {arXiv},
       eprint = {2506.09395},
 primaryClass = {astro-ph.HE},
       adsurl = {https://ui.adsabs.harvard.edu/abs/2025A&A...699A.228M}
}

@ARTICLE{ngc4151_xrism,
       author = {{Xrism Collaboration} and {Audard}, Marc and {Awaki}, Hisamitsu and {Ballhausen}, Ralf and {Bamba}, Aya and {Behar}, Ehud and {Boissay-Malaquin}, Rozenn and {Brenneman}, Laura and {Brown}, Gregory V. and {Corrales}, Lia and {Costantini}, Elisa and {Cumbee}, Renata and {Diaz Trigo}, Maria and {Done}, Chris and {Dotani}, Tadayasu and {Ebisawa}, Ken and {Eckart}, Megan E. and {Eckert}, Dominique and {Enoto}, Teruaki and {Eguchi}, Satoshi and {Ezoe}, Yuichiro and {Foster}, Adam and {Fujimoto}, Ryuichi and {Fujita}, Yutaka and {Fukazawa}, Yasushi and {Fukushima}, Kotaro and {Furuzawa}, Akihiro and {Gallo}, Luigi and {Garc{\'\i}a}, Javier A. and {Gu}, Liyi and {Guainazzi}, Matteo and {Hagino}, Kouichi and {Hamaguchi}, Kenji and {Hatsukade}, Isamu and {Hayashi}, Katsuhiro and {Hayashi}, Takayuki and {Hell}, Natalie and {Hodges-Kluck}, Edmund and {Hornschemeier}, Ann and {Ichinohe}, Yuto and {Ishida}, Manabu and {Ishikawa}, Kumi and {Ishisaki}, Yoshitaka and {Kaastra}, Jelle and {Kallman}, Timothy and {Kara}, Erin and {Katsuda}, Satoru and {Kanemaru}, Yoshiaki and {Kelley}, Richard and {Kilbourne}, Caroline and {Kitamoto}, Shunji and {Kobayashi}, Shogo and {Kohmura}, Takayoshi and {Kubota}, Aya and {Leutenegger}, Maurice and {Loewenstein}, Michael and {Maeda}, Yoshitomo and {Markevitch}, Maxim and {Matsumoto}, Hironori and {Matsushita}, Kyoko and {McCammon}, Dan and {McNamara}, Brian and {Mernier}, Fran{\c{c}}ois and {Miller}, Eric D. and {Miller}, Jon M. and {Mitsuishi}, Ikuyuki and {Mizumoto}, Misaki and {Mizuno}, Tsunefumi and {Mori}, Koji and {Mukai}, Koji and {Murakami}, Hiroshi and {Mushotzky}, Richard and {Nakajima}, Hiroshi and {Nakazawa}, Kazuhiro and {Ness}, Jan-Uwe and {Nobukawa}, Kumiko and {Nobukawa}, Masayoshi and {Noda}, Hirofumi and {Odaka}, Hirokazu and {Ogawa}, Shoji and {Ogorzalek}, Anna and {Okajima}, Takashi and {Ota}, Naomi and {Paltani}, Stephane and {Petre}, Robert and {Plucinsky}, Paul and {Porter}, Frederick S. and {Pottschmidt}, Katja and {Sato}, Kosuke and {Sato}, Toshiki and {Sawada}, Makoto and {Seta}, Hiromi and {Shidatsu}, Megumi and {Simionescu}, Aurora and {Smith}, Randall and {Suzuki}, Hiromasa and {Szymkowiak}, Andrew and {Takahashi}, Hiromitsu and {Takeo}, Mai and {Tamagawa}, Toru and {Tamura}, Keisuke and {Tanaka}, Takaaki and {Tanimoto}, Atsushi and {Tashiro}, Makoto and {Terada}, Yukikatsu and {Terashima}, Yuichi and {Tsuboi}, Yohko and {Tsujimoto}, Masahiro and {Tsunemi}, Hiroshi and {Tsuru}, Takeshi and {Uchida}, Hiroyuki and {Uchida}, Nagomi and {Uchida}, Yuusuke and {Uchiyama}, Hideki and {Ueda}, Yoshihiro and {Uno}, Shinichiro and {Vink}, Jacco and {Watanabe}, Shin and {Williams}, Brian J. and {Yamada}, Satoshi and {Yamada}, Shinya and {Yamaguchi}, Hiroya and {Yamaoka}, Kazutaka and {Yamasaki}, Noriko and {Yamauchi}, Makoto and {Yamauchi}, Shigeo and {Yaqoob}, Tahir and {Yoneyama}, Tomokage and {Yoshida}, Tessei and {Yukita}, Mihoko and {Zhuravleva}, Irina and {Xiang}, Xin and {Minezaki}, Takeo and {Buhariwalla}, Margaret and {Gerolymatou}, Dimitra and {Hagen}, Scott},
        title = "{XRISM Spectroscopy of the Fe K{\ensuremath{\alpha}} Emission Line in the Seyfert Active Galactic Nucleus NGC 4151 Reveals the Disk, Broad-line Region, and Torus}",
      journal = {\apjl},
         year = 2024,
        month = sep,
       volume = {973},
       number = {1},
          eid = {L25},
        pages = {L25},
          doi = {10.3847/2041-8213/ad7397},
archivePrefix = {arXiv},
       eprint = {2408.14300},
 primaryClass = {astro-ph.HE},
       adsurl = {https://ui.adsabs.harvard.edu/abs/2024ApJ...973L..25X}
}

@ARTICLE{young+2005,
       author = {{Young}, A.~J. and {Lee}, J.~C. and {Fabian}, A.~C. and {Reynolds}, C.~S. and {Gibson}, R.~R. and {Canizares}, C.~R.},
        title = "{A Chandra HETGS Spectral Study of the Iron K Bandpass in MCG -6-30-15: A Narrow View of the Broad Iron Line}",
      journal = {\apj},
         year = 2005,
        month = oct,
       volume = {631},
       number = {2},
        pages = {733-740},
          doi = {10.1086/432607},
archivePrefix = {arXiv},
       eprint = {astro-ph/0506082},
 primaryClass = {astro-ph},
       adsurl = {https://ui.adsabs.harvard.edu/abs/2005ApJ...631..733Y}
}

@ARTICLE{fabian+2002,
       author = {{Fabian}, A.~C. and {Vaughan}, S. and {Nandra}, K. and {Iwasawa}, K. and {Ballantyne}, D.~R. and {Lee}, J.~C. and {De Rosa}, A. and {Turner}, A. and {Young}, A.~J.},
        title = "{A long hard look at MCG-6-30-15 with XMM-Newton}",
      journal = {\mnras},
         year = 2002,
        month = sep,
       volume = {335},
       number = {1},
        pages = {L1-L5},
          doi = {10.1046/j.1365-8711.2002.05740.x},
archivePrefix = {arXiv},
       eprint = {astro-ph/0206095},
 primaryClass = {astro-ph},
       adsurl = {https://ui.adsabs.harvard.edu/abs/2002MNRAS.335L...1F}
}

@ARTICLE{marinucci+2014,
       author = {{Marinucci}, A. and {Matt}, G. and {Miniutti}, G. and {Guainazzi}, M. and {Parker}, M.~L. and {Brenneman}, L. and {Fabian}, A.~C. and {Kara}, E. and {Arevalo}, P. and {Ballantyne}, D.~R. and {Boggs}, S.~E. and {Cappi}, M. and {Christensen}, F.~E. and {Craig}, W.~W. and {Elvis}, M. and {Hailey}, C.~J. and {Harrison}, F.~A. and {Reynolds}, C.~S. and {Risaliti}, G. and {Stern}, D.~K. and {Walton}, D.~J. and {Zhang}, W.},
        title = "{The Broadband Spectral Variability of MCG-6-30-15 Observed by NuSTAR and XMM-Newton}",
      journal = {\apj},
         year = 2014,
        month = may,
       volume = {787},
       number = {1},
          eid = {83},
        pages = {83},
          doi = {10.1088/0004-637X/787/1/83},
archivePrefix = {arXiv},
       eprint = {1404.3561},
 primaryClass = {astro-ph.HE},
       adsurl = {https://ui.adsabs.harvard.edu/abs/2014ApJ...787...83M}
}

@ARTICLE{reynolds+2004,
       author = {{Reynolds}, Christopher S. and {Wilms}, J{\"o}rn and {Begelman}, Mitchell C. and {Staubert}, R{\"u}diger and {Kendziorra}, Eckhard},
        title = "{On the deep minimum state in the Seyfert galaxy MCG-6-30-15}",
      journal = {\mnras},
         year = 2004,
        month = apr,
       volume = {349},
       number = {4},
        pages = {1153-1166},
          doi = {10.1111/j.1365-2966.2004.07596.x},
archivePrefix = {arXiv},
       eprint = {astro-ph/0401305},
 primaryClass = {astro-ph},
       adsurl = {https://ui.adsabs.harvard.edu/abs/2004MNRAS.349.1153R}
}

@ARTICLE{dauser+2014,
       author = {{Dauser}, T. and {Garcia}, J. and {Parker}, M.~L. and {Fabian}, A.~C. and {Wilms}, J.},
        title = "{The role of the reflection fraction in constraining black hole spin.}",
      journal = {\mnras},
         year = 2014,
        month = oct,
       volume = {444},
        pages = {L100-L104},
          doi = {10.1093/mnrasl/slu125},
archivePrefix = {arXiv},
       eprint = {1408.2347},
 primaryClass = {astro-ph.HE},
       adsurl = {https://ui.adsabs.harvard.edu/abs/2014MNRAS.444L.100D}
}

@ARTICLE{kaastra_bleeker,
       author = {{Kaastra}, J.~S. and {Bleeker}, J.~A.~M.},
        title = "{Optimal binning of X-ray spectra and response matrix design}",
      journal = {\aap},
         year = 2016,
        month = mar,
       volume = {587},
          eid = {A151},
        pages = {A151},
          doi = {10.1051/0004-6361/201527395},
archivePrefix = {arXiv},
       eprint = {1601.05309},
 primaryClass = {astro-ph.IM},
       adsurl = {https://ui.adsabs.harvard.edu/abs/2016A&A...587A.151K}
}

@ARTICLE{fabian+2014,
       author = {{Fabian}, A.~C. and {Parker}, M.~L. and {Wilkins}, D.~R. and {Miller}, J.~M. and {Kara}, E. and {Reynolds}, C.~S. and {Dauser}, T.},
        title = "{On the determination of the spin and disc truncation of accreting black holes using X-ray reflection}",
      journal = {\mnras},
         year = 2014,
        month = apr,
       volume = {439},
       number = {3},
        pages = {2307-2313},
          doi = {10.1093/mnras/stu045},
archivePrefix = {arXiv},
       eprint = {1401.1615},
 primaryClass = {astro-ph.HE},
       adsurl = {https://ui.adsabs.harvard.edu/abs/2014MNRAS.439.2307F}
}

@ARTICLE{fabian+2012_cygx1,
       author = {{Fabian}, A.~C. and {Wilkins}, D.~R. and {Miller}, J.~M. and {Reis}, R.~C. and {Reynolds}, C.~S. and {Cackett}, E.~M. and {Nowak}, M.~A. and {Pooley}, G.~G. and {Pottschmidt}, K. and {Sanders}, J.~S. and {Ross}, R.~R. and {Wilms}, J.},
        title = "{On the determination of the spin of the black hole in Cyg X-1 from X-ray reflection spectra}",
      journal = {\mnras},
         year = 2012,
        month = jul,
       volume = {424},
       number = {1},
        pages = {217-223},
          doi = {10.1111/j.1365-2966.2012.21185.x},
archivePrefix = {arXiv},
       eprint = {1204.5854},
 primaryClass = {astro-ph.HE},
       adsurl = {https://ui.adsabs.harvard.edu/abs/2012MNRAS.424..217F}
}

@ARTICLE{jiang_highden,
       author = {{Jiang}, Jiachen and {Fabian}, Andrew C. and {Dauser}, Thomas and {Gallo}, Luigi and {Garc{\'\i}a}, Javier A. and {Kara}, Erin and {Parker}, Michael L. and {Tomsick}, John A. and {Walton}, Dominic J. and {Reynolds}, Christopher S.},
        title = "{High Density Reflection Spectroscopy - II. The density of the inner black hole accretion disc in AGN}",
      journal = {\mnras},
         year = 2019,
        month = nov,
       volume = {489},
       number = {3},
        pages = {3436-3455},
          doi = {10.1093/mnras/stz2326},
archivePrefix = {arXiv},
       eprint = {1908.07272},
 primaryClass = {astro-ph.HE},
       adsurl = {https://ui.adsabs.harvard.edu/abs/2019MNRAS.489.3436J}
}

@ARTICLE{matt+91,
       author = {{Matt}, G. and {Perola}, G.~C. and {Piro}, L.},
        title = "{The iron line and high energy bump as X-ray signatures of cold matter in Seyfert 1 galaxies.}",
      journal = {\aap},
         year = 1991,
        month = jul,
       volume = {247},
        pages = {25},
       adsurl = {https://ui.adsabs.harvard.edu/abs/1991A&A...247...25M}
}

@ARTICLE{ogorzalek+2022,
       author = {{Ogorzalek}, A. and {King}, A.~L. and {Allen}, S.~W. and {Raymond}, J.~C. and {Wilkins}, D.~R.},
        title = "{A deep, multi-epoch Chandra HETG study of the ionized outflow from NGC 4051}",
      journal = {\mnras},
         year = 2022,
        month = nov,
       volume = {516},
       number = {4},
        pages = {5027-5051},
          doi = {10.1093/mnras/stac2389},
archivePrefix = {arXiv},
       eprint = {2208.08457},
 primaryClass = {astro-ph.HE},
       adsurl = {https://ui.adsabs.harvard.edu/abs/2022MNRAS.516.5027O}
}

@ARTICLE{cloudy,
       author = {{Ferland}, G.~J. and {Korista}, K.~T. and {Verner}, D.~A. and {Ferguson}, J.~W. and {Kingdon}, J.~B. and {Verner}, E.~M.},
        title = "{CLOUDY 90: Numerical Simulation of Plasmas and Their Spectra}",
      journal = {\pasp},
         year = 1998,
        month = jul,
       volume = {110},
       number = {749},
        pages = {761-778},
          doi = {10.1086/316190},
       adsurl = {https://ui.adsabs.harvard.edu/abs/1998PASP..110..761F}
}

@ARTICLE{cloudy2023,
       author = {{Chatzikos}, M. and {Bianchi}, S. and {Camilloni}, F. and {Chakraborty}, P. and {Gunasekera}, C.~M. and {Guzm{\'a}n}, F. and {Milby}, J.~S. and {Sarkar}, A. and {Shaw}, G. and {van Hoof}, P.~A.~M. and {Ferland}, G.~J.},
        title = "{The 2023 Release of Cloudy}",
      journal = {\rmxaa},
         year = 2023,
        month = oct,
       volume = {59},
        pages = {327-343},
          doi = {10.22201/ia.01851101p.2023.59.02.12},
archivePrefix = {arXiv},
       eprint = {2308.06396},
 primaryClass = {astro-ph.GA},
       adsurl = {https://ui.adsabs.harvard.edu/abs/2023RMxAA..59..327C}
}

@ARTICLE{mytorus,
       author = {{Murphy}, Kendrah D. and {Yaqoob}, Tahir},
        title = "{An X-ray spectral model for Compton-thick toroidal reprocessors}",
      journal = {\mnras},
         year = 2009,
        month = aug,
       volume = {397},
       number = {3},
        pages = {1549-1562},
          doi = {10.1111/j.1365-2966.2009.15025.x},
archivePrefix = {arXiv},
       eprint = {0905.3188},
 primaryClass = {astro-ph.HE},
       adsurl = {https://ui.adsabs.harvard.edu/abs/2009MNRAS.397.1549M}
}

@ARTICLE{wavelet_paper,
       author = {{Wilkins}, D.~R.},
        title = "{Wavelet spectral timing: X-ray reverberation from a dynamic black hole corona hidden beneath ultrafast outflows}",
      journal = {\mnras},
         year = 2023,
        month = dec,
       volume = {526},
       number = {3},
        pages = {3441-3460},
          doi = {10.1093/mnras/stad2936},
archivePrefix = {arXiv},
       eprint = {2309.13107},
 primaryClass = {astro-ph.HE},
       adsurl = {https://ui.adsabs.harvard.edu/abs/2023MNRAS.526.3441W}
}

@ARTICLE{kara_mcg6,
       author = {{Kara}, E. and {Fabian}, A.~C. and {Marinucci}, A. and {Matt}, G. and {Parker}, M.~L. and {Alston}, W. and {Brenneman}, L.~W. and {Cackett}, E.~M. and {Miniutti}, G.},
        title = "{The changing X-ray time lag in MCG-6-30-15}",
      journal = {\mnras},
         year = 2014,
        month = nov,
       volume = {445},
       number = {1},
        pages = {56-65},
          doi = {10.1093/mnras/stu1750},
archivePrefix = {arXiv},
       eprint = {1408.5051},
 primaryClass = {astro-ph.HE},
       adsurl = {https://ui.adsabs.harvard.edu/abs/2014MNRAS.445...56K}
}

@article{reverb_review,
	adsurl = {https://ui.adsabs.harvard.edu/abs/2014A\&ARv..22...72U},
	archiveprefix = {arXiv},
	author = {{Uttley}, P. and {Cackett}, E.~M. and {Fabian}, A.~C. and {Kara}, E. and {Wilkins}, D.~R.},
	doi = {10.1007/s00159-014-0072-0},
	eid = {72},
	eprint = {1405.6575},
	journal = {\aapr},
	month = aug,
	pages = {72},
	primaryclass = {astro-ph.HE},
	title = {{X-ray reverberation around accreting black holes}},
	volume = {22},
	year = 2014
}

@article{done_jin,
	adsurl = {https://ui.adsabs.harvard.edu/abs/2012MNRAS.420.1848D},
	archiveprefix = {arXiv},
	author = {{Done}, Chris and {Davis}, S.~W. and {Jin}, C. and {Blaes}, O. and {Ward}, M.},
	doi = {10.1111/j.1365-2966.2011.19779.x},
	eprint = {1107.5429},
	journal = {\mnras},
	month = mar,
	number = {3},
	pages = {1848--1860},
	primaryclass = {astro-ph.HE},
	title = {{Intrinsic disc emission and the soft X-ray excess in active galactic nuclei}},
	volume = {420},
	year = 2012
}

@article{1zw1_nature,
	adsurl = {https://ui.adsabs.harvard.edu/abs/2021Natur.595..657W},
	archiveprefix = {arXiv},
	author = {{Wilkins}, D.~R. and {Gallo}, L.~C. and {Costantini}, E. and {Brandt}, W.~N. and {Blandford}, R.~D.},
	doi = {10.1038/s41586-021-03667-0},
	eprint = {2107.13555},
	journal = {\nat},
	month = jul,
	number = {7869},
	pages = {657--660},
	primaryclass = {astro-ph.HE},
	title = {{Light bending and X-ray echoes from behind a supermassive black hole}},
	volume = {595},
	year = 2021
}

@article{return_radiation_paper,
	adsurl = {https://ui.adsabs.harvard.edu/abs/2020MNRAS.498.3302W},
	archiveprefix = {arXiv},
	author = {{Wilkins}, D.~R. and {Garc{\'\i}a}, J.~A. and {Dauser}, T. and {Fabian}, A.~C.},
	doi = {10.1093/mnras/staa2566},
	eprint = {2008.08083},
	journal = {\mnras},
	month = nov,
	number = {3},
	pages = {3302--3319},
	primaryclass = {astro-ph.HE},
	title = {{Returning radiation in strong gravity around black holes: reverberation from the accretion disc}},
	volume = {498},
	year = 2020
}

@article{cash,
	adsurl = {https://ui.adsabs.harvard.edu/abs/1979ApJ...228..939C},
	author = {{Cash}, W.},
	doi = {10.1086/156922},
	journal = {\apj},
	month = mar,
	pages = {939--947},
	title = {{Parameter estimation in astronomy through application of the likelihood ratio.}},
	volume = {228},
	year = 1979
}

@article{kass+1995,
	author = {Kass, Robert E. and Raftery, Adrian E.},
	issn = {01621459},
	journal = {Journal of the American Statistical Association},
	number = {430},
	pages = {773--795},
	publisher = {[American Statistical Association, Taylor \& Francis, Ltd.]},
	title = {{Bayes Factors}},
	url = {http://www.jstor.org/stable/2291091},
	volume = {90},
	year = {1995}
}

@article{dic,
	author = {Spiegelhalter, David J. and Best, Nicola G. and Carlin, Bradley P. and {Van Der Linde}, Angelika},
	doi = {10.1111/1467-9868.00353},
	eprint = {https://rss.onlinelibrary.wiley.com/doi/pdf/10.1111/1467-9868.00353},
	journal = {Journal of the Royal Statistical Society: Series B (Statistical Methodology)},
	number = {4},
	pages = {583--639},
	title = {{Bayesian measures of model complexity and fit}},
	url = {https://rss.onlinelibrary.wiley.com/doi/abs/10.1111/1467-9868.00353},
	volume = {64},
	year = {2002}
}

@article{gonzalez+2017,
	adsurl = {https://ui.adsabs.harvard.edu/abs/2017MNRAS.472.1932G},
	archiveprefix = {arXiv},
	author = {{Gonzalez}, A.~G. and {Wilkins}, D.~R. and {Gallo}, L.~C.},
	doi = {10.1093/mnras/stx2080},
	eprint = {1708.03205},
	journal = {\mnras},
	month = dec,
	number = {2},
	pages = {1932--1945},
	primaryclass = {astro-ph.HE},
	title = {{Probing the geometry and motion of AGN coronae through accretion disc emissivity profiles}},
	volume = {472},
	year = 2017
}

@article{yuan_fluxtubes_2,
	adsurl = {https://ui.adsabs.harvard.edu/abs/2019MNRAS.487.4114Y},
	archiveprefix = {arXiv},
	author = {{Yuan}, Yajie and {Spitkovsky}, Anatoly and {Blandford}, Roger D. and {Wilkins}, Dan R.},
	doi = {10.1093/mnras/stz1599},
	eprint = {1901.02834},
	journal = {\mnras},
	month = aug,
	number = {3},
	pages = {4114--4127},
	primaryclass = {astro-ph.HE},
	title = {{Black hole magnetosphere with small-scale flux tubes - II. Stability and dynamics}},
	volume = {487},
	year = 2019
}

@article{fabian+2015,
	adsurl = {https://ui.adsabs.harvard.edu/abs/2015MNRAS.451.4375F},
	archiveprefix = {arXiv},
	author = {{Fabian}, A.~C. and {Lohfink}, A. and {Kara}, E. and {Parker}, M.~L. and {Vasudevan}, R. and {Reynolds}, C.~S.},
	doi = {10.1093/mnras/stv1218},
	eprint = {1505.07603},
	journal = {\mnras},
	month = aug,
	number = {4},
	pages = {4375--4383},
	primaryclass = {astro-ph.HE},
	title = {{Properties of AGN coronae in the NuSTAR era}},
	volume = {451},
	year = 2015
}

@article{plunging_region_paper,
	adsurl = {https://ui.adsabs.harvard.edu/abs/2020MNRAS.493.5532W},
	archiveprefix = {arXiv},
	author = {{Wilkins}, D.~R. and {Reynolds}, C.~S. and {Fabian}, A.~C.},
	doi = {10.1093/mnras/staa628},
	eprint = {2003.00019},
	journal = {\mnras},
	month = apr,
	number = {4},
	pages = {5532--5550},
	primaryclass = {astro-ph.HE},
	title = {{Venturing beyond the ISCO: detecting X-ray emission from the plunging regions around black holes}},
	volume = {493},
	year = 2020
}

@article{gallo_nls1,
	adsurl = {http://adsabs.harvard.edu/abs/2006MNRAS.368..479G},
	author = {{Gallo}, L.~C.},
	doi = {10.1111/j.1365-2966.2006.10137.x},
	eprint = {astro-ph/0602145},
	journal = {\mnras},
	month = may,
	pages = {479--486},
	title = {{Investigating the nature of narrow-line Seyfert 1 galaxies with high-energy spectral complexity}},
	volume = 368,
	year = 2006
}

@article{parker_pca,
	adsurl = {http://adsabs.harvard.edu/abs/2015MNRAS.447...72P},
	archiveprefix = {arXiv},
	author = {{Parker}, M.~L. and {Fabian}, A.~C. and {Matt}, G. and {Koljonen}, K.~I.~I. and {Kara}, E. and {Alston}, W. and {Walton}, D.~J. and {Marinucci}, A. and {Brenneman}, L. and {Risaliti}, G.},
	doi = {10.1093/mnras/stu2424},
	eprint = {1411.4054},
	journal = {\mnras},
	month = feb,
	pages = {72--96},
	primaryclass = {astro-ph.HE},
	title = {{Revealing the X-ray variability of AGN with principal component analysis}},
	volume = 447,
	year = 2015
}

@article{propagating_lag_paper,
	adsurl = {http://adsabs.harvard.edu/abs/2016MNRAS.458..200W},
	archiveprefix = {arXiv},
	author = {{Wilkins}, D.~R. and {Cackett}, E.~M. and {Fabian}, A.~C. and {Reynolds}, C.~S.},
	doi = {10.1093/mnras/stw276},
	eprint = {1602.00022},
	journal = {\mnras},
	month = may,
	pages = {200--225},
	primaryclass = {astro-ph.HE},
	title = {{Towards modelling X-ray reverberation in AGN: piecing together the extended corona}},
	volume = 458,
	year = 2016
}

@article{mrk335_flare_paper,
	adsurl = {http://adsabs.harvard.edu/abs/2015MNRAS.454.4440W},
	archiveprefix = {arXiv},
	author = {{Wilkins}, D.~R. and {Gallo}, L.~C. and {Grupe}, D. and {Bonson}, K. and {Komossa}, S. and {Fabian}, A.~C.},
	doi = {10.1093/mnras/stv2130},
	eprint = {1510.07656},
	journal = {\mnras},
	month = dec,
	pages = {4440--4451},
	primaryclass = {astro-ph.HE},
	title = {{Flaring from the supermassive black hole in Mrk 335 studied with Swift and NuSTAR}},
	volume = 454,
	year = 2015
}

@article{1h0707_jan11,
	adsurl = {http://adsabs.harvard.edu/abs/2012MNRAS.419..116F},
	archiveprefix = {arXiv},
	author = {{Fabian}, A.~C. and {Zoghbi}, A. and {Wilkins}, D. and {Dwelly}, T. and {Uttley}, P. and {Schartel}, N. and {Miniutti}, G. and {Gallo}, L. and {Grupe}, D. and {Komossa}, S. and {Santos-Lle{\'o}}, M.},
	doi = {10.1111/j.1365-2966.2011.19676.x},
	eprint = {1108.5988},
	journal = {\mnras},
	month = jan,
	pages = {116--123},
	primaryclass = {astro-ph.HE},
	title = {{1H 0707-495 in 2011: an X-ray source within a gravitational radius of the event horizon}},
	volume = 419,
	year = 2012
}

@article{1h0707_var_paper,
	adsurl = {http://adsabs.harvard.edu/abs/2014MNRAS.443.2746W},
	archiveprefix = {arXiv},
	author = {{Wilkins}, D.~R. and {Kara}, E. and {Fabian}, A.~C. and {Gallo}, L.~C.},
	doi = {10.1093/mnras/stu1273},
	eprint = {1406.6658},
	journal = {\mnras},
	month = sep,
	pages = {2746--2756},
	primaryclass = {astro-ph.HE},
	title = {{Caught in the act: measuring the changes in the corona that cause the extreme variability of 1H 0707-495}},
	volume = 443,
	year = 2014
}

@article{parker_mrk335,
	adsurl = {http://adsabs.harvard.edu/abs/2014MNRAS.443.1723P},
	archiveprefix = {arXiv},
	author = {{Parker}, M.~L. and {Wilkins}, D.~R. and {Fabian}, A.~C. and {Grupe}, D. and {Dauser}, T. and {Matt}, G. and {Harrison}, F.~A.},
	doi = {10.1093/mnras/stu1246},
	eprint = {1407.8223},
	journal = {\mnras},
	month = sep,
	pages = {1723--1732},
	primaryclass = {astro-ph.HE},
	title = {{The NuSTAR spectrum of Mrk 335: extreme relativistic effects within two gravitational radii of the event horizon?}},
	volume = 443,
	year = 2014
}

@article{mrk335_corona_paper,
	author = {{Wilkins}, D.~R. and {Gallo}, L.~C.},
	doi = {10.1093/mnras/stv162},
	eprint = {1501.05302},
	journal = {\mnras},
	pages = {129--146},
	primaryclass = {astro-ph.HE},
	title = {{Driving extreme variability: the evolving corona and evidence for jet launching in Markarian 335 }},
	volume = 449,
	year = 2015
}

@article{nustar,
	adsurl = {http://adsabs.harvard.edu/abs/2013ApJ...770..103H},
	archiveprefix = {arXiv},
	author = {{Harrison}, F.~A. and {Craig}, W.~W. and {Christensen}, F.~E. and {Hailey}, C.~J. and {Zhang}, W.~W. and {Boggs}, S.~E. and {Stern}, D.},
	doi = {10.1088/0004-637X/770/2/103},
	eid = {103},
	eprint = {1301.7307},
	journal = {\apj},
	month = jun,
	pages = {103},
	primaryclass = {astro-ph.IM},
	title = {{The Nuclear Spectroscopic Telescope Array (NuSTAR) High-energy X-Ray Mission}},
	volume = 770,
	year = 2013
}

@article{garcia+2013,
	adsurl = {http://adsabs.harvard.edu/abs/2013ApJ...768..146G},
	archiveprefix = {arXiv},
	author = {{Garc{\'\i}a}, J. and {Dauser}, T. and {Reynolds}, C.~S. and {Kallman}, T.~R. and {McClintock}, J.~E. and {Wilms}, J. and {Eikmann}, W.},
	doi = {10.1088/0004-637X/768/2/146},
	eid = {146},
	eprint = {1303.2112},
	journal = {\apj},
	month = may,
	pages = {146},
	primaryclass = {astro-ph.HE},
	title = {{X-Ray Reflected Spectra from Accretion Disk Models. III. A Complete Grid of Ionized Reflection Calculations}},
	volume = 768,
	year = 2013
}

@article{ghisellini+04,
	adsurl = {http://adsabs.harvard.edu/abs/2004A\%26A...413..535G},
	author = {{Ghisellini}, G. and {Haardt}, F. and {Matt}, G.},
	doi = {10.1051/0004-6361:20031562},
	eprint = {astro-ph/0310106},
	journal = {\aap},
	month = jan,
	pages = {535--545},
	title = {{Aborted jets and the X-ray emission of radio-quiet AGNs}},
	volume = 413,
	year = 2004
}

@article{xmm,
	adsurl = {http://adsabs.harvard.edu/abs/2001A\%26A...365L...1J},
	author = {{Jansen}, F. and {Lumb}, D. and {Altieri}, B. and {Clavel}, J. and {Ehle}, M. and {Erd}, C. and {Gabriel}, C. and {Guainazzi}, M. and {Gondoin}, P. and {Much}, R. and {Munoz}, R. and {Santos}, M. and {Schartel}, N. and {Texier}, D. and {Vacanti}, G.},
	doi = {10.1051/0004-6361:20000036},
	journal = {\aap},
	month = jan,
	pages = {L1--L6},
	title = {{XMM-Newton observatory. I. The spacecraft and operations}},
	volume = 365,
	year = 2001
}

@article{suzaku,
	adsurl = {http://adsabs.harvard.edu/abs/2007PASJ...59S...1M},
	author = {{Mitsuda}, K. and {Bautz}, M. and {Inoue}, H. and {Kelley}, R.~L. and {Koyama}, K. and {Kunieda}, H. and {Makishima}, K. and {Ogawara}, Y. and {Petre}, R. and {Takahashi}, T. and {Tsunemi}, H. and {White}, N.~E.},
	doi = {10.1093/pasj/59.sp1.S1},
	journal = {\pasj},
	month = jan,
	pages = {1--7},
	title = {{The X-Ray Observatory Suzaku}},
	volume = 59,
	year = 2007
}

@article{goodman_weare,
	author = {{Goodman}, J. and {Weare}, J.},
	journal = {Comm. App. Math. and Comp. Sci},
	pages = {65},
	volume = {5},
	year = {2010}
}

@article{dauser+13,
	adsurl = {http://adsabs.harvard.edu/abs/2013MNRAS.430.1694D},
	archiveprefix = {arXiv},
	author = {{Dauser}, T. and {Garcia}, J. and {Wilms}, J. and {B{\"o}ck}, M. and {Brenneman}, L.~W. and {Falanga}, M. and {Fukumura}, K. and {Reynolds}, C.~S.},
	doi = {10.1093/mnras/sts710},
	eprint = {1301.4922},
	journal = {\mnras},
	month = apr,
	pages = {1694--1708},
	primaryclass = {astro-ph.HE},
	title = {{Irradiation of an accretion disc by a jet: general properties and implications for spin measurements of black holes}},
	volume = 430,
	year = 2013
}

@article{beloborodov,
	adsurl = {http://adsabs.harvard.edu/abs/1999ApJ...510L.123B},
	author = {{Beloborodov}, A.~M.},
	doi = {10.1086/311810},
	eprint = {astro-ph/9809383},
	journal = {\apjl},
	month = jan,
	pages = {L123--L126},
	title = {{Plasma Ejection from Magnetic Flares and the X-Ray Spectrum of Cygnus X-1}},
	volume = 510,
	year = 1999
}

@article{miniutti+04,
	adsurl = {http://adsabs.harvard.edu/abs/2004MNRAS.349.1435M},
	author = {{Miniutti}, G. and {Fabian}, A.~C.},
	doi = {10.1111/j.1365-2966.2004.07611.x},
	eprint = {astro-ph/0309064},
	journal = {\mnras},
	month = apr,
	pages = {1435--1448},
	title = {{A light bending model for the X-ray temporal and spectral properties of accreting black holes}},
	volume = 349,
	year = 2004
}

@article{dauser+10,
	adsurl = {http://adsabs.harvard.edu/abs/2010MNRAS.tmp.1460D},
	archiveprefix = {arXiv},
	author = {{Dauser}, T. and {Wilms}, J. and {Reynolds}, C.~S. and {Brenneman}, L.~W.},
	doi = {10.1111/j.1365-2966.2010.17393.x},
	eprint = {1007.4937},
	journal = {\mnras},
	month = sep,
	pages = {1460--+},
	primaryclass = {astro-ph.HE},
	title = {{Broad emission lines for a negatively spinning black hole}},
	year = 2010
}

@article{halpern-84,
	adsurl = {http://adsabs.harvard.edu/abs/1984ApJ...281...90H},
	author = {{Halpern}, J.~P.},
	doi = {10.1086/162077},
	journal = {\apj},
	month = jun,
	pages = {90--94},
	title = {{Variable X-ray absorption in the QSO MR 2251 - 178}},
	volume = 281,
	year = 1984
}

@article{haardt+91,
	adsurl = {http://adsabs.harvard.edu/abs/1991ApJ...380L..51H},
	author = {{Haardt}, F. and {Maraschi}, L.},
	doi = {10.1086/186171},
	journal = {\apjl},
	month = oct,
	pages = {L51--L54},
	title = {{A two-phase model for the X-ray emission from Seyfert galaxies}},
	volume = 380,
	year = 1991
}

@article{tanaka+95,
	adsurl = {http://adsabs.harvard.edu/abs/1995Natur.375..659T},
	author = {{Tanaka}, Y. and {Nandra}, K. and {Fabian}, A.~C. and {Inoue}, H. and {Otani}, C. and {Dotani}, T. and {Hayashida}, K. and {Iwasawa}, K. and {Kii}, T. and {Kunieda}, H. and {Makino}, F. and {Matsuoka}, M.},
	doi = {10.1038/375659a0},
	journal = {\nat},
	month = jun,
	pages = {659--661},
	title = {{Gravitationally redshifted emission implying an accretion disk and massive black hole in the active galaxy MCG-6-30-15}},
	volume = 375,
	year = 1995
}

@article{miniutti+03,
	adsurl = {http://ukads.nottingham.ac.uk/abs/2003MNRAS.344L..22M},
	author = {{Miniutti}, G. and {Fabian}, A.~C. and {Goyder}, R. and {Lasenby}, A.~N.},
	doi = {10.1046/j.1365-8711.2003.06988.x},
	eprint = {arXiv:astro-ph/0307163},
	journal = {\mnras},
	month = sep,
	pages = {L22--L26},
	title = {{The lack of variability of the iron line in MCG-6-30-15: general relativistic effects}},
	volume = 344,
	year = 2003
}

@article{ross_fabian,
	adsurl = {http://ukads.nottingham.ac.uk/abs/2005MNRAS.358..211R},
	author = {{Ross}, R.~R. and {Fabian}, A.~C.},
	doi = {10.1111/j.1365-2966.2005.08797.x},
	eprint = {arXiv:astro-ph/0501116},
	journal = {\mnras},
	month = mar,
	pages = {211--216},
	title = {{A comprehensive range of X-ray ionized-reflection models}},
	volume = 358,
	year = 2005
}

@article{fabian+89,
	adsurl = {http://ukads.nottingham.ac.uk/abs/1989MNRAS.238..729F},
	author = {{Fabian}, A.~C. and {Rees}, M.~J. and {Stella}, L. and {White}, N.~E.},
	journal = {\mnras},
	month = may,
	pages = {729--736},
	title = {{X-ray fluorescence from the inner disc in Cygnus X-1}},
	volume = 238,
	year = 1989
}

@article{george_fabian,
	adsurl = {http://ukads.nottingham.ac.uk/abs/1991MNRAS.249..352G},
	author = {{George}, I.~M. and {Fabian}, A.~C.},
	journal = {\mnras},
	month = mar,
	pages = {352--367},
	title = {{X-ray reflection from cold matter in active galactic nuclei and X-ray binaries}},
	volume = 249,
	year = 1991
}

@article{1h0707_emis_paper,
	adsurl = {http://adsabs.harvard.edu/abs/2011MNRAS.414.1269W},
	archiveprefix = {arXiv},
	author = {{Wilkins}, D.~R. and {Fabian}, A.~C.},
	doi = {10.1111/j.1365-2966.2011.18458.x},
	eprint = {1102.0433},
	journal = {\mnras},
	month = jun,
	pages = {1269--1277},
	primaryclass = {astro-ph.HE},
	title = {{Determination of the X-ray reflection emissivity profile of 1H 0707-495}},
	volume = 414,
	year = 2011
}

@article{ross_fabian_ballantyne,
	adsurl = {http://adsabs.harvard.edu/abs/2002MNRAS.336..315R},
	author = {{Ross}, R.~R. and {Fabian}, A.~C. and {Ballantyne}, D.~R.},
	doi = {10.1046/j.1365-8711.2002.05758.x},
	eprint = {arXiv:astro-ph/0206170},
	journal = {\mnras},
	month = oct,
	pages = {315--318},
	title = {{Multiple X-ray reflection from ionized slabs}},
	volume = 336,
	year = 2002
}

@article{brenneman_reynolds,
	adsurl = {http://adsabs.harvard.edu/abs/2006ApJ...652.1028B},
	author = {{Brenneman}, L.~W. and {Reynolds}, C.~S.},
	doi = {10.1086/508146},
	eprint = {arXiv:astro-ph/0608502},
	journal = {\apj},
	month = dec,
	pages = {1028--1043},
	title = {{Constraining Black Hole Spin via X-Ray Spectroscopy}},
	volume = 652,
	year = 2006
}

@article{merloni_fabian,
	adsurl = {http://adsabs.harvard.edu/abs/2001MNRAS.328..958M},
	author = {{Merloni}, A. and {Fabian}, A.~C.},
	doi = {10.1046/j.1365-8711.2001.04925.x},
	eprint = {arXiv:astro-ph/0104271},
	journal = {\mnras},
	month = dec,
	pages = {958--968},
	title = {{Thunderclouds and accretion discs: a model for the spectral and temporal variability of Seyfert 1 galaxies}},
	volume = 328,
	year = 2001
}

@article{understanding_emis_paper,
	adsurl = {http://adsabs.harvard.edu/abs/2012MNRAS.424.1284W},
	archiveprefix = {arXiv},
	author = {{Wilkins}, D.~R. and {Fabian}, A.~C.},
	doi = {10.1111/j.1365-2966.2012.21308.x},
	eprint = {1205.3179},
	journal = {\mnras},
	month = aug,
	pages = {1284--1296},
	primaryclass = {astro-ph.HE},
	title = {{Understanding X-ray reflection emissivity profiles in AGN: locating the X-ray source}},
	volume = 424,
	year = 2012
}

@inproceedings{xspec,
	adsurl = {http://adsabs.harvard.edu/abs/1996ASPC..101...17A},
	author = {{Arnaud}, K.~A.},
	booktitle = {{Astronomical Data Analysis Software and Systems V}},
	editor = {{Jacoby}, G.~H. and {Barnes}, J.},
	pages = {17},
	series = {{Astronomical Society of the Pacific Conference Series}},
	title = {{XSPEC: The First Ten Years}},
	volume = 101,
	year = 1996
}

@article{svoboda+12,
	adsurl = {http://adsabs.harvard.edu/abs/2012A\%26A...545A.106S},
	archiveprefix = {arXiv},
	author = {{Svoboda}, J. and {Dov{\v c}iak}, M. and {Goosmann}, R.~W. and {Jethwa}, P. and {Karas}, V. and {Miniutti}, G. and {Guainazzi}, M.},
	doi = {10.1051/0004-6361/201219701},
	eid = {A106},
	eprint = {1208.0360},
	journal = {\aap},
	month = sep,
	pages = {A106},
	primaryclass = {astro-ph.HE},
	title = {{Origin of the X-ray disc-reflection steep radial emissivity}},
	volume = 545,
	year = 2012
}

@article{miller+08,
	adsurl = {http://adsabs.harvard.edu/abs/2008A\%26A...483..437M},
	archiveprefix = {arXiv},
	author = {{Miller}, L. and {Turner}, T.~J. and {Reeves}, J.~N.},
	doi = {10.1051/0004-6361:200809590},
	eprint = {0803.2680},
	journal = {\aap},
	month = may,
	pages = {437--452},
	title = {{An absorption origin for the X-ray spectral variability of MCG-6-30-15}},
	volume = 483,
	year = 2008
}

@article{sunyaev_trumper,
	adsurl = {http://adsabs.harvard.edu/abs/1979Natur.279..506S},
	author = {{Sunyaev}, R.~A. and {Tr{\"u}mper}, J.},
	doi = {10.1038/279506a0},
	journal = {\nat},
	month = jun,
	pages = {506--508},
	title = {{Hard X-ray spectrum of CYG X-1}},
	volume = 279,
	year = 1979
}

@article{xillver_density,
	adsurl = {https://ui.adsabs.harvard.edu/\#abs/2016MNRAS.462..751G},
	author = {{Garc{\'\i}a}, Javier A. and {Fabian}, Andrew C. and {Kallman}, Timothy R. and {Dauser}, Thomas and {Parker}, Michael L. and {McClintock}, Jeffrey E. and {Steiner}, James F. and {Wilms}, J{\"o}rn},
	doi = {10.1093/mnras/stw1696},
	journal = {\mnras},
	month = oct,
	pages = {751--760},
	primaryclass = {astro-ph.HE},
	title = {{The effects of high density on the X-ray spectrum reflected from accretion discs around black holes}},
	volume = {462},
	year = 2016
}

@article{garcia+2014,
	adsurl = {https://ui.adsabs.harvard.edu/abs/2014ApJ...782...76G},
	archiveprefix = {arXiv},
	author = {{Garc{\'\i}a}, J. and {Dauser}, T. and {Lohfink}, A. and {Kallman}, T.~R. and {Steiner}, J.~F. and {McClintock}, J.~E. and {Brenneman}, L. and {Wilms}, J. and {Eikmann}, W. and {Reynolds}, C.~S. and {Tombesi}, F.},
	doi = {10.1088/0004-637X/782/2/76},
	eid = {76},
	eprint = {1312.3231},
	journal = {\apj},
	month = {Feb},
	number = {2},
	pages = {76},
	primaryclass = {astro-ph.HE},
	title = {{Improved Reflection Models of Black Hole Accretion Disks: Treating the Angular Distribution of X-Rays}},
	volume = {782},
	year = {2014}
}

@article{reynolds_spin_review,
	adsurl = {https://ui.adsabs.harvard.edu/abs/2021ARA\&A..59..117R},
	archiveprefix = {arXiv},
	author = {{Reynolds}, Christopher S.},
	doi = {10.1146/annurev-astro-112420-035022},
	eprint = {2011.08948},
	journal = {\araa},
	month = sep,
	pages = {117--154},
	primaryclass = {astro-ph.HE},
	title = {{Observational Constraints on Black Hole Spin}},
	volume = {59},
	year = 2021
}

@article{1zw1_flare_paper,
	adsurl = {https://ui.adsabs.harvard.edu/abs/2022MNRAS.512..761W},
	archiveprefix = {arXiv},
	author = {{Wilkins}, D.~R. and {Gallo}, L.~C. and {Costantini}, E. and {Brandt}, W.~N. and {Blandford}, R.~D.},
	doi = {10.1093/mnras/stac416},
	eprint = {2202.06958},
	journal = {\mnras},
	month = may,
	number = {1},
	pages = {761--775},
	primaryclass = {astro-ph.HE},
	title = {{Acceleration and cooling of the corona during X-ray flares from the Seyfert galaxy I Zw 1}},
	volume = {512},
	year = 2022
}
\bibliographystyle{aasjournalv7}



\end{document}